\documentclass[aps,prb,superscriptaddress,twocolumn,preprintnumbers,amsmath,amssymb,floatfix]{revtex4-1}
\usepackage{color}
\usepackage{array}
\usepackage{amsmath}
\usepackage{amssymb}
\usepackage{graphicx}
\usepackage{epstopdf}
\usepackage{hyperref}
\usepackage{float}
\usepackage{bibentry}
\usepackage{diagbox}
\usepackage{slashed}

\newcommand \be{\begin{equation}}
\newcommand \ee{\end{equation}}
\newcommand \bes{\begin{equation*}} 
\newcommand \ees{\end{equation*}}
\newcommand \bea{\begin{eqnarray}}
\newcommand \eea{\end{eqnarray}}
\newcommand \beas{\begin{eqnarray*}} 
\newcommand \eeas{\end{eqnarray*}}
\newcommand \bfg{\begin{figure}}
\newcommand \efg{\end{figure}}
\newcommand \bfgs{\begin{figure*}} 
\newcommand \efgs{\end{figure*}}
\newcommand \bwt{\begin{widetext}}
\newcommand \ewt{\end{widetext}}

\newcommand{\ii}{\mathrm{i}}
\newcommand{\TT}{\mathcal{T}}
\newcommand{\PP}{\mathcal{P}}

\begin{document}
\title{Anyon condensation and a generic tensor-network construction for symmetry protected topological phases}
\author{Shenghan Jiang}
\author{Ying Ran}

\affiliation{Department of Physics,
Boston College, Chestnut Hill, MA 02467, USA}

\date{\today}

\begin{abstract}
We present systematic constructions of tensor-network wavefunctions for bosonic symmetry protected topological (SPT) phases respecting both onsite and spatial symmetries. From the classification point of view, our results show that in spatial dimensions $d=1,2,3$, the cohomological bosonic SPT phases protected by a general symmetry group $SG$ involving onsite and spatial symmetries are classified by the cohomology group $H^{d+1}(SG,U(1))$, in which both the time-reversal symmetry and mirror reflection symmetries should be treated as anti-unitary operations. In addition, for every SPT phase protected by a discrete symmetry group and some SPT phases protected by continuous symmetry groups, generic tensor-network wavefunctions can be constructed which would be useful for the purpose of variational numerical simulations. As a by-product, our results demonstrate a generic connection between rather conventional symmetry enriched topological phases and SPT phases via an anyon condensation mechanism. 
\end{abstract}

\maketitle

\tableofcontents

\section{Introduction}
Recently the interplay between symmetry and topology in condensed matter physics attract considerable interest both theoretically and experimentally. After the discovery of topological insulators\cite{kane2005quantum,bernevig2006quantum,moore2007topological,fu2007topological,roy2009topological,Qi:2011p1057,Hasan:2010p3045}, it is theoretically recognized that there exist many new types of symmetric topological states of matter. In the absence of topological order, symmetry could protect different topological phases, which are often referred to as symmetry protected topological (SPT) phases\cite{Schuch:2011p165139,Chen:2011p235128,fidkowski2011topological,Pollmann:2012p75125,Chen:2012p1604,Chen:2011p235141,chiu2016classification}. In particular, the bosonic SPT phases require strong interactions to realize. 

Previously SPT phases have been theoretically investigated using various different theoretical frameworks\cite{Chen:2012p1604,lu2012theory,vishwanath2013physics}. In particular, a wide range of SPT phases protected by onsite symmetry groups have been systematically classified and investigated\cite{Chen:2012p1604}, based on a definition of short-range-entangled quantum phases. These SPT phases are found to be directly related to the group cohomology theory, which we will refer to as cohomological SPT phases. 

Generally in condensed matter systems spatial symmetries (e.g., lattice space group) are present. It is known that such symmetries could protect topological phases such as the topological crystalline insulators in fermionic systems\cite{chiu2013classification,fu2011topological}. In bosonic systems, analogous but correlation-driven SPT phases protected by spatial symmetries have been investigated recently, for instance, using topological field theory analysis\cite{cho2015topological,yoshida2015bosonic} and dimension reduction techniques\cite{song2016topological}. However, so far the systematic understanding of spatial-symmetry-protected SPT phases is still lacking. 

Apart from classification problems, it is certainly very important to understand whether these SPT phases can be realized in experimental systems. However, although it is known that there exist a vast number of correlation-driven SPT phases in two and higher spatial dimensions, very few of them are shown to be realized in more or less simple and realistic quantum models\cite{senthil2013integer}. 

The challenge here, at least to some extent, is due to the lack of physical guidelines and suitable numerical methods. In history, the successful discovery of topological insulators very much benefits from the band-inversion picture\cite{fu2007topological}, which is a very useful physical guideline. In this sense, it is highly desirable to develop more physical guidelines for realizing correlation-driven SPT phases. 

In addition, in order to search for SPT phases in correlated models, intensive numerical simulations are inevitable. It is also desirable to develop new numerical methods suitable for simulating SPT phases. In particular, for realistic models, one usually has to perform variational simulations based on certain choice of variational wavefunctions. Can one construct generic wavefunctions for SPT phases that are suitable for numerical simulations?

In this paper, we further develop a symmetric tensor-network theoretical framework that is powerful to address the conceptual and practical issues raised above. Let us firstly describe the results of this paper. We mainly focus on the bosonic cohomological SPT phases. The major new results of this work are two-fold. First, we identify the interpretation of cohomological SPT phases in a general tensor-network formulation, which allows us to construct generic tensor-network wavefunctions for SPT phases protected by onsite symmetries and/or spatial symmetries\,(see Sec.\ref{sec:tensor_construction}). Such generic tensor-network wavefunctions are suitable to perform variational numerical simulations in searching for SPT phases in practical model systems. Second, this interpretation shows that, for a general symmetry group $SG$, which may involve both onsite symmetries and spatial symmetries, these cohomological SPT phases can be classified by $H^{d+1}(SG,U(1))$. Here the $(d+1)$-th cohomology group $H^{d+1}(SG,U(1))$ are defined such that \emph{the time-reversal symmetry and any mirror reflection symmetries act on the $U(1)$ group in the anti-unitary fashion, while other symmetries act on the $U(1)$ group in the unitary fashion. } 

We would like to point out that the cohomological SPT phases classified by $H^{d+1}(SG,U(1))$ may or may not host gapless boundary states, related to whether one can choose a physical edge such that the symmetry protecting the SPT phase is still preserved along the boundary. For instance, in 2+1D, the inversion symmetry (equivalent to $180^\circ$ spatial rotation) generate a $Z_2$ unitary group. Because $H^3(Z_2,U(1))=Z_2$, according to our main result, there is one nontrivial SPT phase protected by inversion symmetry alone in 2+1D. However, near the edge the inversion symmetry is always broken and gapless edge states are not expected to present. This phenomenon is similar to the inversion symmetry protected topological insulators in weakly interacting fermionic systems, e.g., axion insulators\cite{Turner:2012p165120}.

Previously progresses on analytically understanding SPT phases with onsite symmetries based on the tensor-network formulation in 2+1D were made\cite{Williamson:2014p}. Comparing with earlier results, the current construction captures general spatial symmetries and applies in one, two and three spatial dimensions, and therefore is more general. In addition, in the current construction, the information of the SPT phases are encoded in certain \emph{local} constraints on the building block tensors, i.e., the local tensors are living inside certain specific sub-Hilbert spaces. Such local constraints can be easily implemented in practical numerical simulations. We will provide some concrete examples of such SPT tensor-network wavefunctions in Sec.\ref{sec:tensor_examples}.

There are several by-products of this paper that are related to the special cases of the more general results above. For instance, when $SG$ involves translation symmetries in two and higher spatial dimensions $d$, our construction related to $H^{d+1}(SG,U(1))$ clearly demonstrates so-called ``weak topological indices'', whose physical origin is related to lower dimensional SPT phases. As a concrete example, previously we demonstrated that there are 4 distinct featureless Mott insulators on the honeycomb lattice at half-filling\cite{kim2016featureless}. These distinct featureless Mott insulators now can be nicely interpreted as the consequence of two weak topological indices.

An more important by-product of this paper is a generic relation in 2+1D between the SPT phases and symmetry enriched topological (SET) phases via an anyon condensation mechanism, which provides new physical guidelines realizing SPT phases. SET phases are symmetric phases featuring topological order and anyon excitations. The interplay between symmetry and the topological order gives rise to so-called symmetry enriched phenomena such as symmetry fractionalization\cite{Wen:2002p165113,Essin:2013p104406,Mesaros:2013p155115,hung2013quantized,Hung:2013p195103,Lu:2013classification,Barkeshli:2014p,qi2016classification,Tarantino2016symmetry,teo2016globally}. 

One can consider an SET phase characterized by a usual abelian discrete gauge theory, in which gauge charges feature nontrivial symmetry fractionalizations. Such an SET phase can be quite conventional in the sense that there is no robust gapless edge states, and can be realized in rather simple model systems\cite{moessner2001resonating,balents2002fractionalization}.  It turns out that after the gauge fluxes boson-condense and destroy the topological order, the resulting confined phase must be SPT phase if the condensed gauge fluxes carry nontrivial quantum numbers and certain Criterion (see Sec.\ref{sec:anyon_cond}) is satisfied. 

This by-product signals that the traditional treatment on confinement-deconfinement phase transitions\cite{fradkin1979phase} may worth being revisited when physical symmetries are implemented. Although the general Criterion on the relation between SPT and SET phases is obtained using the tensor-network formulation in Sec.\ref{sec:tensor_construction}, a major advantage of this by-product is that it can be understood using more conventional formulations which we will discuss below.

\section{The connection between SET phases and SPT phases via anyon condensation}\label{sec:anyon_cond}
In this section we discuss a by-product of our general results obtained in Sec.\ref{sec:tensor_construction}. Instead of using tensor-network formulation, here we use (topological) field theoretical languages, which does not require the readers to be familiar with tensor-network formulations. The discussions in this section suggest that the confinement-deconfinement phase transitions of gauge theories, e.g. a usual $Z_2$ gauge theory need to be reconsidered when symmetries are present, because different ways to confine the gauge fields may lead to different SPT phases. For instance, it is well-known that valence bond solids(VBS) in quantum spin systems can be viewed as the confined phases of gauge theories. At the end of this section, we discuss the possible realizations of SPT VBS phases. 

Previously a related physical route to realize SPT phases has been discussed\cite{vishwanath2013physics,senthil2013integer,Geraedts2013288}, which states that condensing vortices in superfluid carrying $U(1)$ quantum numbers could lead to SPT phases. The current discussion can be viewed as analogous phenomena but in the context of topologically ordered phases. In addition, in the current work, general spatial and onsite symmetries are considered and systematic results are obtained.

\subsection{A criterion to generate general cohomological SPT phases via anyon condensation}
The connection between SET phases and SPT phases via anyon condensation can be quite general. In fact, the original study understanding the so-called $E_8$ state was achieved by condensing bosonic anyons coupled with multi-layers of $p+ip$ topological superconductors\cite{kitaev2006anyons}. Later on it was understood that quite systematically, starting from a fermionic SPT phase, after coupling with a dynamical gauge field and condense the appropriate bosonic anyon, one could confine the fermionic degrees of freedom and obtain a bosonic SPT phase\cite{you2015Bridging}. 

However, in those previous constructions of SPT phases, before anyon condensation, the SET phases themselves already feature gapless edge states. Indeed, before coupling to the dynamical gauge fields, the systems are already in fermionic SPT phases. In this paper, we study a different type of generic connections between SET and SPT phases via anyon condensations. Namely, the SET phases themselves contain \emph{no} symmetry protected edge states. In fact we will consider particularly simple SET phases: the usual discrete abelian gauge theories with certain symmetries. Here by ``usual'' we mean that, for instance, for a $Z_2$ gauge theory we only consider the toric-code type topological order and do not consider the double-semion topological order. At the superficial level, it is unclear how these simple SET phases are connected with SPT phases.

We will state a Criterion to obtain cohomological SPT phases via condensing (self-statistics) bosonic anyons in these simple SET phases. A proof of this Criterion based on tensor-network construction will be given in Sec.\ref{sec:tensor_anyon_condensation}. Before providing this tensor-network based argument, in Sec.\ref{sec:K_matrix} we present several examples demonstrating the application of this criterion using the $K$-matrix Chern-Simons effective theories\cite{wen1992classification}.

The topological quasiparticles in a usual $Z_n$ gauge theory include the gauge charges and the gauge fluxes, both are self-statistics bosonic. They can generate all other quasiparticles via fusion. Let's consider a $Z_{n_1}\times Z_{n_2}\times...\times Z_{n_k}$ finite abelian gauge theory, in the presence of a symmetry group $SG$ that could be a combination of onsite symmetries and spatial symmetries. In the following discussion, we denote a general gauge flux as an $m$-quasiparticle, and a general gauge charge as an $e$-quasiparticle (they do not have to be unit gauge charge/flux).  $SG$ can be a combination of onsite and spatial symmetries. It turns out that $SG$ may transform the topological quasiparticles according to certain projective representations --- a phenomenon that has been called symmetry fractionalization.

It is known that the symmetry fractionalization pattern in the above SET phase can characterized by the following mathematical expression:
\begin{align}
    \Omega_{g_1} \Omega_{g_2} = \lambda(g_1, g_2)\Omega_{g_1g_2},\label{eq:sym_frac}
\end{align}
where $g_1,g_2\in SG$, and $\Omega_{g}$ is the symmetry transformation on the quasiparticles, while $\lambda(g_1, g_2)$ is an abelian quasiparticle in the theory. Physically, it means that the operation $\Omega_{g_1} \Omega_{g_2}$ on some quasiparticle-$a$ are different from the operation $\Omega_{g_1g_2}$ on quasiparticle-$a$ by a full braiding phase between quasiparticle-$a$ and $\lambda(g_1, g_2)$. The associative condition of symmetry operations dictates the following fusing relation:
\begin{align}
 \lambda(g_1, g_2)\lambda(g_1g_2,g_3)=\lambda(g_2, g_3)\lambda(g_1,g_2g_3).\label{eq:2_cocycle}
\end{align}
Here we particularly focus on situations in which symmetry operations would \emph{not} change anyon types of $\lambda(g_1, g_2)$. Because $\Omega_{g}$ can be redefined by a braiding phase factor with a quasiparticle $b_g$, $\lambda(g_1, g_2)$ is well-defined up to a fusion with the quasiparticle $b_{g_1}b_{g_2}b_{g_1g_2}^{-1}$ (inverse means antiparticle.). Mathematically Eq.(\ref{eq:2_cocycle}) indicates that $\lambda(g_1, g_2)$ is a 2-cocycle in the second-cohomology group $H^2(SG,\mathcal{A})$, where $\mathcal{A}$ is the fusion group of the abelian quasiparticles in the SET phase. 

For instance, consider a $Z_2$ gauge theory with an onsite Ising symmetry group $Z_2^{onsite}=\{I,g\}$, in which only the $e$-particle features nontrivial symmetry fractionalization: although $g^2=I$, when acting on the $e$-particle $g(e)^2=-1$. The $-1$ phase factor here can be interpreted as the braiding phase between the $e$ particle with an $m$-particle. Consequently this SET phase can be described using the formulation in Eq.(\ref{eq:sym_frac}) by $\lambda(g, g)=m$, while all other $\lambda$'s are trivial.

Starting from the SET phase, \emph{our goal is to destroy the topological order completely by boson-condensing all the $m$-particles, while leaving the physical symmetry unbroken.} It is straightforward to show that as long as one of the condensed $m$-particles hosts non-trivial symmetry fractionalization, the $m$-condensed phase would spontaneously break the symmetry. \footnote{One way to see this is that the nontrivial projective representations can always fuse into nontrivial representations of the identity particle. Consequently one can always construct gauge invariant order parameters breaking symmetry in the boson condensed phase, if the bosons feature nontrivial symmetry fractionalization.} Therefore, in order to be able to preserve the symmetry, all the $m$-particles must have trivial symmetry fractionalization. \emph{Namely $\lambda(g_1, g_2)$ in Eq.(\ref{eq:sym_frac}) can be chosen such that all $\lambda(g_1, g_2)$ do not contain $e$-quasiparticles, while they may contain $m$-particles and their bound states} (meaning that the $e$-particles could have non-trivial symmetry fractionalization).

All the condensed $m$-quasiparticles have trivial symmetry fractionalization, but they may or may not carry non-trivial usual symmetry representations (i.e., usual quantum numbers). One may worry that condensing bosons carrying non-trivial quantum numbers would also break the physical symmetry. However, because the $m$-quasiparticles are topological excitations, symmetry breaking does not have to happen. In fact, as long as the quantum numbers carried by the condensed $m$-quasiparticles are such that the identity quasiparticles generated by fusing them (a local physical excitation) always carry trivial quantum number, the symmetry is preserved even after the $m$-condensation. 

Consequently, if we try to preserve the symmetry in the $m$-condensation, the quantum numbers carried by condensed $m$-particles cannot be arbitrary. First, they needs to be one-dimensional representations of the symmetry since higher dimensional representations can always fuse into nontrivial representations for the identity quasiparticle. Let us denote the one-dimensional representation for an $m$-quasiparticle by $\chi_m$, and $\forall g\in SG$, $\chi_m(g)\in U(1)$. We have:
\begin{align}
 \chi_m(g_1g_2)=\chi_m(g_1)\cdot \chi_m(g_2)^{s(g_1)}, \forall g_1,g_2\in SG.\label{eq:qn}
\end{align}
Here $s(g)=1$ if $g$ is a unitary symmetry and $s(g)=-1$ if $g$ is an anti-unitary symmetry.

In order to preserve symmetry in the $m$-condensate (i.e., all condensed identity particles carry trivial quantum numbers), we have the following constraint on $\chi$: if two gauge-flux quasiparticles $m$ and $m'$ fuse into the quasiparticle $m\cdot m'$, then the quantum numbers carried by all the three quasiparticles must satisfy 
\begin{align}
 \chi_m(g)\cdot \chi_{m'}(g)=\chi_{m\cdot m'}(g), \forall g\in SG.  \label{eq:preserving_symm}
\end{align}
For example, this condition dictates that $\chi_m(g)\in Z_n$ if $m$ is the gauge flux in the $Z_n$ gauge theory. 

The question is, what is the symmetric phase after the $m$-condensation?

\textbf{Criterion: } The above $m$-condensed phase is a cohomological SPT phase characterized by a 3-cocycle:
\begin{align}
 \omega_\lambda^\chi(g_1,g_2,g_3)\equiv \chi_{\lambda(g_2,g_3)}(g_1)\in H^3(SG,U(1)) \label{eq:criterion}
\end{align}

From Eq.(\ref{eq:criterion}), in order to realize a nontrivial SPT phase, two ingredients are required in this anyon-condensation mechanism: (1)the $e$-quasiparticles have some nontrivial symmetry fractionalizations so that $\lambda$'s are formed by nontrivial $m$-quasiparticles; and (2) the quantum numbers carried by the condensed $m$-particles $\chi$ are nontrivial. We will justify this Criterion using tensor-network formulation in \ref{sec:tensor_construction}. Here, let us only show three facts confirming that the Criterion is self-consistent. These facts are also useful to keep in mind in our discussions on examples.

\textbf{(i)}: $\omega_\lambda^\chi(g_1,g_2,g_3)$ is necessarily a 3-cocycle, which means that it satisfies:
\begin{align}
&\omega_\lambda^\chi(g_1g_2,g_3,g_4)\cdot\omega_\lambda^\chi(g_1,g_2,g_3g_4)\notag\\
=&\omega_\lambda^\chi(g_2,g_3,g_4)^{s(g_1)}\cdot\omega_\lambda^\chi(g_1,g_2g_3,g_4)\cdot\omega_\lambda^\chi(g_1,g_2,g_3).\label{eq:3_cocycle}
\end{align}
But this 3-cocycle condition directly follows from the fusion rule Eq.(\ref{eq:2_cocycle}), Eq.(\ref{eq:qn}), and the symmetry-preserving condition Eq.(\ref{eq:preserving_symm}).

\textbf{(ii)}: Choosing equivalent 2-cocycle $\lambda(g_1,g_2)$ in Eq.(\ref{eq:2_cocycle}) to represent the same physical symmetry fractionalization would at most modify $\omega_\lambda^\chi(g_1,g_2,g_3)$ by a 3-coboundary and thus would not change its equivalence class. This fact is straightforward to show realizing  $\lambda(g_1,g_2)$ in Eq.(\ref{eq:2_cocycle}) is well defined only up to a 2-coboundary, i.e.:
\begin{align}
 \lambda(g_1,g_2)\rightarrow \lambda(g_1,g_2)\cdot \epsilon(g_1)\cdot \epsilon(g_2)\cdot \epsilon^{-1}(g_1g_2).
\end{align}


\textbf{(iii)}: The quantum number $\chi_m(g)$ in Eq.(\ref{eq:qn}) is also well-defined up to a 1-coboundary: $\chi_m(g)\rightarrow \chi_m(g)\cdot\frac{\alpha_m^{s(g)}}{\alpha_m}$, where $\alpha_m$ like a gauge choice. It is straightforward to also show that, if this modification of $\chi_m(g)$ preserve the relation Eq.(\ref{eq:preserving_symm}), then it can only induce a change of $\omega_\lambda^\chi(g_1,g_2,g_3)$ by a 3-coboundary.

\textbf{Remark-I}: \emph{Time-reversal symmetry, mirror symmetries and the anti-unitary transformation.} The above Criterion need to be used with the following caution in mind. The Criterion has a straightforward interpretation when $SG$ only involves unitary symmetries, including usual onsite symmetries, translational/rotational spatial symmetries and their combinations. However, the time-reversal $\mathcal{T}$ and mirror symmetries $\mathcal{P}$ need to be treated as anti-unitary transformations. Namely, $s(g)=-1$ if $g=\mathcal{T}$ or $g=\mathcal{P}$. And generally if one counts the total number of $\mathcal{T}$ operation \emph{and} mirror symmetry operations in $g$, then $s(g)=-1$ iff this total number is an odd number. For instance, the product of two different mirror planes is a rotational symmetry and should be treated as a unitary transformation.

More precisely, if we consider the creation operator of an $m$-particle as $m^{\dag}\sim e^{i\phi_m}$, then in order to use the Criterion, we \emph{assume} that the transformation rules for the phase variable $\phi_m$ as: $g: \phi_m\rightarrow -\phi_m +\theta_g$ if $g=\mathcal{T}$ or $g=\mathcal{P}$, where $e^{i\theta_g}$ is a $U(1)$ phase. Because $\mathcal{T}$ involves the complex conjugation while $\mathcal{P}$ does not, this leads to: $\mathcal{T}: m^{\dag}\rightarrow e^{-i\theta_{\mathcal{T}}}m^{\dagger}$, and $\mathcal{P}: m^{\dag}\rightarrow e^{i\theta_{\mathcal{P}}}m$.

Clearly, with these transformation rules, the $\mathcal{T}$ quantum number $\chi_m(\mathcal{T})$ carried by an $m$-particle alone is only a gauge choice and is not well-defined.  But, for instance, the combination of the two transformations: $\mathcal{T}\cdot\mathcal{P}$ should be treated as a unitary transformation and its quantum number carried by an $m$-particle is well-defined.

These transformation rules can be physically interpreted as follows. In the usual discrete Abelian gauge theories, the $e$-particles and $m$-particles are dual variables, and it is a matter of choice to call which particles as gauge charges(fluxes). However, if one treats $e$'s as particles, then the $m$'s need to be treated as vortices. Under either $\mathcal{T}$ or $\mathcal{P}$, if a particle transforms into a particle (an anti-particle), then its vortex transforms into an anti-vortex (a vortex). We assign the above transformation rules for the $m$-particles in order for the $e$-particles to have well-defined symmetry fractionalizations. We will come back to this issue with a detailed field-theoretical discussion shortly in Sec.\ref{sec:K_matrix}.

\textbf{Remark-II}: \emph{Definition of quantum numbers carried by $m$-particles.} In Eq.(\ref{eq:qn},\ref{eq:criterion}) we introduce the quantum numbers carried by an $m$-particle $\chi_m(g),\forall g\in SG$. We firstly emphasize the fact that, apart from the antiunitary transformations like $\mathcal{T},\mathcal{P}$, these quantum numbers are numerically measurable for a low energy $m$-particle using tensor-network algorithms (see Sec.\ref{sec:tensor_construction} for details). However, it would be useful to sharply define these quantum numbers in a way that is independent of the tensor-network formulation. Below we provide such a definition using a symmetry-defect argument for on-site unitary symmetries only.

The subtleties to define these quantum numbers for a given $m$-particle arise from the fact that an anyon $m$ is not a local excitation. To define how an $m$-particle transforms under a symmetry $g$, one has to find a way to define an local symmetry operator $\Omega_g$ acting on a finite region $A$ covering the $m$-particle. It has been argued that \cite{Barkeshli:2014p,PhysRevX.5.041013}, \emph{for an onsite unitary $g$}, $\Omega_g$ can be interpreted as the following physical transformation of the wavefunction: (1) creating a pair of symmetry-$g$ defects; (2) adiabatically braiding one of the symmetry defect around the $m$-particle and finally annihilating with the other symmetry defect (the path of the moving symmetry defect encloses of a region $A$ covering the $m$-particle); (3) applying the symmetry transformation $g$ for the physical degrees of freedom within $A$ only. The quantum number carried by the $m$-particle is the Berry's phase accumulated over this process, relative to the Berry's phase obtained via the same process in the ground state.

The ambiguity in defining quantum numbers of the $m$-particle using the above symmetry-defect argument can now be understood. The symmetry defects created in pair may or may not contain other anyons, e.g., an $e$-particle, which have nontrivial braiding statistics with the $m$-particle being studied. Different choices of the symmetry defects used in the above process may lead to different quantum numbers due to braiding statistics between the $e$-particle in the symmetry defects and the $m$-particle being studied. Therefore, to well-define the $\chi_m(g)$ quantum number, one needs to make a particular choice of the symmetry defects. As will be proved in Sec.\ref{sec:tensor_anyon_condensation} and \ref{sec:measure_qn}, it turns out that \emph{the quantum numbers $\chi_m(g)$ in the Criterion are defined such that the symmetry defects in the above process have trivial symmetry fractionalizations.} We denote this choice of the symmetry defect as the canonical choice of symmetry defect. The canonical choice of symmetry defects rules out the possibility that the $g_1$-symmetry-defects contain extra $e$-particles having nontrivial statistics with $\lambda(g_2,g_3)$ in Eq.(\ref{eq:criterion}), and thus well-define the $\chi_{\lambda(g_2,g_3)}(g_1)$. 

However, for spatial symmetries and the time-reversal symmetry, it is unclear how to systematically create symmetry defects. For these symmetries, unfortunately we currently do not know to define the quantum numbers $\chi_m(g)$'s independent of the tensor-network formulation. We will provide the measurable meaning of these quantum numbers in the tensor-network language in Sec.\ref{sec:measure_qn}.

\subsection{Examples: anyon condensation induced SPT phases in the Chern-Simons $K$-matrix formalism}\label{sec:K_matrix}
The purpose of this subsection is to demonstrate the application of the Criterion Eq.(\ref{eq:criterion}) in some simple examples, within a convenient field-theory description: the multi-component Chern-Simons theory, or the K-matrix formulation. In particular, this formulation has been further developed by Lu and Vishwanath to successfully describe the SPT phases and their gapless edge states\cite{lu2012theory}. All the SPT phases studied here can be realized by condensing visons in a usual $Z_2$ gauge theory, which may be useful to motivate microscopic model realizations of them.

The topological Lagrangian of a general multi-component Chern-Simons theory is:
\begin{align}
 \mathcal{L}=-\frac{1}{4\pi}\sum_{I,J}K_{IJ}\epsilon^{\mu\nu\lambda}a^I_\mu\partial_{\nu}a^J_{\lambda}+\sum_I a^I_{\mu}j_I^{\mu},\label{eq:Kmatrix}
\end{align}
where $j_I^{\mu}$ for $I=1,2,..N$ are the currents of quasiparticles coupling with gauge fields $a^I_\mu$. 
For the usual $Z_2$ gauge theory, the $K$-matrix can be chosen to be: $K^{Z_2}=\begin{pmatrix}0&2\\2&0\end{pmatrix}$.

Physically, this mutual-Chern-Simons theory can be interpreted as follows. Let us start from a boson superfluid phase, formed by boson $b$, and consider the vortices. For the purpose of physical arguments below, it is convenient to introduce the boson number conservation $U(1)$ symmetry which can be removed later. The well-known boson-vortex duality states that one can describe the system as:
\begin{align}
 \mathcal{L}=-\frac{1}{2\kappa}(\epsilon^{\mu\nu\lambda}\partial_{\nu} a_{\lambda})^2- a_{\mu}j_v^{\mu},\label{eq:SF_dual}
\end{align}
where $j_v^{\mu}$ is the current of the vortices. We will use $\Psi_v$ to denote the single vortex operator. The gauge flux of $a_\mu$ is the density of the original boson $b$: $j_b^{\mu}=\frac{1}{2\pi}\epsilon^{\mu\nu\lambda}\partial_{\nu}a_{\lambda}$. In the superfluid phase the vortices are gapped and the $U(1)$ Goldstone mode is described by the photon mode of $a_{\mu}$ (i.e., the Maxwell-like dynamics in the first term in Eq.(\ref{eq:SF_dual})). 

Now let us consider the vortex condensed phase (i.e., the Mott insulator phase of the boson $b$). One way to describe the vortex condensation is to introduce an additional gauge field $a^v$ to describe the vortex current: $j_v^\mu=\frac{1}{2\pi}\epsilon^{\mu\nu\lambda}\partial_{\nu}a^v_{\lambda}$. In order to have vortex condensation captured, the dynamics of $a^v$ should be Maxwell-like. Consequently the vortex condensed phase is described by:
\begin{align}
 \mathcal{L_{\mbox{v-cond.}}}=&-\frac{1}{2\kappa}(\epsilon^{\mu\nu\lambda}\partial_{\nu} a_{\lambda})^2-\frac{1}{2\kappa^v}(\epsilon^{\mu\nu\lambda}\partial_{\nu} a^v_{\lambda})^2\notag\\
 &-\frac{1}{2\pi}\epsilon^{\mu\nu\lambda} a_{\mu}\partial_{\nu}a^v_{\lambda}
\end{align}
If one ignores the higher order Maxwell dynamics, and only focus on the topological terms, the Chern-Simons description of the vortex condensate is found to have the form of Eq.(\ref{eq:Kmatrix}) with $K^{triv.}=\begin{pmatrix}0&1\\1&0\end{pmatrix}$. The two component gauge fields can be identified: $a^1_{\mu}=a_{\mu}$ and $a^2_{\mu}=a^v_{\mu}$. Equations  of motion tell that the quasiparticle current $j_1^{\mu}$ should be identified with that of $2\pi$-$a^v_{\mu}$-flux (i.e., vortex $\Psi_v$), and the quasiparticle current $j^{\mu}_2$ is that of the $2\pi$-$a_{\mu}$-flux (the original boson $b$). As explained in Ref.\onlinecite{lu2012theory}, these quasiparticles could transform nontrivially under global symmetry, and many SPT phases can be described by this $K^{triv.}$ effective theory by demonstrating the existence of symmetry protected gapless edge states.

One can now view a $Z_2$ topologically ordered state described by $K^{Z_2}=\begin{pmatrix}0&2\\2&0\end{pmatrix}$ as an intermediate phase between the superfluid phase and the vortex condensed phase. Instead of directly condensing $\Psi_v$, one could firstly condense the double-vortices $\Psi_v^2$. Such double-vortex condensate can be again formulated by introducing the double-vortex current $j^{\mu}_{dv}=\frac{1}{2\pi}\epsilon^{\mu\nu\lambda}\partial_{\nu}a^{dv}_{\lambda}$ carrying two unit $a_{\mu}$ gauge charges (a term $-2a_{\mu}j^{\mu}_{dv}$ in the Lagrangian), and add some Maxwell dynamics for $a^{dv}$,
\begin{align}
 \mathcal{L_{\mbox{dv-cond.}}}=-\frac{1}{\pi}\epsilon^{\mu\nu\lambda} a_{\mu}\partial_{\nu}a^{dv}_{\lambda}+...
\end{align}
where $...$ include Maxwell dynamics for $a_{\mu}$ and $a^{dv}_{\mu}$. The mutual Chern-Simons term here is just the $K^{Z_2}$ in the $K$-matrix formulations. In such a gauge-charge-2 condensate, the bosonic topological quasiparticles include the unpaired single-vortex: $\Psi_v$, or the $\pi$-flux of $a^{dv}_{\mu}$  (labelled as quasiparticle-$m$), and the quantized $\pi$-flux vortex of $a_{\mu}$ (labelled as quasiparticle-$e$). Note that in this continuum theory, the $\pi$-flux and $-\pi$-flux are microscopically distinct, and we label $e^{\dag}$ as the creation operator the $\pi$-flux of $a_{\mu}$. Consequently $e$ is the operator creating the $-\pi$-flux. In addition, $e^{\dag}e^{\dag}=b^{\dag}$.

\textbf{Remark-III:} In this formulation, the relation between the symmetry transformation laws of the quasiparticles $e,m$ in the double-vortex condensate and the quasiparticles $\Psi_v, b$ the single-vortex condensate is now established: \emph{the quantum numbers carried by $\Psi_v$ is the same as those carried by $m$, and the quantum numbers carried by $b$ is \emph{twice} of those carried by $e$.} \footnote{The first half of this statement is in fact implicitly related to our definition of the quantum numbers carried by the $m$-particle as explained in Remark-II. The canonical symmetry defects in measuring these quantum numbers for onsite unitary symmetries do not contain $e$-particles, and consequently would not be affected by the confinement phase transition.}

The bulk Chern-Simons effective theory Eq.\ref{eq:Kmatrix} is accompanied with an effective edge theory:
\begin{align}
 S_{edge}=\sum_{I,J}\int\frac{dtdx}{4\pi}K_{IJ}\partial_t \phi_I\partial_x \phi_J-V_{IJ}\partial_x\phi_I\partial_x\phi_J+...\label{eq:Kmatrix_edge}
\end{align}
where the $K_{IJ}$ term is the universal Berry's phase, leading to the Kac-Moody algebra $[\partial_x\phi_I(x),\partial_y\phi_J(y)]=2\pi i K^{-1}_{IJ}\partial_x\delta(x-y)$. The $V_{IJ}$ term is non-universal and depends on details of the edge, and  ``$...$'' represents other symmetry allowed terms describing local dynamics. 

The phase variables $\phi_I$'s in Eq.(\ref{eq:Kmatrix_edge}) can be interpreted as the phases of quasiparticles: $e^{i\phi_I}$ can be identified with the quasiparticle creation operator for the current $j^{\mu}_I$ in Eq.(\ref{eq:Kmatrix}). For example, in the double-vortex condensate, one has $K=K^{Z_2}$,  $\phi_1=\phi_m$ and $\phi_2=\phi_e$, where $m^{\dagger}\sim e^{i\phi_m},e^{\dagger}\sim e^{i\phi_e}$. On the other hand, in the single-vortex condensate, we have $K=K^{triv.}$, $\phi_1=\phi_v$ and $\phi_2=\phi_b$, where $\Psi_v^{\dagger}\sim e^{i\phi_v},b^{\dagger}\sim e^{i\phi_b}$.

As explained in Ref.\onlinecite{lu2012theory,Lu:2013classification}, in the absence of symmetry, cosine terms describing local dynamics $\sum_I C_I \cos(\sum_J K_{IJ}\phi_J+\chi_I)$ are allowed in the ``$...$'' in Eq.(\ref{eq:Kmatrix_edge}) (we only consider bosonic systems in this paper). And when these terms are large, often the edge states can be fully gapped by pinning the phase variables to their classical minima. However, in the presence of symmetry,the transformation rules of $\phi_I$ sometimes dictate that the edge states can only be gapped out after spontaneously breaking the symmetry. When this happens for systems without topological order, i.e. $K=K^{triv.}$, the bulk state can be identified as an SPT phase with symmetry protected edge states. 

We will apply the Criterion Eq.(\ref{eq:criterion}) for the symmetry groups ($SG$) in Table \ref{tb:Z2_examples} in 2+1D.
\begin{table}[h]
\begin{tabular}{|c|c|}
  \hline
  $SG$&$H^3(SG,U(1))$\\
  \hline
  $Z^{onsite}_2\equiv\{I,\sigma\}$&$Z_2$\\
  \hline
  $Z^{TP}_2\equiv\{I,\mathcal{T}\cdot\mathcal{P}\}$& $Z_2$\\
  \hline
  $Z^{onsite}_2\times Z_2^T\equiv \{I,\sigma\}\times \{I,\mathcal{T}\}$& $Z_2^2$\\
  \hline
  $Z^{onsite}_2\times Z_2^P\equiv \{I,\sigma\}\times \{I,\mathcal{P}\}$& $Z_2^2$\\
  \hline
  $Z_2^{TP}\times Z_2^T\simeq Z_2^P \times Z_2^T$ & $Z_2^2$\\
  \hline
\end{tabular}
\caption{Five examples of SPT phases studied in this section.}\label{tb:Z2_examples}
\end{table}
Here $\sigma$ is an onsite unitary Ising symmetry, $\mathcal{T}$ is the time-reversal, $\mathcal{P}$ is a mirror reflection symmetry, and $\mathcal{T}\cdot\mathcal{P}$ is their combination. According to the Criterion and Remark-I, $\mathcal{T}$ and $\mathcal{P}$ should be both treated as anti-unitary, but $\mathcal{T}\cdot\mathcal{P}$ is unitary. One can see that although the $SG$'s of the former two examples (latter three examples) in Table \ref{tb:Z2_examples} are physically very different, at the mathematical group theoretical level, they are identical.

The explicit forms of the inequivalent 3-cocycles can be obtained by direct calculations. In these simple examples, it turns out that one can always choose the 3-cocycle $\omega$ such that $\omega(g_1,g_2,g_3)=-1$ for certain $g_1,g_2,g_3$, while all other $\omega(g_1,g_2,g_3)=1$. We list the nontrivial cocycles in Table \ref{tb:Z2_cocycles},\ref{tb:Z2_Z2T_cocycles}. The trivial cocycle can be chosen such that $\omega(g_1,g_2,g_3)=1$, $\forall g_1,g_2,g_3$.

\begin{table}
 \begin{tabular}{|c|c|}
 \hline
 cocycle $\omega$ & $\omega(g_1,g_2,g_3)=-1$ iff\\
 \hline
 $\omega_{1}$&$g_1=g_2=g_3=u$\\
 \hline
 \end{tabular}
 \caption{$SG=Z^{onsite}_2=\{I,\sigma\}$ or $SG=Z^{TP}_2=\{I,\mathcal{T}\cdot\mathcal{P}\}$. Denoting $Z^{onsite}_2/Z^{TP}=\{I,u\}$, two inequivalent 3-cocycles $\omega_0$(trivial) and $\omega_1$ form a $Z_2$ group.}
 \label{tb:Z2_cocycles}
\end{table}

\begin{table}
 \begin{tabular}{|c|c|}
 \hline
 cocycle $\omega$ & $\omega(g_1,g_2,g_3)=-1$ iff\\\hline
 $\omega_{[1,0]}$&$g_1,g_2,g_3$ all contain $u$\\\hline
 $\omega_{[0,1]}$&$g_1$ contains $u$ and $g_2,g_3$ both contain $\eta$.\\\hline
 $\omega_{[1,1]}$&$g_1$ contains $u$ and $g_2,g_3$ both contain \\
 &either $u$ or $\eta$ except for $g_2=g_3=u\cdot\eta$. \\\hline
 \end{tabular}
 \caption{$SG=Z^{onsite}_2\times Z_2^T$, or $SG=Z^{onsite}_2\times Z_2^P$, or $SG=Z^{TP}_2\times Z_2^T$. Denoting $Z^{onsite}_2/Z^{TP}=\{I,u\}$ and $Z_2^T$/$ Z_2^P=\{I,\eta\}$, the four inequivalent 3-cocycles $\omega_{[0,0]}$(trivial), $\omega_{[1,0]}$, $\omega_{[0,1]}$, $\omega_{[1,1]}$ form a $Z_2^2$ group. Note that $u$ is a unitary transformation and $\eta$ is an anti-unitary transformation.}
 \label{tb:Z2_Z2T_cocycles}
\end{table}


\textbf{Remark-IV:} \emph{time-reversal and mirror symmetries} In order for the 2-component mutual Chern-Simons theories of either $K^{triv.}$ or $K^{Z_2}$ to be symmetric under $\mathcal{T}$ or $\mathcal{P}$, it is required that the $a^1_{\mu}$ and $a^2_{\mu}$ to transform oppositely under these symmetries. Consequently, denoting the densities of the two types of quasiparticles coupled with $a^1_{\mu}$($a^2_{\mu}$) as $\rho_1$($\rho_2$), if one has $\mathcal{T}:\rho_1\rightarrow\rho_1$ ($\mathcal{P}:\rho_1\rightarrow\rho_1$), one must also have $\mathcal{T}:\rho_2\rightarrow-\rho_2$ ($\mathcal{P}:\rho_2\rightarrow-\rho_2$), and vice versa. 

For instance, if one requires $\mathcal{P}: e^{\dagger}\rightarrow e^{i\alpha_e} e^{\dagger}$ , then $\mathcal{P}: m^{\dagger}\rightarrow e^{i\alpha_m} m$, where $e^{i\alpha_e},e^{i\alpha_m}$ are phase factors. After choosing a $\mathcal{P}$ symmetric edge along the $x$-direction, these leads to the following rules in the effective theory Eq.(\ref{eq:Kmatrix_edge}): $\mathcal{P}: \phi_e(t,x,y)\rightarrow \phi_e(t,-x,y)+\alpha_e; \phi_m(t,x,y)\rightarrow -\phi_m(t,-x,y)+\alpha_m$. As discussed in Remark-I, to use the Criterion, we always require that under either $\mathcal{P}$ or $\mathcal{T}$, $\phi_m$ flips sign but $\phi_e$ does not.

All SPT phase examples discussed in this section can be realized via the anyon condensation Criterion starting from a SET phase with usual $Z_2$ topological order. Our strategy is two-step. For a given SPT 3-cocycle $\omega(g_1,g_2,g_3)$, using the Criterion, we look for the $Z_2$ topologically ordered SET phase with desired symmetry properties $\chi_m(g_1)$ and $\lambda(g_2,g_3)$. Second, we condense the $m$-particle and demonstrate the resulting phase is indeed an SPT phase by studying its edge effective theory Eq.(\ref{eq:Kmatrix_edge}).
\subsubsection{$SG=Z^{onsite}_2$}
As the simplest example of the Criterion, let us consider the SPT phase corresponds to the 3-cocycle $\omega_1$ for $SG=Z^{onsite}_2=\{I,g\}$ in Table \ref{tb:Z2_cocycles}. The desired $Z_2$ topologically ordered SET phase can be easily identified:
\begin{align}
 &\chi_{\lambda(g_2,g_3)}(g_1)=\omega_1(g_1,g_2,g_3)\notag\\
 \Rightarrow\;\;
 &\chi_m(g)=-1, \lambda(g,g)=m,
\end{align}
while all other $\chi,\lambda$'s are trivial. Namely this is an SET phase in which the gauge charge $e$ features nontrivial symmetry fractionalization: $g(e)^2=-1$, and the gauge flux $m$ has no nontrivial symmetry fractionalization but carries a nontrivial Ising quantum number $\chi_m(g)=-1$.

These symmetry transformation properties can be implemented in the $K$-matrix formulation with $K=K^{Z_2}$ and $g: m^{\dagger}\rightarrow -m^{\dagger};e^{\dagger}\rightarrow i\cdot e^{\dagger}$. In the corresponding edge theory Eq.(\ref{eq:Kmatrix_edge}), these lead to:
\begin{align}
 g:\phi_m\rightarrow \phi_m+\pi;\ \;\;\phi_e\rightarrow \phi_e+\pi/2
\end{align}

In this SET phase, it is perfectly fine to have a gapped edge without breaking physical symmetry. For example, symmetry allows $C\cdot\cos(2\phi_m+\chi_m)$ term in the ``$...$''. When this term is large enough the edge states will be gapped out by pinning $2\phi_m$ to a semiclassical minimum, which does \emph{not} break the physical symmetry. Note that $e^{i\phi_m}$ itself is an anyon operator and does not correspond to a local order parameter. 

Next, we condense the $m$-particles (the remaining single-vortices) to destroy the topological order without breaking the symmetry. The resulting single-vortex condensate is described by $K=K^{triv.}$. According to Remark-III, we have $g: \Phi_v^{\dagger}\rightarrow -\Phi_v^{\dagger};b^{\dagger}\rightarrow -b^{\dagger}$. In the corresponding edge theory Eq.(\ref{eq:Kmatrix_edge}), these lead to: 
\begin{align}
 g: \phi_v\rightarrow \phi_v+\pi; \;\;\phi_b\rightarrow \phi_b+\pi.\label{eq:Z2_SPT_edge_rules}
\end{align}
This is exactly the symmetry properties of the $Z_2^{onsite}$ SPT phase studied in Ref.\onlinecite{lu2012theory}, where it is shown that it is impossible to gap out the edge states without spontaneously breaking the $Z_2^{onsite}$ symmetry. In Ref.\onlinecite{lu2012theory}, Eq.(\ref{eq:Z2_SPT_edge_rules}) was obtained by systematically investigating all possible self-consistent transformation rules and searching for symmetry protected gapless edge states. But here, with the help of the Criterion and knowledge of the 3-cocycle $\omega_1$, Eq.(\ref{eq:Z2_SPT_edge_rules}) is directly obtained. These results are summarized in Table \ref{tb:Z2_SPT}.

\begin{table}
 \begin{tabular}{|c|c|c|}
  \hline
  3-cocycle&SET bulk&SPT edge\\\hline
  $\omega_1$
  &$\begin{aligned}[t] g: m^{\dagger}&\rightarrow -m^{\dagger}\\
  e^{\dagger}&\rightarrow i\cdot e^{\dagger}\end{aligned}$
  &$\begin{aligned}[t]g: \phi_v&\rightarrow \phi_v+\pi\\
    \phi_b&\rightarrow \phi_b+\pi\end{aligned}$
    \\\hline
 \end{tabular}
 \caption{The symmetry properties of the nontrivial SPT phase protected by $SG=Z^{onsite}_2=\{I,g\}$, and the SET phase before the anyon condensation.}
 \label{tb:Z2_SPT}
\end{table}

\subsubsection{$SG=Z^{onsite}_2\times Z_2^T$}
There are three nontrivial cohomological SPT phases protected by $SG=Z^{onsite}_2\times Z_2^T=\{I,g\}\times\{I,\mathcal{T}\}$, whose corresponding nontrivial 3-cocycles are listed in Table \ref{tb:Z2_Z2T_cocycles}. We discuss them separately:

\textbullet $\omega_{[1,0]}$: We need $\chi_m(g)=-1$ and $\lambda(g,g)=m$ in the SET phase (all other $\lambda$'s are trivial). After condensing $m$-particles gapless edge states are protected by $g$ alone, as already discussed in Eq.(\ref{eq:Z2_SPT_edge_rules}).

\textbullet $\omega_{[0,1]}$: We again need an SET phase with $\chi_m(g)=-1$, but $\lambda(\mathcal{T},\mathcal{T})=m$ (all other $\lambda$'s are trivial). The latter condition dictates that the $e$-particles are Kramer doublets because they form projective representations under time reversal: $\mathcal{T}(e)^2=-1$. The symmetry transformation rules in the bulk effective theory can be implemented as: $g: m^{\dagger}\rightarrow -m^{\dagger};\;e^{\dagger}\rightarrow e^{\dagger}$, while
 $\mathcal{T}:m^{\dagger}\rightarrow m^{\dagger};\;e^{\dagger}\rightarrow -i\cdot e.$. In the corresponding edge theory:
 \begin{align}
  g: & \phi_m\rightarrow\phi_m+\pi;\;\;\phi_e\rightarrow \phi_e,\notag\\
  \mathcal{T}:&\phi_m\rightarrow -\phi_m;\;\;\phi_e\rightarrow \phi_e+\pi/2.
 \end{align}
More precisely, for example, the first rule should be interpreted as $\phi_m(t,x,y)\rightarrow\phi_m(-t,x,y,)+\pi$ and we have been ignoring the space-time coordinates to save notations. After condensing $m$-particles, the resulting phase is described by $K=K^{triv.}$ with the following symmetry transformations on the edge degrees of freedom:
\begin{align}
g: & \phi_v\rightarrow\phi_v+\pi;\;\;\phi_b\rightarrow \phi_b,\notag\\
  \mathcal{T}:&\phi_v\rightarrow -\phi_v;\;\;\phi_b\rightarrow \phi_b+\pi.
\end{align}
Clearly the cosine terms $\cos(\phi_v+\chi_v)$ and $\cos(\phi_b+\chi_b)$ are not allowed by symmetry and gapless edge states are protected. This is indeed the symmetry properties of another $SG=Z^{onsite}_2\times Z_2^T$ SPT phase studied in Ref.\onlinecite{lu2012theory}. 

\textbullet $\omega_{[1,1]}$: We need an SET phase in which $\chi_m(g)=-1$, and both $\lambda(g,g)=\lambda(\mathcal{T},\mathcal{T})=m$ (i.e. both $g(e)^2=\mathcal{T}(e)^2=-1$). In the edge theory of this SET phase:
\begin{align}
  g: & \phi_m\rightarrow\phi_m+\pi;\;\;\phi_e\rightarrow \phi_e+\pi/2,\notag\\
  \mathcal{T}:&\phi_m\rightarrow -\phi_m;\;\;\phi_e\rightarrow \phi_e+\pi/2.
 \end{align}
After condensing $m$-particles, the resulting phase is described by $K=K^{triv.}$ with the following symmetry transformations on the edge degrees of freedom:
\begin{align}
g: & \phi_v\rightarrow\phi_v+\pi;\;\;\phi_b\rightarrow \phi_b+\pi,\notag\\
  \mathcal{T}:&\phi_v\rightarrow -\phi_v;\;\;\phi_b\rightarrow \phi_b+\pi.
\end{align}
The edge theory of this SPT phase was also pointed out in Ref.\onlinecite{lu2012theory}. Again in the current paper, using the Criterion, all these SPT phases are directly obtained. The results of this part are summarized in Table \ref{tb:Z2_Z2T_SPT}.

\begin{table}
 \begin{tabular}{|c|c|c|}
  \hline
  3-cocycle&SET bulk&SPT edge\\\hline
  $\omega_{[1,0]}$
  &$\begin{aligned}[t] g: m^{\dagger}&\rightarrow -m^{\dagger}\\
  e^{\dagger}&\rightarrow i\cdot e^{\dagger}\\
  \mathcal{T}:m^{\dagger}&\rightarrow m^{\dagger}\\
  e^{\dagger}&\rightarrow e\end{aligned}$
  &$\begin{aligned}[t]g: \phi_v&\rightarrow \phi_v+\pi\\
    \phi_b&\rightarrow \phi_b+\pi\\
     \mathcal{T}:\phi_v&\rightarrow -\phi_v\\
     \phi_b&\rightarrow \phi_b
    \end{aligned}$
    \\\hline
  $\omega_{[0,1]}$
  &$\begin{aligned}[t] g: m^{\dagger}&\rightarrow -m^{\dagger}\\
  e^{\dagger}&\rightarrow e^{\dagger}\\
  \mathcal{T}:m^{\dagger}&\rightarrow m^{\dagger}\\
  e^{\dagger}&\rightarrow -i\cdot e\end{aligned}$
  &$\begin{aligned}[t]g: \phi_v&\rightarrow \phi_v+\pi\\
    \phi_b&\rightarrow \phi_b\\
     \mathcal{T}:\phi_v&\rightarrow -\phi_v\\
     \phi_b&\rightarrow \phi_b+\pi
    \end{aligned}$
  \\\hline
  $\omega_{[1,1]}$
  &$\begin{aligned}[t] g: m^{\dagger}&\rightarrow -m^{\dagger}\\
  e^{\dagger}&\rightarrow i\cdot e^{\dagger}\\
  \mathcal{T}:m^{\dagger}&\rightarrow m^{\dagger}\\
  e^{\dagger}&\rightarrow -i\cdot e \end{aligned}$
  &$\begin{aligned}[t]g: \phi_v&\rightarrow \phi_v+\pi\\
    \phi_b&\rightarrow \phi_b+\pi\\
     \mathcal{T}:\phi_v&\rightarrow -\phi_v\\
     \phi_b&\rightarrow \phi_b+\pi
    \end{aligned}$
  \\\hline
 \end{tabular}
\caption{The symmetry properties of the three nontrivial SPT phases protected by $SG=Z^{onsite}_2\times Z_2^T=\{I,g\}\times\{I,\mathcal{T}\}$, together with those of the corresponding SET phases before anyon condensations.}\label{tb:Z2_Z2T_SPT}
\end{table}

\subsubsection{$SG=Z^{onsite}_2\times Z_2^P$}
Again there are three nontrivial cohomological SPT phases as listed in Table \ref{tb:Z2_Z2T_cocycles}. Because the analysis is similar to the previous case, we only list the results in Table \ref{tb:Z2_Z2P_SPT}. Note that we will choose a $\mathcal{P}$ symmetric edge along the $x$-direction, and will again ignore the space-time coordinates to save notations: e.g., $\mathcal{P}: \phi\rightarrow \pm\phi+\alpha$ really means $\mathcal{P}: \phi(t,x,y)\rightarrow \pm\phi(t,-x,y)+\alpha$. We find that the three nontrivial SPT phases obtained here are consistent with earlier results in Ref.\onlinecite{yoshida2015bosonic} obtained by directly studying the symmetry transformations in the $K^{triv}$ effective theory without resorting to group cohomology.

\begin{table}
 \begin{tabular}{|c|c|c|}
  \hline
  3-cocycle&SET bulk&SPT edge\\\hline
  $\omega_{[1,0]}$
  &$\begin{aligned}[t] g: m^{\dagger}&\rightarrow -m^{\dagger}\\
  e^{\dagger}&\rightarrow i\cdot e^{\dagger}\\
  \mathcal{P}:m^{\dagger}&\rightarrow m\\
  e^{\dagger}&\rightarrow e^{\dagger}\end{aligned}$
  &$\begin{aligned}[t]g: \phi_v&\rightarrow \phi_v+\pi\\
    \phi_b&\rightarrow \phi_b+\pi\\
     \mathcal{P}:\phi_v&\rightarrow -\phi_v\\
     \phi_b&\rightarrow \phi_b
    \end{aligned}$
    \\\hline
  $\omega_{[0,1]}$
  &$\begin{aligned}[t] g: m^{\dagger}&\rightarrow -m^{\dagger}\\
  e^{\dagger}&\rightarrow e^{\dagger}\\
  \mathcal{P}:m^{\dagger}&\rightarrow m\\
  e^{\dagger}&\rightarrow i\cdot e^{\dagger}\end{aligned}$
  &$\begin{aligned}[t]g: \phi_v&\rightarrow \phi_v+\pi\\
    \phi_b&\rightarrow \phi_b\\
     \mathcal{P}:\phi_v&\rightarrow -\phi_v\\
     \phi_b&\rightarrow \phi_b+\pi
    \end{aligned}$
  \\\hline
  $\omega_{[1,1]}$
  &$\begin{aligned}[t] g: m^{\dagger}&\rightarrow -m^{\dagger}\\
  e^{\dagger}&\rightarrow i\cdot e^{\dagger}\\
  \mathcal{P}:m^{\dagger}&\rightarrow m\\
  e^{\dagger}&\rightarrow i\cdot e^{\dagger}\end{aligned}$
  &$\begin{aligned}[t]g: \phi_v&\rightarrow \phi_v+\pi\\
    \phi_b&\rightarrow \phi_b+\pi\\
     \mathcal{P}:\phi_v&\rightarrow -\phi_v\\
     \phi_b&\rightarrow \phi_b+\pi
    \end{aligned}$
  \\\hline
 \end{tabular}
\caption{The symmetry properties of the three nontrivial SPT phases protected by $SG=Z^{onsite}_2\times Z_2^P=\{I,g\}\times\{I,\mathcal{P}\}$, together with those of the corresponding SET phases before anyon condensations.}\label{tb:Z2_Z2P_SPT}
\end{table}

\subsubsection{$SG=Z^{TP}_2\times Z_2^T\simeq Z^{P}_2\times Z_2^T$ and $SG=Z^{TP}_2$}
As mentioned before, both $\mathcal{T},\mathcal{P}$ send $\phi_m$ to $-\phi_m$ up to phase shifts. These phase shifts are changing under gauge transformation  $\phi_m\rightarrow \phi_m+\delta$ and are not well-defined. But their combination $\mathcal{T}\cdot\mathcal{P}$ should be treated as a unitary transformation sending $\phi_m$ to $\phi_m$ up to a well-defined phase shift, whose possible values are limited to $0$ and $\pi$ since $(\mathcal{T}\cdot\mathcal{P})^2=I$ assuming $m$-particles have trivial symmetry fractionalization. Using the anyon condensation mechanism (the Criterion) and the cocycles listed in Table \ref{tb:Z2_Z2T_cocycles} and Table \ref{tb:Z2_cocycles}, one can straightforwardly obtain the three nontrivial SPT phases protected by $SG=Z^{TP}_2\times Z_2^T\simeq Z^{P}_2\times Z_2^T$ and the one nontrivial SPT phase protected by $SG=Z^{TP}_2=\{I,\mathcal{T}\cdot\mathcal{P}\}$. After choosing a $\mathcal{P}$ symmetric edge along the $x$-direction, we list the results in Table \ref{tb:Z2P_Z2T_SPT} and \ref{tb:Z2TP_SPT}. One can easily check that indeed the cosine terms $\cos(\phi_v+\chi_v)$ or $\cos(\phi_b+\chi_b)$ are forbidden by symmetry, and the symmetry allowed terms like $\cos(2\phi_v+\chi_v)$ or $\cos(2\phi_b+\chi_b)$ would spontaneously break the symmetry after gapping out the edge modes. These SPT phases, to our knowledge, have not been pointed out before.

\begin{table}
 \begin{tabular}{|c|c|c|}
  \hline
  3-cocycle&SET bulk&SPT edge\\\hline
  $\omega_{[1,0]}$
  &$\begin{aligned}[t] \mathcal{P}: m^{\dagger}&\rightarrow -m\\
  e^{\dagger}&\rightarrow i\cdot e^{\dagger}\\
  \mathcal{T}:m^{\dagger}&\rightarrow m^{\dagger}\\
  e^{\dagger}&\rightarrow e\end{aligned}$
  &$\begin{aligned}[t]\mathcal{P}: \phi_v&\rightarrow -\phi_v+\pi\\
    \phi_b&\rightarrow \phi_b+\pi\\
     \mathcal{T}:\phi_v&\rightarrow -\phi_v\\
     \phi_b&\rightarrow \phi_b
    \end{aligned}$
    \\\hline
  $\omega_{[0,1]}$
  &$\begin{aligned}[t] \mathcal{P}: m^{\dagger}&\rightarrow -m\\
  e^{\dagger}&\rightarrow i\cdot e^{\dagger}\\
  \mathcal{T}:m^{\dagger}&\rightarrow m^{\dagger}\\
  e^{\dagger}&\rightarrow -i\cdot e\end{aligned}$
  &$\begin{aligned}[t]\mathcal{P}: \phi_v&\rightarrow -\phi_v+\pi\\
    \phi_b&\rightarrow \phi_b+\pi\\
     \mathcal{T}:\phi_v&\rightarrow -\phi_v\\
     \phi_b&\rightarrow \phi_b+\pi
    \end{aligned}$
  \\\hline
  $\omega_{[1,1]}$
  &$\begin{aligned}[t] \mathcal{P}: m^{\dagger}&\rightarrow -m\\
  e^{\dagger}&\rightarrow  e^{\dagger}\\
  \mathcal{T}:m^{\dagger}&\rightarrow m^{\dagger}\\
  e^{\dagger}&\rightarrow -i\cdot e \end{aligned}$
  &$\begin{aligned}[t]\mathcal{P}: \phi_v&\rightarrow -\phi_v+\pi\\
    \phi_b&\rightarrow \phi_b\\
     \mathcal{T}:\phi_v&\rightarrow -\phi_v\\
     \phi_b&\rightarrow \phi_b+\pi
    \end{aligned}$
  \\\hline
 \end{tabular}
\caption{The symmetry properties of the three nontrivial SPT phases protected by $SG=Z^{TP}_2\times Z_2^T\simeq Z^{P}_2\times Z_2^T=\{I,\mathcal{P}\}\times\{I,\mathcal{T}\}$, together with those of the corresponding SET phases before anyon condensations.}
\label{tb:Z2P_Z2T_SPT}
\end{table}

\begin{table}
 \begin{tabular}{|c|c|c|}
  \hline
  3-cocycle&SET bulk&SPT edge\\\hline
  $\omega_1$
  &$\begin{aligned}[t] \mathcal{T}\cdot\mathcal{P}: m^{\dagger}&\rightarrow -m^{\dagger}\\
  e^{\dagger}&\rightarrow -i\cdot e\end{aligned}$
  &$\begin{aligned}[t]\mathcal{T}\cdot\mathcal{P}: \phi_v&\rightarrow \phi_v+\pi\\
    \phi_b&\rightarrow \phi_b+\pi\end{aligned}$
    \\\hline
 \end{tabular}
 \caption{The symmetry properties of the nontrivial SPT phase protected by $SG=Z^{TP}_2$, and the SET phase before the anyon condensation.}
 \label{tb:Z2TP_SPT}
\end{table}

\subsection{Possible realizations --- SPT Valence Bond Solids}
Valence Bond Solids(VBS) can be realized in quantum spin-$1/2$ model systems\cite{PhysRevLett.62.1694,moessner2001resonating,Senthil:2004p1490,PhysRevLett.98.227202}. They spontaneously break the lattice translational symmetry but preserve the spin-rotational symmetry/time-reversal symmetry. The characteristic of a VBS phase is the long-range bond-bond correlation function. It is quite popular to visualize these phases as if the neighboring spin-$1/2$'s form static spin-singlet valence bond patterns, which suggests that they may be adiabatically connected to a limit in which the global wavefunctions are simply direct products of all the valence bonds. 

However , from a general point of view, this picture of VBS may be misleading: the long-range bond-bond correlation function does not imply that the wavefunction can be always adiabatically connected to a direct product state. Motivated by the examples studied in Table \ref{tb:Z2P_Z2T_SPT}, below we propose new types of SPT-VBS phases protected by a mirror symmetry $\mathcal{P}$ and the time-reversal symmetry $\mathcal{T}$. In fact, it is even unclear whether these SPT-VBS phases are already realized in existing models featuring VBS phases. 

One could understand a VBS phase in spin models with a half-integer spin per unit-cell by starting from a $Z_2$ quantum spin liquid(QSL) phase. Quite generally, in a $Z_2$ QSL, the $e$-particles are the Kramer-doublet spinons, and the $m$-particles are the spinless visons. Namely the fact that the $e$-particles are Kramer-doublets basically comes for free. It is well-known that the half-integer spin per unit-cell would dictate that the visons have nontrivial translational symmetry fractionalization. Consequently condensing the visons would break translational symmetry but preserve the spin-rotational symmetry, resulting in a VBS phase. But the VBS phase can be still symmetric under certain mirror reflection. For instance, the columnar VBS pattern on the square lattice is symmetric under the mirror reflection around the line crossing the bond centers along a column. The vison would certainly have trivial symmetry fractionalization under the $\mathcal{T}$ and $\mathcal{P}$ defined here.

Let us particularly pay attention to the two SPT phases characterized by $\omega_{[0,1]}$ and $\omega_{[1,1]}$ in Table \ref{tb:Z2P_Z2T_SPT}. Before the $m$-particle condensation, the corresponding two SET phases both have Kramer-doublet $e$-particles, and their difference lies in the presence/absence of symmetry fractionalization of $\mathcal{P}$. In both case, one could realize the corresponding SPT phases by condensing the $m$-particle (vison) which is odd under the combination $\mathcal{T}\cdot\mathcal{P}$: $m^{\dagger}\rightarrow -m$. 

Namely, whether the topological trivial VBS or the SPT-VBS is realized completely depends on which vison is condensed: the $\mathcal{T}\cdot\mathcal{P}$ even vison or the $\mathcal{T}\cdot\mathcal{P}$ odd vison. This is an energetic question and one need to numerically measure this quantum number for the low energy visons near the condensation. However, as mentioned before, such measurement is nontrivial to perform and we currently only know how to do it using tensor-network-based algorithms (see Sec.\ref{sec:tensor_anyon_condensation} for details).

Note that although we propose the SPT-VBS phases using the anyon-condensation mechanism from $Z_2$ QSLs, one does not have to realize the $Z_2$ QSL in spin models in order to realize the SPT-VBS phases. The anyon-condensation mechanism is simply one route to ensure that SPT-VBS phase can be obtained. As stable phases, SPT-VBS phases may be obtained via other routes\footnote{for instance, the VBS phase in the context of the easy-plane deconfined criticality is obtained by condensing magnetic vortices coupling with a $U(1)$ gauge field. It would be interesting to understand whether the $\mathcal{T}\cdot\mathcal{P}$ quantum number discussed here can be generalized to these vortex-like objects.}, or even first-order phase transitions, which do not involve QSLs.

\section{Symmetric tensor-network constructions in 2+1D}\label{sec:2D_tensor_SPT}
In this section, we develop a general formulation to construct/classify 2+1D cohomological bosonic SPT phases protected by both on-site symmetries as well as spatial symmetries by Projected Entangled Pair States\,(PEPS). For each class we provide generic tensor wavefunctions, which are useful for numerical simulations.

\subsection{A simple example: $Z_2$ SPT}\label{sec:Z2_SPT_example}
Before developing a general formulation, we will study a simple example: the SPT phase protected by onsite $Z_2$ symmetry\cite{Chen:2011p235141}.

\begin{figure}
    \includegraphics[width=0.45\textwidth]{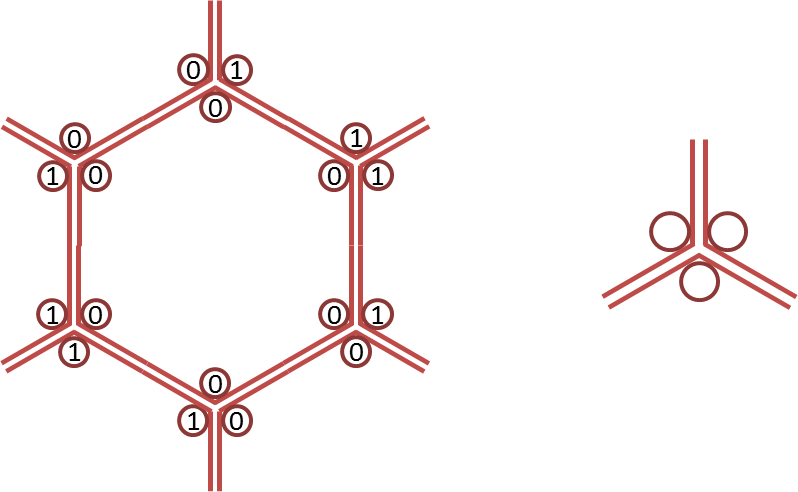}
    \caption{The $Z_2$ symmetric wavefunction on the honeycomb lattice. Each site contains three qubits. The six qubits around each plaquette are all in the same spin state. The $Z_2$ symmetry flips spins, which acts as $\sigma_x$.}
    \label{fig:honeycomb_Z2_PEPS}
\end{figure}

Let us first focus on the fixed point wavefunction of the nontrivial $Z_2$ SPT phase. Here, we follow the convention in Ref.\onlinecite{chen2015towards}. The system lives on a honeycomb lattice, where each lattice site contains three qubits, as shown in Fig.\,\ref{fig:honeycomb_Z2_PEPS} as three circles.  The six spin $\frac{1}{2}$'s around a plaquette are either all in the $|0\rangle$ state or all in the $|1\rangle$ state, forming $Z_2$ domains. The fixed point wavefunction for the nontrivial $Z_2$ SPT phase is
\begin{align}
    |\psi\rangle=\sum_\mathcal{C}(-1)^{N_\mathcal{C}}|\mathcal{C}\rangle
    \label{eq:Z2_SPT_wavefunction}
\end{align}
where $\mathcal{C}$ denotes $Z_2$ domain configurations and $N_\mathcal{C}$ is the number of domain walls in $\mathcal{C}$.

\begin{figure}
    \includegraphics[width=0.45\textwidth]{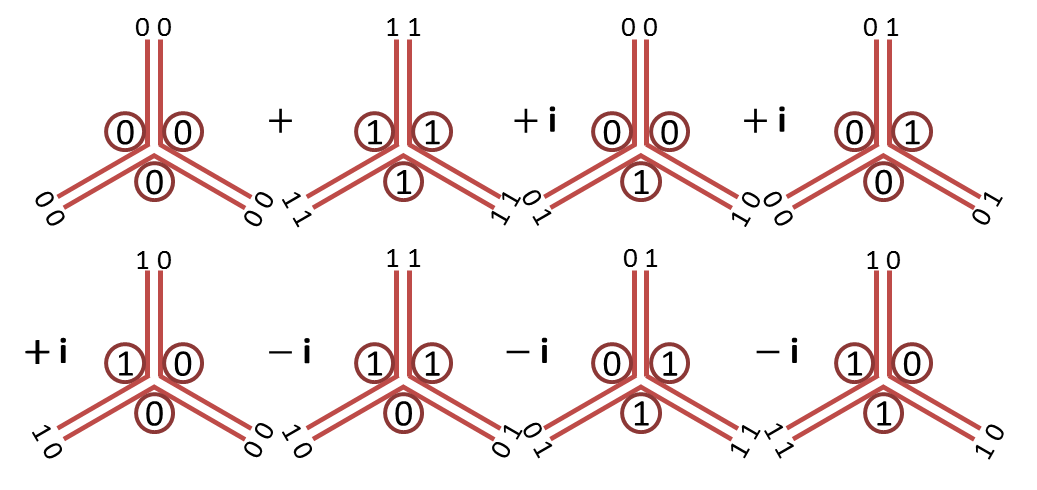}
    \caption{The tensor state representing the nontrivial $Z_2$ SPT wavefunction defined in Eq.(\ref{eq:Z2_SPT_wavefunction}). An internal leg support two dimensional Hilbert space. Physical states are labeled by numbers in the circle, while virtual states are labeled by numbers at the end of internal legs. }
    \label{fig:honeycomb_Z2_SPT_tensors}
\end{figure}

The nontrivial SPT state can be represented with tensors given in Fig. \ref{fig:honeycomb_Z2_SPT_tensors}. A site tensor has six internal\,(virtual) legs, where each internal leg represents a qubit. Here, we choose tensors to be the same for both sub-lattices. One can easily check that the tensor network state indeed represents the wavefunction defined in Eq.(\ref{eq:Z2_SPT_wavefunction}).

\begin{figure}
    \includegraphics[width=0.45\textwidth]{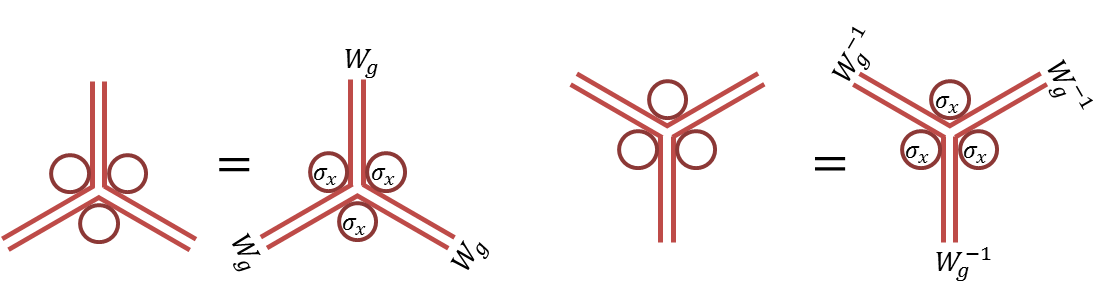}
    \caption{Symmetry conditions for the $Z_2$ symmetric state. $W_g$\,($W_g^{-1}$) denotes the associated gauge transformation. For the wavefunction defined in Eq.(\ref{eq:Z2_SPT_wavefunction}), $W_g=|11\rangle\langle00|+\ii|10\rangle\langle01|+\ii|01\rangle\langle10|+|00\rangle\langle11|$.}
    \label{fig:honeycomb_Z2_SPT_symmetry}
\end{figure}

\begin{figure}
    \includegraphics[width=0.45\textwidth]{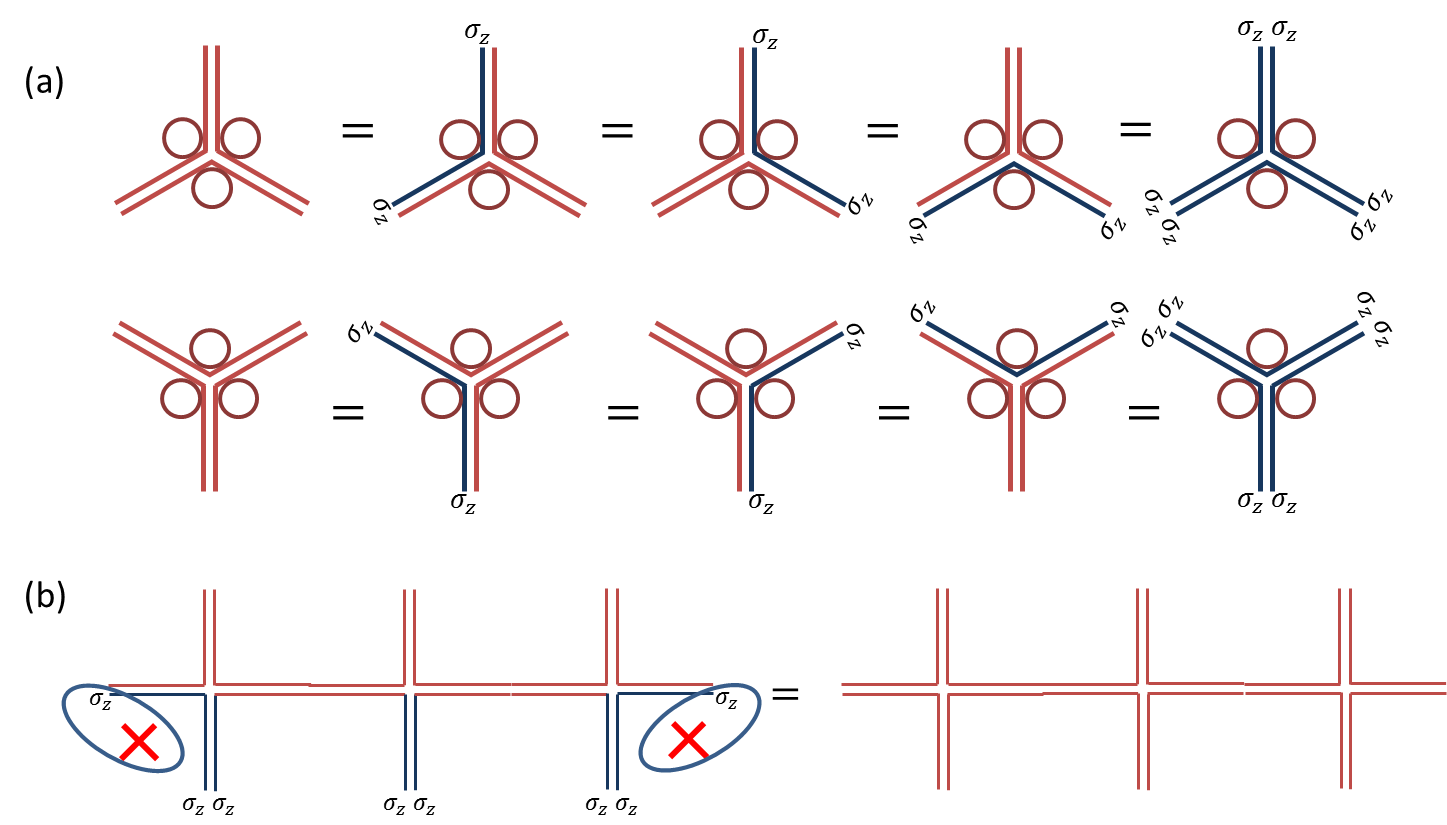}
    \caption{(a) Gauge transformations which leave local tensors invariant. (b) The condensation of visons carrying $Z_2$ symmetry charge.}
    \label{fig:honeycomb_Z2_SPT_IGG}
\end{figure}

It is instructive to see how the $Z_2$ symmetry acts on local tensors. A local tensor is not invariant under $g$ action, but the transformed tensor differ from the original one by some gauge transformation on internal legs, labeled as $W_g$\,($W_g^{-1}$), as shown in Fig. \ref{fig:honeycomb_Z2_SPT_symmetry}. For tensors defined on Fig. \ref{fig:honeycomb_Z2_SPT_tensors}), we obtain that 
\begin{align}
    W_g=|11\rangle\langle00|+\ii|10\rangle\langle01|+\ii|01\rangle\langle10|+|00\rangle\langle11|
    \label{}
\end{align}
We point out here, $W_g$ does not form a $Z_2$ group. Instead, we have
\begin{align}
    W_g^2=\sigma_z\otimes\sigma_z
    \label{eq:Z2_SPT_chargon_fractionalization}
\end{align}
So, after applying Ising symmetry twice, we are left with the $\sigma_z$ action on all internal legs, and trivial action on all physical legs. Notice, the $\sigma_z$ action on every internal leg is a special kind of gauge transformation, which leaves every single tensor invariant, as indicated by tensor equations on Fig. \ref{fig:honeycomb_Z2_SPT_IGG}(a). This kind of gauge transformations form a group, named as the \emph{invariant gauge group} (IGG). IGG is essential for tensor network constructions of nontrivial phases. 

Here, IGG is a $Z_2$ group, since $\sigma_z^2=\mathrm{I}$. In general, a nontrivial $Z_2$ IGG leads to the $Z_2$ toric code topological order\cite{Swingle:2010p,Schuch:2010p2153,jiang2015symmetric}. However, we claim that the $Z_2$ topological order is killed due to tensor equations in Fig. \ref{fig:honeycomb_Z2_SPT_IGG}. To see this, we first point out that a site tensor is invariant under single-leg $\sigma^z$ action on internal legs of one plaquette. Notice that the single-leg $\sigma_z$ action anticommutes with $W_g$, while double-leg $\sigma_z\otimes\sigma_z$ action commutes with $W_g$: 
\begin{align}
    W_g\sigma_z=-\sigma_z W_g
    \label{eq:Z2_SPT_lambda_Z2_charge}
\end{align}
The physical meaning of the single-leg $\sigma_z$ action is to create a (topologically-trivial) $Z_2$ symmetry charge excitation. To see this, we first point out that action of $Z_2$ symmetry $g$ on a local patch $\mathcal{R}$ is naturally defined as acting $g$ on physical sites of $R$ and $W_g$ on the boundary virtual legs of $R$. If $R$ contains one tensor with a single-leg $\sigma_z$ action, we get an extra minus sign due to Eq.(\ref{eq:Z2_SPT_lambda_Z2_charge}), which is interpreted as a $Z_2$ symmetry charge inside $R$.

The fact that a site tensor is invariant under two single-leg $\sigma_z$ action indicates the existence of a particular sub-group of IGG -- the ``plaquette IGG'', whose elements only have nontrivial action on internal legs within one plaquette. By multiplying all nontrivial plaquette IGG elements of all plaquettes, we recover the nontrivial element of the original $Z_2$ IGG, which is double-leg $\sigma_z$ action on every internal leg. The decomposition of IGG element into plaquette IGG elements is essential for the construction of generic wavefunctions of SPT phases. 

As we will see, the toric code topological order is killed due to the presence of the plaquette IGG. We put the system on a torus. The topological degenerate ground states are captured by inserting the non-contractible $\sigma_z$ loops. Since every tensor is invariant under two single-leg $\sigma_z$ actions, the wavefunction with non-contractable $\sigma_z$ loop turns out to be the same as the original wavefunction.  So, there is no topological ground state degeneracy, and the state has no topological order.

The physical reason can be interpreted as vison\,($m$) condensation. A pair of $m$-particles are created at two ends of a double-leg $\sigma_z\otimes\sigma_z$ string. As indicated in Fig. \ref{fig:honeycomb_Z2_SPT_IGG}(b), the creation of a pair of bond states of $Z_2$ symmetry charges and visons leaves the wavefunction invariant. In other words, these bound states\,($m$-particles carrying $Z_2$ odd quantum number) are condensed, thus killing the topological order.

There remains one question to be answered: what is the SET phase\,($Z_2$ topological order with $Z_2$ symmetry) before condensation? To see this, let us re-examine Eq.(\ref{eq:Z2_SPT_chargon_fractionalization}): two $Z_2$ symmetry defects $W_g$ fuse to a vison, which means $e$ carries fractional $Z_2$ quantum number and $m$ has the trivial symmetry fractionalization pattern. 

Let us summarise the previous discussion. We start from an SET phase with $Z_2^g$ topological order, where $e$-particles carry fractional $Z_2^s$ quantum number, as indicated in Eq.(\ref{eq:Z2_SPT_chargon_fractionalization}). Eq.(\ref{eq:Z2_SPT_lambda_Z2_charge}) tells us that the single-leg $\sigma_z$ action creates nontrivial $Z_2^s$ symmetry charge\footnote{One may wonder whether the local $Z_2^s$ charge of an $m$-particle is well defined, since we can always attach $e$ particle to the symmetry defect $W_g$, which will change the result of local symmetry action due to the nontrivial braiding phase between $e$ and $m$. However, if we always require that \emph{symmetry defects have the trivial symmetry fractionalization pattern}, quantum numbers of $m$-particles are well defined}.
The plaquette IGG defined on Fig. \ref{fig:honeycomb_Z2_SPT_IGG}(a) leads to the condensation of visons carrying nontrivial $Z_2$ charges. In the following, we show that any state satisfying these tensor equations is either a nontrivial $Z_2$ SPT phase, or a spontaneously symmetry breaking phase in the thermodynamic limit. 

\begin{figure}
    \includegraphics[width=0.45\textwidth]{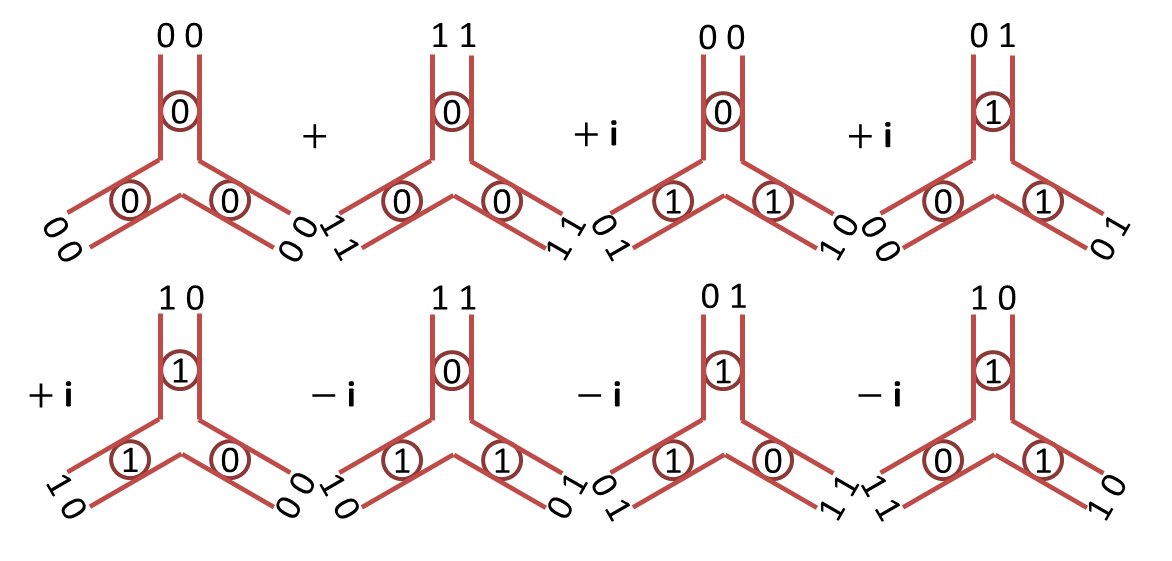}
    \caption{Tensors representing the double semion fixed point wavefunction}
    \label{fig:honeycomb_double_semion_tensors}
\end{figure}

\begin{figure}
    \includegraphics[width=0.45\textwidth]{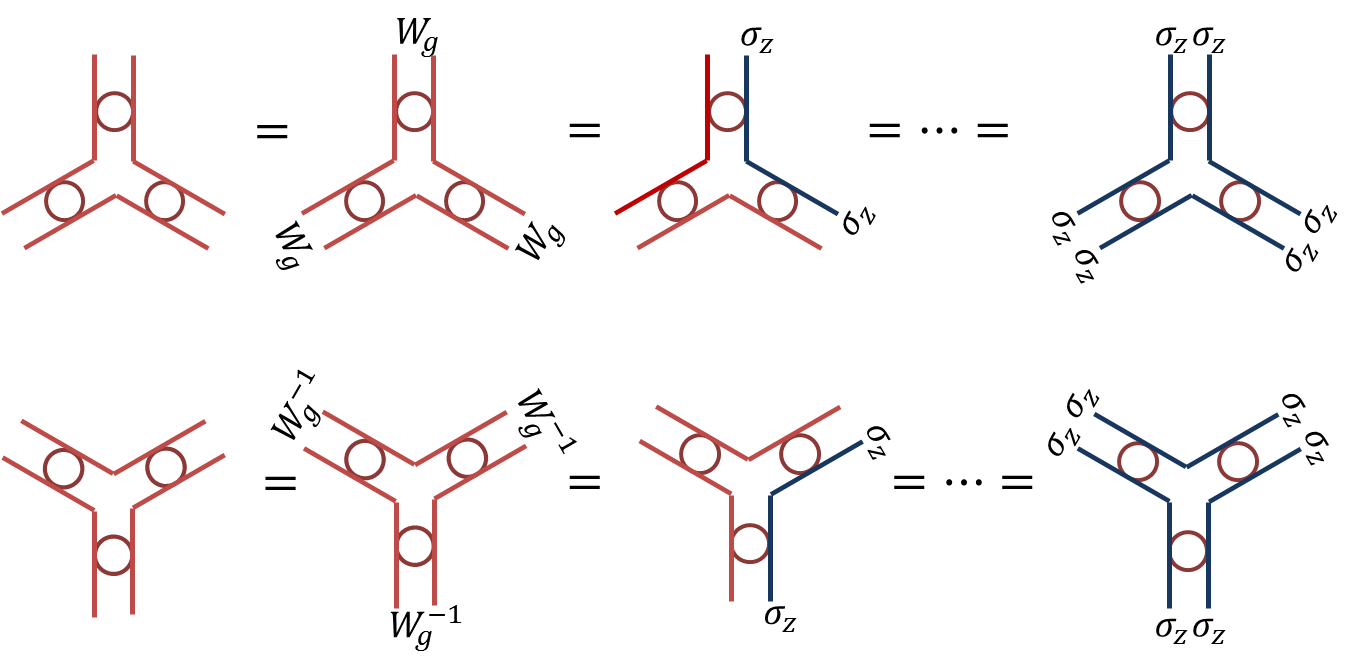}
    \caption{IGG for double semion topological order}
    \label{fig:honeycomb_double_semion_IGG}
\end{figure}

One way to see this is to gauge the $Z_2^s$ symmetry. It is known that gauging the nontrivial $Z_2^s$ SPT phase gives us the double semion topological order\cite{levin2012braiding}. Let us verify it in the tensor network formulation. As shown in Fig.\,\ref{fig:honeycomb_double_semion_tensors}, for the gauged $Z_2$ SPT state, physical degrees of freedom live on links. The physical state on the link is determined by  the ``difference'' of the two internal legs. The $Z_2$ symmetric condition for $g$ and $W_g$ in Fig.\,\ref{fig:honeycomb_Z2_SPT_symmetry} becomes a new IGG element, as indicated in Fig.\,\ref{fig:honeycomb_double_semion_IGG}. Similar to the ungauged theory, $W_g$ also satisfies Eq.(\ref{eq:Z2_SPT_chargon_fractionalization}) and Eq.(\ref{eq:Z2_SPT_lambda_Z2_charge}).

According to Eq.(\ref{eq:Z2_SPT_chargon_fractionalization}), the gauged tensor state actually holds an $Z_4$ global IGG: $\{\mathrm{I},W_g,\sigma_z\otimes\sigma_z,W_g\cdot(\sigma_z\otimes\sigma_z)\}$. $Z_4$ flux, labeled as $m_0$\,($m_0^\dagger$), are created at ends of $W_g$ strings. And ends of $\sigma_z\otimes\sigma_z$ strings are double $Z_4$ flux, labeled as $m_0^2$. To see the physical meaning of single leg action of $\sigma_z$, we first note that it is a self boson. And braiding $m_0$ around it, one obtain $\pi$ phase according to Eq.(\ref{eq:Z2_SPT_lambda_Z2_charge}). So, the single leg action of $\sigma_z$ corresponds to a double $Z_4$ charge $e_0^2$. Due to the existence of nontrivial plaquette IGG elements, bound states of $m_0^2$ and $e_0^2$ are condensed, as shown in Fig. \ref{fig:honeycomb_Z2_SPT_IGG}(b). And all other particles sharing nontrivial braiding statistics with $m_0^2e_0^2$ are confined. Then, the remaining topological order can be determined by the following table:
\begin{center}
    \begin{tabular}{|c|c|c|c|c|}
        \hline
        \diagbox{charge}{flux} & 0 & 1 & 2 & 3\\\hline
        0 & $\mathrm{I}$ & $\times$ & $b$ & $\times$\\\hline
        1 & $\times$ & s & $\times$ & $\bar{s}$\\\hline
        2 & $b$ & $\times$ & $\mathrm{I}$ & $\times$\\\hline
        3 & $\times$ & $\bar{s}$ & $\times$ & $s$\\\hline
    \end{tabular}
\end{center}
Here, $s$ and $\bar{s}$ are semions and $b$ is a self boson. The fusion and braiding rules of the remaining quasiparticles are the same as the double semion topological order. So, the condensed phase holds an double semion topological order.

Then, we conclude that the ungauged phase is the nontrivial $Z_2$ SPT. Notice that $b$ boson may condense in the long wavelength, thus kill the double semion topological order. In the ungauged theory, this corresponds to the spontaneously symmetry breaking phase.

\subsection{General Framework}\label{sec:tensor_construction}
Let us summarize what we have learned from the above simple example. To construct the SPT state on tensor networks, we require that
\begin{itemize}
    \item the tensor network state is symmetric, as shown in Fig. \ref{fig:honeycomb_Z2_SPT_symmetry};
    \item tensors have some nontrivial IGG structure, as shown in Fig. \ref{fig:honeycomb_Z2_SPT_IGG};
    \item the symmetry transformation rules and IGG elements are interplaying with each other, as given in Eq.(\ref{eq:Z2_SPT_chargon_fractionalization}) and Eq.(\ref{eq:Z2_SPT_lambda_Z2_charge}).
\end{itemize}
We will follow the above strategy in this part and develop a general framework for SPT phases on tensor networks. The three cohomology classification naturally emerges from tensor equations. 

\subsubsection{Symmetries}
Let us first discuss how to impose symmetries on tensor networks\cite{Perez-Garcia:2010p25010,Zhao:2010p174411,Singh:2010p50301,Singh:2011p115125,Singh:2012p195114,Bauer:2011p125106,Weichselbaum:2012p2972,jiang2015symmetric}. We focus on the case where the state is a 1D representation of symmetry group $SG$: 
\begin{align}
    g\circ|\Psi\rangle=\mathrm{e}^{\ii\theta_g}|\Psi\rangle,\forall g\in SG 
    \label{}
\end{align}
Here $SG$ includes both onsite symmetries as well as lattice symmetries. 

Consider a PEPS state formed by site tensors. We assume that \emph {for a symmetric PEPS state, the symmetry transformed tensors and the original tensors are related by a gauge transformation (up to a $U(1)$ phase factor)}:
\begin{align}
    \Theta_gW_gg\circ \mathbb{T}=\mathbb{T}
    \label{}
\end{align}
Here, $\mathbb{T}$ represents the tensor states \emph{with all internal legs uncontracted}. Namely $\mathbb{T}=\bigotimes_aT^a$, where $T_a$ represents a local tensor at site $a$. $W_g$ is a gauge transformation, which acts on all internal legs of the tensor network:
\begin{align}
    W_g=\bigotimes_{(a,i)}W_g(a,i)
    \label{}
\end{align}
where $(a,i)$ labels a leg of site $a$. If leg $(a,u)$ and $(b,v)$ are connected, according to the definition of gauge transformation, $W_g(a,u)\cdot W_g^{\mathrm{t}}(b,v)=\mathrm{I}$. $\Theta_g$ is a tensor-dependent $U(1)$ phase. In the following, we will focus on systems defined on an infinite lattice, for which we can always absorb $\Theta_g$ to $W_g$. So, the symmetric condition for a tensor wavefunction can be expressed as
\begin{align}
    W_gg\circ \mathbb{T}=\mathbb{T}
    \label{eq:symmetric_tensor_condition}
\end{align}
To be more clear, we can write the above equation explicitly as
\begin{align}
    &(W_g(a,u))_{\alpha\alpha'}.(W_g(a,v))_{\beta\beta'}\dots g\circ(T^a_{uv\dots})^{i}_{\alpha'\beta'\dots}\notag\\
    =&(T^a_{uv\dots})^i_{\alpha'\beta'\dots}
    \label{}
\end{align}
where $T_a$ labels a tensor at site $a$, and $u,v\dots$ labels legs of tensor $T^a$. 

\subsubsection{Invariant gauge group}
The invariant gauge group (IGG) is a sub-group of gauge transformations, \emph{whose element leaves every tensor -- or equivalently the tensor state before contraction ($\mathbb{T}$) -- completely invariant}\cite{Swingle:2010p,Schuch:2010p2153,jiang2015symmetric}. Notice that a general gauge transformation only leaves the physical wavefunction invariant, while could transform the site tensors nontrivially. To make the discussion below clear, we denote any element in IGG as a \emph{global} IGG element, since by definition this element is a gauge transformation involving all virtual legs on the tensor network. 

\begin{figure}
    \includegraphics[width=0.45\textwidth]{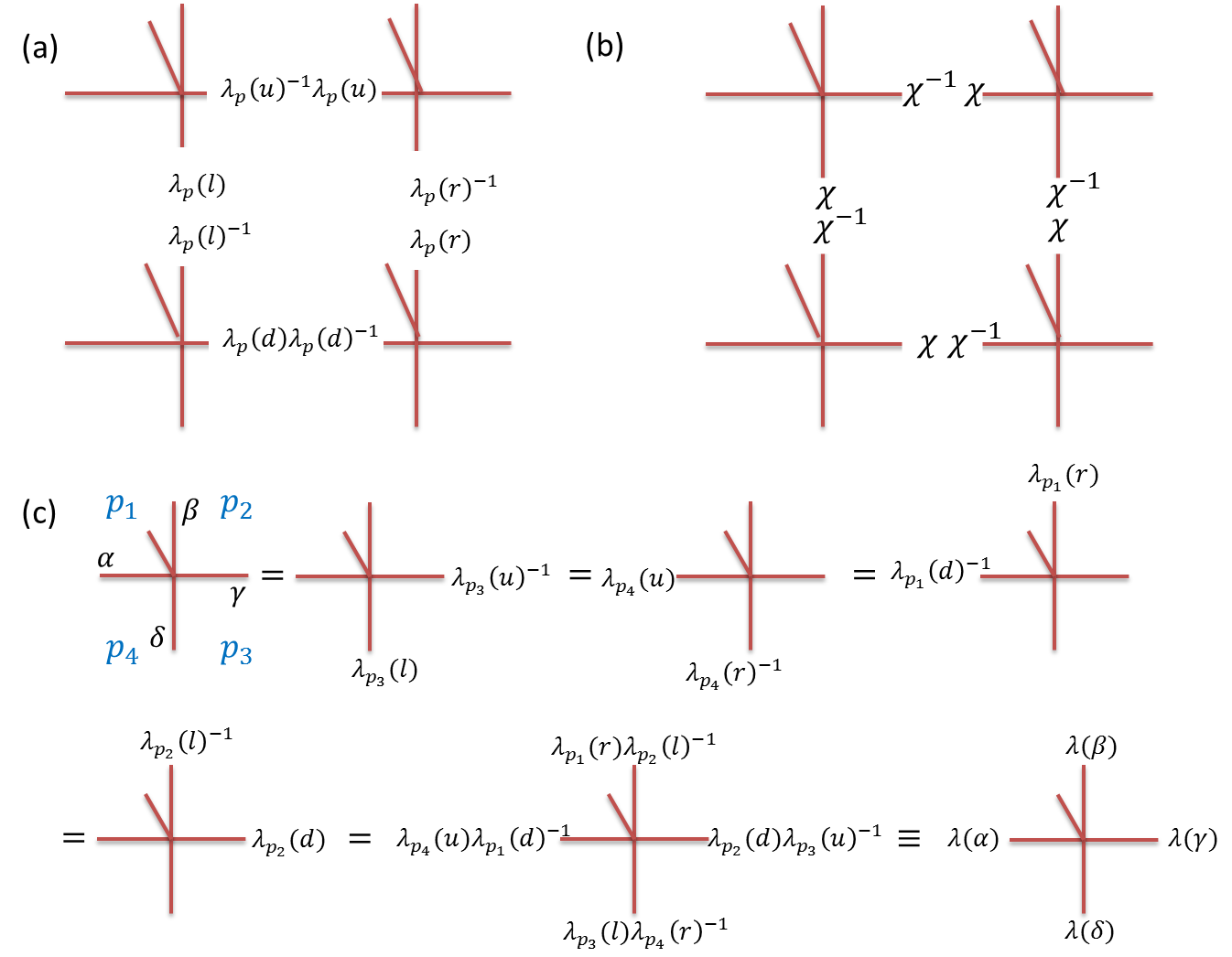}
    \caption{(a) An example of the plaquette IGG element $\lambda_p$. (b) The plaquette IGG element formed by complex number $\chi$ and $\chi^{-1}$. This kind of plaquette IGG exists for any PEPS state. (c) A site tensor lives on the subspace which is invariant under action of IGGs. Here, $p_1,p_2,p_3,p_4$ are four neighbouring plaquettes around the tensor and $\alpha,\beta,\gamma,\delta$ denote legs of the tensor. The last equation indicates that a global IGG element is obtained from multiplication of plaquette IGG elements.}
    \label{fig:plaquette_IGG}
\end{figure}

We also introduce a special type of IGG elements -- the plaquette IGG element $\lambda_p$, where $\lambda_p$ acts nontrivially only on internal legs of plaquette $p$, as shown in Fig. \ref{fig:plaquette_IGG}(a). The plaquette IGG is a generalization of the single leg action of $\sigma_z$ in Fig.\,\ref{fig:honeycomb_Z2_SPT_IGG}. For any given plaquette $p$, the collection of plaquette IGG elements $\{\lambda_p\}$ acting on $p$ forms a subgroup of IGG. To construct SPT, we further assume that \emph{any global IGG element can always be decomposed into the product of plaquette IGG elements, $\lambda=\prod_p\lambda_p$.} Namely, plaquette IGG elements can generate the full IGG. 


\begin{figure}
    \includegraphics[width=0.45\textwidth]{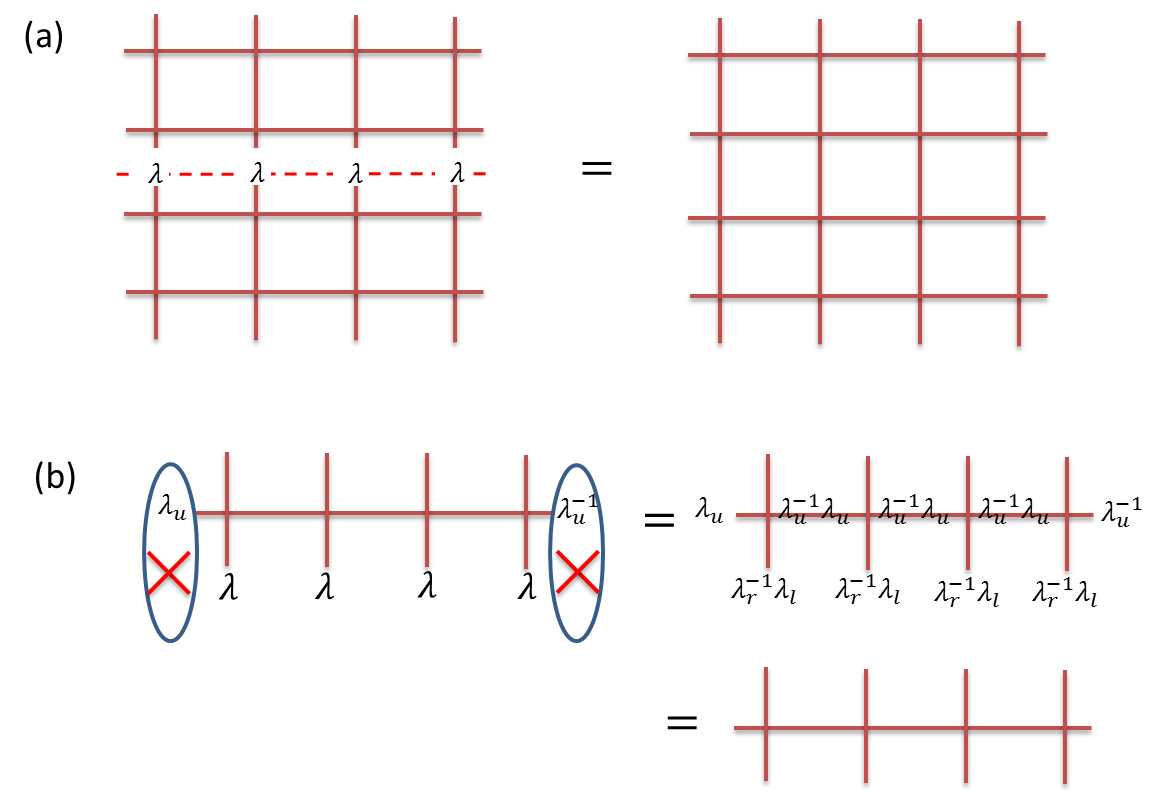}
    \caption{Since any global IGG element can be decomposed to plaquette IGG elements, (a) there is no topological ground state degeneracy; (b) bound states of $\lambda$-fluxes and $\lambda_p$ objects are condensed. Here, $\lambda_i$ is short for $\lambda_p(i)$.}
    \label{fig:anyon_condensation}
\end{figure}

For SPT tensor wavefunctions, we have assumed that the decomposition from a global IGG element to the product of plaquette IGG elements always exist. One may ask whether the decomposition is unique. The answer is no. To see this, we consider the decomposition of the trivial action $\mathrm{I}$ on all internal legs. There is a special kind of plaquette IGG element: for every plaquette $\lambda_l=\lambda_u=\lambda_r=\lambda_d=\chi$, where $\chi$ is a complex number, as shown in Fig. \ref{fig:plaquette_IGG}(b). We also label this IGG element as $\chi_p$. Then, $\prod_p\chi_p=\mathrm{I}$. We assume that \emph{this is the only way to decompose $\mathrm{I}$}. Notice that the identity $\prod_p\chi_p=\mathrm{I}$ directly leads to the fact that the phase factor $\chi$ in any plaquette is the same. So, for any global IGG element, there is only one global phase ambiguity to decompose into the plaquette IGG elements $\lambda_p$ reads
\begin{align}
    \lambda=\prod_p\lambda_p=\prod_p\chi_p\lambda_p
    \label{eq:IGG_phase_ambiguity}
\end{align}
It turns out that this phase ambiguity is essential to get SPT phases, and naturally gives 3-cohomology classification. 

\subsubsection{Cohomology from symmetry equations on PEPS}
For group elements $g_1,g_2$, we have
\begin{align}
    \mathbb{T}=W_{g_1}g_1W_{g_2}g_2\circ \mathbb{T} = W_{g_1g_2}g_1g_2\circ \mathbb{T}, 
    \label{}
\end{align}
Since $W_{g_1}g_1W_{g_2}g_2$ and $W_{g_1g_2}g_1g_2$ only differ by a gauge transformation, and they both leave $\mathbb{T}$ invariant. So, they should differ up to an IGG element, which we label as $\lambda(g_1,g_2)$,
\begin{align}
    W_{g_1}g_1W_{g_2}g_2=\lambda(g_1,g_2)W_{g_1g_2}g_1g_2
    \label{eq:Wg_proj_rep}
\end{align}
which generalize Eq.(\ref{eq:Z2_SPT_chargon_fractionalization}). According to associativity
\begin{align}
    (W_{g_1}g_1W_{g_2}g_2)W_{g_3}g_3=W_{g_1}g_1(W_{g_2}g_2W_{g_3}g_3)
    \label{}
\end{align}
we get
\begin{align}
    \lambda(g_1,g_2)\lambda(g_1g_2,g_3)=\,{}^{W_{g_1}g_1}\!\lambda(g_2,g_3)\lambda(g_1,g_2g_3)
    \label{eq:IGG_two_cocycle}
\end{align}
where we define ${}^a\!b\equiv a\cdot b\cdot a^{-1}$. Particularly, for a leg $i$, we have 
\begin{align}
    \left(\,{}^{W_{g}g}\!\lambda\right)(i)=W_g(i)\cdot\lambda^{s(g)}(g^{-1}(i))\cdot \left[W_g(i)\right]^{-1}
    \label{}
\end{align}
where $s(g)$ is complex conjugate if $g$ contains time reversal action.

One can decompose $\lambda$'s into $\lambda_p$'s, and due to the phase ambiguity Eq.(\ref{eq:IGG_phase_ambiguity}), $\lambda_p$'s satisfy 
\begin{align}
    &\lambda_p(g_1,g_2)\lambda_p(g_1g_2,g_3)=\notag\\
    &\omega_p(g_1,g_2,g_3)\,{}^{W_{g_1}g_1}\!\lambda_p(g_2,g_3)\lambda_p(g_1,g_2g_3)
    \label{eq:plq_IGG_twist_two_cocycle}
\end{align}
where $\omega_p(g_1,g_2,g_3)$ is the phase IGG satisfying $\mathrm{I}=\prod_p\omega_p(g_1,g_2,g_3)$.

In Appendix \ref{app:three_cohomology}, we prove $\omega_p$ satisfies three cocycle condition: 
\begin{align}
    &\omega_p(g_1,g_2,g_3)\omega_p(g_1,g_2g_3,g_4)\,{}^{g_1}\omega_p(g_2,g_3,g_4)\notag\\
    &=\omega_p(g_1g_2,g_3,g_4)\omega_p(g_1,g_2,g_3g_4)
    \label{eq:omega_three_cocycle}
\end{align}
And $\omega_p$ is defined up to a coboundary:
\begin{align}
    \omega_p(g_1,g_2,g_3)\sim\omega_p(g_1,g_2,g_3)\frac{\chi_p(g_1,g_2)\chi_p(g_1g_2,g_3)}{{}^{g_1}\!\chi_p(g_2,g_3)\chi_p(g_1,g_2g_3)}
    \label{}
\end{align}

The action of $g$ on $\omega_p$\,($\chi_p$) follows a very simple rule: for a leg $i$, we have $({}^g\!\omega_p)(i)=\omega^{s(g)}_{g^{-1}(p)}(g^{-1}(i))$, where $s(g)$ is complex conjugate if $g$ contains time reversal. Then, consider $\omega_p$, we have
\begin{itemize}
    \item For unitary onsite symmetry $g$, ${}^g\!\omega_p=\omega_p$
    \item For time reversal symmetry $\mathcal{T}$, ${}^\mathcal{T}\!\omega_p=\omega_p^*$
    \item For translation and/or rotation symmetry $T_i$ and $C_i$, ${}^{T_i}\!\omega_p=\,{}^{C_i}\!\omega_p=\omega_p$
    \item For reflection symmetry $\sigma$, ${}^\sigma\!\omega_p=\omega_p^{-1}$
\end{itemize}

\subsubsection{Methods to construct generic SPT tensor wavefunctions}
Now, we have developed a general way to write down tensor equations for SPT phases: Eq.(\ref{eq:Wg_proj_rep}),Eq.(\ref{eq:IGG_two_cocycle}) and Eq.(\ref{eq:plq_IGG_twist_two_cocycle}). 
The next step is to answer the following question: given a symmetry group $SG$ and a cohomology class $[\omega]$, how do we construct generic SPT wavefunctions from tensor equations? This problem actually can be decomposed to three parts:
\begin{enumerate}
    \item Figure out the group structure for $\lambda$'s, $\lambda_p$'s and $W_g$'s to realize the SPT phase.
    \item Obtain the representation of the IGG and symmetry on tensor networks.
    \item Find subspace of tensors, which are invariant under IGG action on internal legs as well as symmetry actions on both physical legs and virtual legs.
\end{enumerate}
The second part and the third part are relatively easy to solve, and we give examples in Sec.\,\ref{sec:tensor_examples}. Here, we focus on the first part, and we provide two methods in the following.

The first way is to start from exact solvable models. If there exists an exact solvable model realizing some SPT phase, one can construct a fixed point wavefunction by PEPS. Then, one can extract tensor equations as well as the group structure for $\lambda$'s $\lambda_p$'s and $W_g$'s. For example, as we show in Sec.\,\ref{sec:Z2_SPT_example}, to realize a nontrivial $Z_2^s$ SPT, $\lambda$'s form a $Z_2^g$ group. $\lambda_p$'s form group $Z_2\times U(1)$ for any plaquette $p$. And $W_g$ is a projective representation with coefficient in $Z_2$, which anticommutes with nontrivial $\lambda_p$. 

Notice that the group structure for IGG and $W_g$ does not depend on whether $SG$ is onsite or spatial. So, we are also able to figure out IGG and $W_g$ for spatial SPT phases. For example, as we will show in Sec.\,\ref{sec:tensor_examples}, for the nontrivial inversion SPT phase, $\lambda$'s form a $Z_2^g$ group, which is the same as the case for $Z_2^s$ onsite SPT phase. The only difference is that for the inversion SPT and $Z_2$ onsite SPT, the IGGs have distinct representations on internal legs.

For every SPT phase protected by a discrete symmetry group and also some SPT phases protected by continuous symmetry groups, one can write down exact solvable models. So one is able to realize those generic SPT wavefunctions by tensors. 

The second way is related to a mathematical object named as crossed module extensions. It is known in mathematical literatures that crossed module extensions of $SG$ by $U(1)$ are classified by $H^3(SG,U(1))$. And as we show in Appendix\,\ref{app:three_cohomology}, our tensor constructions can be viewed as a representation of crossed module extensions. So, given a crossed module extension, we are able to figure out the group structure for IGG and $W_g$'s.

\subsection{A by-product: the general anyon condensation mechanism for realizing SPT phases}\label{sec:tensor_anyon_condensation}
Using the above results, here we prove the Criterion of the anyon condensation mechanism. We will start from an SET phase with discrete Abelian topological order and condense $m$-particles to confine the gauge field, and demonstrate the Criterion to realize SPT phases. For the purpose of presentation, we will consider $Z_N$ topologically ordered SET phases with the symmetry group $SG$, but one can straightforwardly generalize the discussion below for SET phases with any discrete Abelian gauge groups $Z_{N_1}\times Z_{N_2}...$.

In order to represent a regular $Z_N$ topological order in the tensor-network formulation, one needs to introduce a nontrivial \emph{global} IGG\cite{Swingle:2010p,Schuch:2010p2153,jiang2015symmetric}, labeled as $H$. In particular, there is a nontrivial global IGG element $J\in H$ satisfying $J^N=\mathrm{I}$, and representing the $Z_N$ gauge transformation. Here $J$ is nontrivial means that it is not $U(1)$-phase multiplications on the virtual legs. A $J$ string is interpreted as a $Z_N$ flux line, while the $Z_N$ gauge flux and its antiparticle are created at two ends of the $J$ string. Besides the nontrivial $Z_N$ IGG, there is always ``trivial'' IGG $X$, whose elements are loops of phases. So, we start from tensor states with an abelian IGG $H\times X$. 

In the presence of symmetry $SG$ and IGG $H\times X$, the tensor equations read
\begin{align}
    W_{g_1}g_1W_{g_2}g_2=\xi(g_1,g_2)\eta(g_1,g_2)W_{g_1g_2}g_1g_2,\,\forall g_1,g_2\in SG
    \label{}
\end{align}
where $\xi(g_1,g_2)\in X$, and $\eta(g_1,g_2)\in H$. $\xi$'s and $\eta$'s both satisfy the two-cocycle condition:
\begin{align}
    &\xi(g_1,g_2)\xi(g_1g_2,g_3)=\,{}^{g_1}\!\xi(g_2,g_3)\xi(g_1,g_2g_3)\notag\\
    &\eta(g_1,g_2)\eta(g_1g_2,g_3)=\,{}^{W_{g_1}g_1}\!\eta(g_2,g_3)\eta(g_1,g_2g_3)
    \label{eq:anyon_cond_two_cocycle}
\end{align}
We point out that $\eta$'s label the symmetry fractionalization pattern of $Z_N$ charges. 

How about the symmetry properties for fluxes? To see this, let us study the symmetry action on $Z_N$ flux line $J$: $\,{}^{W_{g}g}\!J\in H\times X$. Since we are studying phases featuring symmetry fractionalizations, we require that the anyon types are invariant under symmetry action:
\begin{align}
    \,{}^{W_{g}g}\!J=\chi_J(g)\cdot J
    \label{eq:chi_J}
\end{align}
where $\chi_J(g)\in X$, and $\left(\chi_J(g)\right)^N=1$. Further, $\chi_J: SG\rightarrow Z_N$ is a representation of $SG$, since
\begin{align}
    &\,{}^{W_{g_1}g_1W_{g_2}g_2}\!J=\chi_J(g_1)\,{}^{g_1}\!\chi_J(g_2)\cdot J\notag\\
    =&\,{}^{\xi(g_1,g_2)\eta(g_1,g_2)W_{g_1g_2}g_1g_2}\!J=\chi_J(g_1g_2)\cdot J
    \label{}
\end{align}
where we use the fact $\xi(g_1,g_2)\eta(g_1,g_2)$ commute with $J$. So,
\begin{align}
    \chi_J(g_1g_2)=\chi_J(g_1)\,{}^{g_1}\!\chi_J(g_2)
    \label{eq:flux_string_qn}
\end{align}
Notice that both time reversal $\TT$ and reflection $P$ should be treated as antiunitary operations. 

To proceed, we point out that the building blocks for $X$ are plaquette phase IGGs:
\begin{align}
    \chi=\prod_p\chi_p, \forall \chi\in X
    \label{}
\end{align}
Here $\chi_p\in X_p$, where $X_p\subset X$ is the plaquette IGG of $p$, whose elements are loops of phases along virtual legs of plaquette $p$. As before, the decomposition to plaquette phase IGG elements has a single phase ambiguity:
\begin{align}
    \chi=\prod_p\chi_p=\prod_p\epsilon_p\chi_p
    \label{eq:loop_phase_plq_IGG_ambiguity}
\end{align}
Here $\epsilon_p(i)=\epsilon^{\pm1}$, where $\pm1$ pattern follows as Fig.\,\ref{fig:plaquette_IGG}(b).

Then, according to Eq.(\ref{eq:flux_string_qn}) and Eq.(\ref{eq:loop_phase_plq_IGG_ambiguity}), we obtain
\begin{align}
    \chi_{p,J}(g_1)\,{}^{g_1}\!\chi_{p,J}(g_2)=\omega_{p,J}(g_1,g_2)\chi_{p,J}(g_1g_2)
    \label{eq:flux_sf}
\end{align}
where $\omega_{p,J}(g_1,g_2)(i)=\omega_J(g_1,g_2)^{\pm1}$. Because $\chi_J(g)^N=1$, clearly phase factors $\chi_{p,J}(g)$ and $\omega_{p,J}(g_1,g_2)$ can be chosen to be $Z_N$ elements. It is straightforward to check that $\omega_{p,J}$ satisfies the two-cocycle condition:
\begin{align}
    \omega_{p,J}(g_1,g_2)\omega_{p,J}(g_1g_2,g_3)=\,{}^{g_1}\!\omega_{p,J}(g_2,g_3)\omega_{p,J}(g_1,g_2g_3)
    \label{}
\end{align}
It turns out that $\omega_{J}(g_1,g_2)\in Z_N$ labels the symmetry fractionalization pattern of $Z_N$ fluxes. 

For onsite symmetries, we can restrict to one internal leg $i$. Then, Eq.(\ref{eq:flux_sf}) becomes a relation for phase factors. We can always tune $\omega_{p,J}$ to be trivial by redefining $\chi_{p,J}(g)\rightarrow\epsilon_J(g)\cdot\chi_{p,J}(g)$. In other words, onsite symmetry fractionalization patterns for fluxes are always trivial for the case IGG equals $H\times X$. Notice, fluxes can carry fractional spatial symmetry quantum numbers in general.

Now, let us derive the Criterion to obtain SPT phases by condensing fluxes. In this tensor formulation, we require nontrivial plaquette IGG for every plaquette. And the plaquette IGG for $p$ is labeled as $H_p\times X_p$. 

To kill the topological order, we require the decomposition of $J$ as 
\begin{align}
    J=\prod_p J_p=\prod_p\epsilon_p J_p
    \label{}
\end{align}
where $J_p$ is a nontrivial plaquette IGG element for plaquette $p$. Again, the decomposition has an $U(1)$ ambiguity $\epsilon_{p,J}$. As shown in Fig.\,\ref{fig:plaquette_IGG}(d), the bound state of $Z_N$ fluxes and $J_p$ is condensed according to the above equation. 
Notice that there is a canonical choice for $J_p$ such that $J_p^N=I$. So we can choose $H_p\cong Z_N$, and $H_p\times X_p$ is an abelian group. Further, as we prove in Appendix\,\ref{app:three_cohomology}, elements of plaquette IGG for different plaquettes commute. Thus, we conclude, the whole IGG is abelian.

To see the symmetry action on $J_p$, or equivalently, the symmetry quantum number carried by $J_p$, we have
\begin{align}
    &\,{}^{W_{g}g}\!J=\prod_p\,{}^{W_{g}g}\!J_p\notag\\
    =&\chi_J(g)\cdot J=\prod_p\chi_{p,J}(g)J_p
    \label{}
\end{align}
Due to the $U(1)$ ambiguity, we conclude
\begin{align}
    \,{}^{W_{g}g}\!J_p=\epsilon_{p,J}(g)\chi_{p,J}(g)J_p
    \label{}
\end{align}
We further have
\begin{align}
    &\,{}^{W_{g_1}g_1W_{g_2}g_2}\!J_p=\epsilon_{p,J}(g_1)\chi_{p,J}(g_1)\,{}^{g_1}\!\epsilon_{p,J}(g_2)\,{}^{g_1}\!\chi_{p,J}(g_2)\cdot J_p\notag\\
    =&\,{}^{\xi(g_1,g_2)\eta(g_1,g_2)W_{g_1g_2}g_1g_2}\!J_p=\epsilon_{p,J}(g_1g_2)\chi_{p,J}(g_1g_2)\cdot J_p
    \label{}
\end{align}
where we use the fact that $\xi\eta$ commutes with $J_p$. Comparing with Eq.(\ref{eq:flux_sf}), we conclude
\begin{align}
    \omega_{p,J}(g_1,g_2)=\frac{\epsilon_{p,J}(g_1g_2)}{\epsilon_{p,J}(g_1)\,{}^{g_1}\!\epsilon_{p,J}(g_2)}
    \label{}
\end{align}
is a two-coboundary. Namely, in this tensor formulation, symmetry-preserving flux-condensation requires fluxes to have no symmetry fractionalization.

In the following, we focus on a simple case: 
\begin{align}
 \chi_J(g)=1,\forall g\in SG.\label{eq:simple_case}
\end{align}

If instead $\chi_J(g)$ is nontrivial phase factor for symmetry $g$, the quantum number carried by the flux will depend on the details of the region of local-symmetry action as well as the flux string configuration. Although this situation is not violating basic principles, it is rather unlikely in usual models. In addition, the main purpose of this section is to derive the Criterion for anyon condensation mechanism, where we assume the quantum numbers of the flux is independent of the details of local symmetry action. Consequently, in this section, we do not consider this situation and focus on the cases given by Eq.(\ref{eq:simple_case}).

We choose a canonical gauge such that $J_p^N=I$, and $\eta_p=J_p^m$ for $\eta=J^m$, $\forall m$. In particular, we have
\begin{align}
    \eta_p\cdot\eta_p'=(\eta\cdot \eta')_p
    \label{}
\end{align}
Then, according to Eq.(\ref{eq:anyon_cond_two_cocycle}), we have
\begin{align}
    \eta_p(g_1,g_2)\eta_p(g_1g_2,g_3)=\eta_p(g_2,g_3)\eta_p(g_1,g_2g_3)
    \label{}
\end{align}
Let us define
\begin{align}
    \omega_1(g_1,g_2,g_3)=\frac{\,{}^{W_{g_1}g_1}\!\eta_p(g_2,g_3)\eta_p(g_1,g_2g_3)}{\eta_p(g_1,g_2)\eta_p(g_1g_2,g_3)}=\frac{\,{}^{W_{g_1}g_1}\!\eta_p(g_2,g_3)}{\eta_p(g_2,g_3)}
\end{align}
which is the quantum number of condensed fluxes. 
We also define
\begin{align}
    \omega_2(g_1,g_2,g_3)=\frac{\,{}^{W_{g_1}g_1}\!\xi_p(g_2,g_3)\xi_p(g_1,g_2g_3)}{\xi_p(g_1,g_2)\xi_p(g_1g_2,g_3)}
    \label{}
\end{align}
Following Appendix\,\ref{app:three_cohomology}, one can prove $\omega_1$ and $\omega_2$ are both three-cocycles. And the obtained SPT phase is characterized by $[\omega]=[\omega_1]\cdot[\omega_2]$, where $[\cdot]$ means equivalent class up to coboundary. Notice that even before anyon condensation (without nontrivial plaquette IGG $H_p$), $\omega_2$ is still present -- it is ``background'' SPT index unaffected by anyon condensation. However, because $\omega_2$ is obtained from the algebra of phase factors (instead of matrices), $\omega_2$ can be nontrivial only due to spatial translational symmetries (i.e. $\omega_2$ is only describing a weak SPT indices). The strong SPT indices can only appear due to $\omega_1$. So we have proved the Criterion as in Sec.\,\ref{sec:anyon_cond}. 

\subsection{Algorithms to measure anyon quantum numbers}\label{sec:measure_qn}

\begin{figure}
    \includegraphics[width=0.45\textwidth]{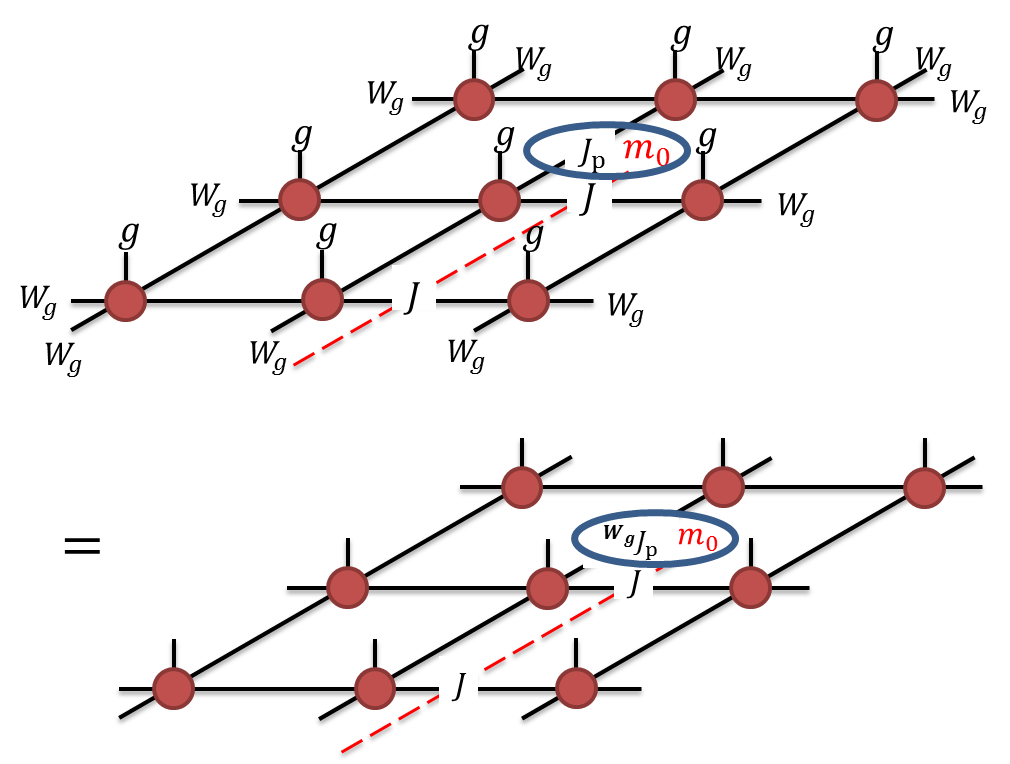}
    \caption{Measurement of the quantum number $\chi_{m_0}(g)$ carried by an $m$-particle for a local unitary symmetry $g$. According to Eq.(\ref{eq:chi_J},\ref{eq:simple_case}), $J$ commute with $W_g$, so we conclude that the quantum number is obtained by $\chi_{m_0}(g)\cdot J_p=\,{}^{W_{g}g}\!J_p$.}
    \label{fig:flux_quantum_number}
\end{figure}

It would be useful to be able to numerically measure the quantum numbers carried by the low energy $m$-particles inside the SET phase near the condensation phase transition. Such measurements, together with the Criterion, would allow one to predict the nature of the resulting symmetric phases. Now let us present several ``conceptual'' algorithms to measure these quantum numbers. Although these algorithms could be implemented in the existing tensor-network algorithms\cite{PhysRevB.92.201111} to practically measure these quantum numbers, here our focus is mainly to clarify conceptual issues. In particular, the quantum numbers introduced in the previous section may appear somewhat formal, and it would be ideal to explicitly demonstrate their measurable meanings.

We again focus on ordinary $Z_N$ gauge theories. As discussed before, the two ends of an open string created by a sequence of $J$ operations on the virtual bonds actually describe an elementary $m$-particle (coined $m_0$) and its anti-particle (coined $m_0^{\dagger}$). In order to simulate the low energy excitations within the topological sectors corresponding to $m_0$ and $m_0^{\dagger}$, one needs to further variationally optimize the tensors over finite regions (about correlation-length size) near the centers of these $m$-particles. Namely, a low energy excitation state $|\Psi_{ex}\rangle$ hosting $m_0$ and $m_0^{\dagger}$ quasiparticles is obtained by only modifying these local tensors (coined excited-state-local-tensors) while leaving all other tensors in the network (coined ground-state-local-tensors) the same as the ground state (apart from multiplying a sequence of $J$ operations on the string).

Our basic scheme is to use the symmetry transformation rules on the ground-state-local-tensors to obtain the symmetry properties of $m_0$ and $m_0^{\dagger}$. Let us start from discussing the measurement of the quantum number of an onsite unitary symmetry $g\in SG$, as shown in Fig.(\ref{fig:flux_quantum_number}). For example, let us focus on $m_0$. The local action of $g$ on $m_0$ is described by applying $W_g$ on a loop of virtual legs enclosing $m_0$ (but \emph{not} enclosing $m_0^{\dagger}$), together with applying the physical transformation $g$ on the physical legs inside the region enclosed by the $W_g$-loop. Physically, such a tensor-network operation corresponds to braiding a $g$-symmetry-defect (described by the end point of the $W_g$-string) around $m_0$. It turns out that the condition ${}^{W_gg}J=J$ (i.e. Eq.(\ref{eq:chi_J},\ref{eq:simple_case})) dictates that \emph{the $g$-symmetry-defect itself has no symmetry fractionalization}. It also dictates that the $m_0$ is transformed by this local action back to the same topological sector.

Now quantum number carried by $m_0$: $\chi_{m_0}(g)$ has direct measurable meaning. After applying the local action of $g$ on $m_0$, one obtains a new physical state $|\Psi'_{ex}\rangle$, corresponding to applying symmetry $g$ only on $m_0$ but not on $m_0^{\dagger}$. Due to symmetry, $|\Psi'_{ex}\rangle$ can at most differ from  $|\Psi_{ex}\rangle$ by a phase factor, which is exactly the measurable meaning of $\chi_{m_0}(g)$. Note that the variationally determined excited-state-local-tensors around $m_0$ only introduces a common global phase ambiguity in the physical state $|\Psi_{ex}\rangle$ and $|\Psi'_{ex}\rangle$, and consequently not affecting their relative phase $\chi_{m_0}(g)$.

Similar discussion can be naturally extended to rotational spatial symmetries, which can be treated as unitary operations. The only modification is that one needs to choose the position $m_0$ to be invariant under the rotations in order to respect these symmetries. 

\begin{figure}
    \includegraphics[width=0.45\textwidth]{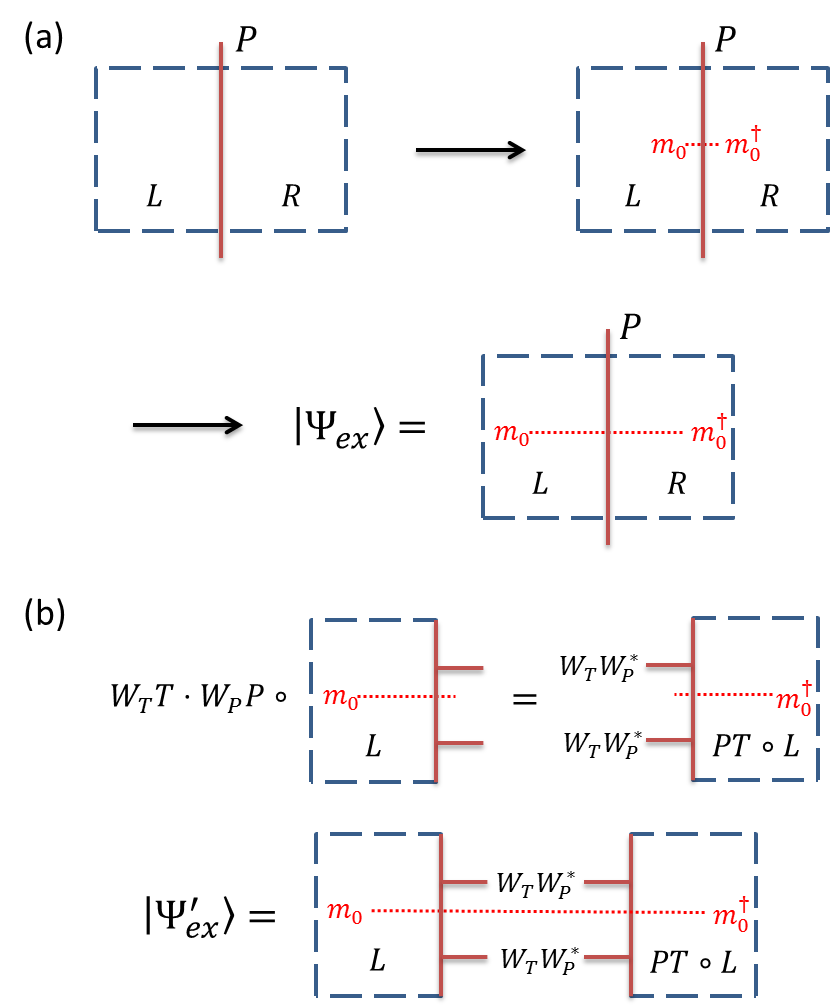}
    \caption{(a) The procedure to create $|\Psi_{ex}\rangle$ which is $\TT\cdot\PP$ invariant. One first creates a pair of $m_0$ and $m_0^\dagger$ from ground state, and then move away from each other. The global phase of $|\Psi_{ex}\rangle$ by requiring the wavefunction overlap between adjacent states to be real and positive. (b) $|\Psi'_{ex}\rangle$ is obtained by gluing between the original left-half of the tensor-network with the $\mathcal{T}\cdot\mathcal{P}$ transformed left-half tensor-network. The $\mathcal{T}\cdot\mathcal{P}$ transformed left-half tensor-network is obtained by transforming the physical legs of the left-half via $\mathcal{T}\cdot\mathcal{P}$, together with applying $W_{\mathcal{T}}\mathcal{T}\cdot W_\mathcal{P}\cdot\mathcal{T}^{-1}$ on all the virtual legs cut by the mirror line. $\chi_{m_0}(\TT\cdot\PP)$ is defined as phase difference between $|\Psi_{ex}\rangle$ and $|\Psi'_{ex}\rangle$.}
    \label{fig:flux_PT_qn}
\end{figure}

The more interesting and nontrivial situation is the time-reversal $\mathcal{T}$ and mirror reflection $\mathcal{P}$. It is straightforward to show that the assumption Eq. (\ref{eq:chi_J},\ref{eq:simple_case}) leads to the following transformation rules: $\mathcal{T}: e\rightarrow e^{\dagger}, m\rightarrow m$, and $\mathcal{P}: e\rightarrow e, m\rightarrow m^{\dagger}$. And the quantum numbers $\chi_{m}(g)$ should be treated as an element in $H^{1}(SG,Z_N)$ but with $\mathcal{T}$ and $\mathcal{P}$ acting anti-unitarily on $Z_N$. However, their combination $\mathcal{T}\cdot\mathcal{P}$ should be treated as unitary and the corresponding quantum number is sharply measurable. Below we present such an algorithm, which is depicted in Fig.\,\ref{fig:flux_PT_qn}.

Let us choose the positions of $m_0$ and $m_0^{\dagger}$ to be $\mathcal{P}$ image of each other. For example, we will consider the situation that $m_0$($m_0^{\dagger}$) is located in the left(right) half of the sample, and the mirror is the vertical line. Consequently $|\Psi_{ex}\rangle$ is $\mathcal{T}\cdot\mathcal{P}$ symmetric. Our goal is to measure the quantum number $\chi_{m_0}(\mathcal{T}\cdot\mathcal{P})$. This quantity may appear to be strange because we know that the combination $\mathcal{T}\cdot\mathcal{P}$ would send $m_0$ to $m_0^{\dagger}$ --- a different quasiparticle. But it turns out that this is exactly what is required to sharply measure $\chi_{m_0}(\mathcal{T}\cdot\mathcal{P})$.

Similar to previous example, our plan is to apply $\mathcal{T}\cdot\mathcal{P}$ only on $m_0$ and obtain a new excited physical state $|\Psi'_{ex}\rangle$. But because of the nature of $\mathcal{P}$, the $|\Psi'_{ex}\rangle$ should be obtained by gluing (i.e. contracting virtual legs) between the original left-half of the tensor-network with the $\mathcal{T}\cdot\mathcal{P}$ transformed left-half tensor-network (which is now on the right-half). Specifically,  the $\mathcal{T}\cdot\mathcal{P}$ transformed left-half tensor-network is obtained by transforming the physical legs of the left-half via $\mathcal{T}\cdot\mathcal{P}$, together with applying $W_{\mathcal{T}}\mathcal{T}\cdot W_\mathcal{P}\cdot\mathcal{T}^{-1}$ on all the virtual legs cut by the mirror line. The procedure to obtain $|\Psi'_{ex}\rangle$ is shown in Fig.\,\ref{fig:flux_PT_qn}(b).

If one naively uses the phase difference between this $|\Psi'_{ex}\rangle$ and $|\Psi_{ex}\rangle$ to measure $\chi_{m_0}(\mathcal{T}\cdot\mathcal{P})$, one will find that it is not well-defined. The reason is that the global phase factor of $|\Psi_{ex}\rangle$ is not properly chosen yet. In order to sharply measure $\chi_{m_0}(\mathcal{T}\cdot\mathcal{P})$, one needs to fully determine the global phase factor of $|\Psi_{ex}\rangle$ relative to the ground state in the following sense. In order to construct $|\Psi_{ex}\rangle$, one can imagine to firstly create a pair of $m_0$ and $m_0^{\dagger}$ near each other, and then further move them away from each other to a large distance, while \emph{maintaining $\mathcal{T}\cdot\mathcal{P}$ over the whole process}, as shown in Fig.\,\ref{fig:flux_PT_qn}(a). This process would create a sequence of states, with ground state as the first one and $|\Psi_{ex}\rangle$ as the last one. The global phase factor of $|\Psi_{ex}\rangle$ is determined by requiring the \emph{wavefunction overlap between adjacent states in this sequence to be positive and real}.

Because the global phase factor of $|\Psi_{ex}\rangle$ is fixed, the only ambiguity in the tensor-network construction of $|\Psi_{ex}\rangle$ is a global phase factor $e^{i\theta}$ on the left-half, and $e^{-i\theta}$ on the right-half. But this relative phase ambiguity would not affect the phase difference between this $|\Psi'_{ex}\rangle$ and $|\Psi_{ex}\rangle$ discussed above. Namely the phase difference between this $|\Psi'_{ex}\rangle$ and $|\Psi_{ex}\rangle$ is now sharply measurable, which is nothing but $\chi_{m_0}(\mathcal{T}\cdot\mathcal{P})$.

\subsection{Examples}\label{sec:tensor_examples}
We present some explicit examples for the 2+1D SPT. Let us consider square lattice with a $d=2$ qubit on each site. For simplicity, we will focus on the case where all tensors are translationally invariant. We label the legs of a site tensor as $\alpha,\beta,\gamma,\delta$, and plaquette IGG elements act as $\lambda_l,\lambda_u,\lambda_r,\lambda_d$, as shown in Fig.\,\ref{fig:plaquette_IGG}.

\subsubsection{SPT phases protected by inversion symmetry}
Consider nontrivial SPT phases protected by inversion symmetry $\mathcal{I}$. According to the discussion in the previous part, the inversion protected SPT phases are classified by $H^2(Z_2^\mathcal{I},U(1))=Z_2$. Namely, there is only one nontrivial phase. 

We start with a tensor network with $Z_2$ global IGG $\left\{ \mathrm{I},\lambda \right\}$. Tensor equations for this nontrivial SPT phase are
\begin{align}
    &W_\mathcal{I}\mathcal{I}\cdot W_\mathcal{I}\mathcal{I}=\lambda\notag\\
    &\,{}^{W_\mathcal{I}\mathcal{I}}\!\lambda_p=-\lambda_p
    \label{}
\end{align}
where $\lambda_p$ is the plaquette IGG element. For a single leg action, we have
\begin{align}
    &\,{}^{\mathcal{I}}\!W_\mathcal{I}(i)=W_I^\mathrm{t}(\mathcal{I}(i)), \quad i=\alpha,\beta,\gamma,\delta\notag\\
    &\,{}^{\mathcal{I}}\!\lambda_j=\lambda_{\mathcal{I}(j)}^\mathrm{t}, \quad j=l,u,r,d
    \label{}
\end{align}
Here, due to translational invariance, we define $\lambda_j\triangleq\lambda_p(j),\,\forall p$.

The simplest solution requires internal bond dimension $D=6$. IGG elements are represented as
\begin{align}
    &\lambda=\sigma_0\oplus(-\sigma_0\otimes\sigma_0)\notag\\
    &\lambda_l=\lambda_u=\sigma_z\oplus(\sigma_z\otimes\sigma_z)\notag\\
    &\lambda_r=\lambda_d=\sigma_z\oplus(-\sigma_z\otimes\sigma_z)
    \label{eq:inversion_SPT_plaquette_IGG}
\end{align}
and the inversion operation on internal legs is
\begin{align}
    W_\mathcal{I}(i)=\sigma_x\oplus(\sigma_y\otimes\sigma_x)
    \label{eq:inversion_SPT_symmetry}
\end{align}

Now, let us determine the constraint Hilbert space for the nontrivial SPT phase. As shown in Fig.\,\ref{fig:plaquette_IGG}(c), we require that the single tensor lives in the subspace which is invariant under action of plaquette IGG elements, where the nontrivial plaquette IGG element in Eq.(\ref{eq:inversion_SPT_plaquette_IGG}). Further, we require the single tensor to be inversion symmetric: $W_\mathcal{I}\mathcal{I}\circ T^a=T^a$, where $W_\mathcal{I}$ is given in Eq.(\ref{eq:inversion_SPT_symmetry}). Then, by solving these linear equations, we obtain a $D_{\mathcal{I}}=74$ dimensional (complex) Hilbert space. We point out that the original Hilbert space for a site tensor is $dD^4=2592$ dimensional.

It is also straightforward to check that the only nontrivial cocycle phase is $\omega(\mathcal{I},\mathcal{I},\mathcal{I})=-1$, which cannot be tuned away.

\subsubsection{SPT phases protected by time reversal and reflection symmetries}
Now, we study a more interesting example: 2D SPT phases protected by $Z_2^P\times Z_2^\mathcal{T}$ (reflection and time reversal) symmetry. The four group elements are $\left\{ \mathrm{I},P,\mathcal{T},P\mathcal{T} \right\}$, where $\mathcal{T}=\sigma_x\mathcal{K}$ and $P$ is the reflection along $y$ axis. As we mentioned above, both $P$ and $\mathcal{T}$ should be treated as ``anti-unitary'' action. Then, $ P\mathcal{T} $ should be treated as a unitary action. Namely, we have
\begin{align}
    H^3(Z_2^P\times Z_2^\mathcal{T},U(1))=H^3(Z_2\times Z_2^\mathcal{T},U(1))=Z_2\times Z_2
    \label{}
\end{align}

The tensor equations for these SPT phases are:
\begin{align}
    &W_\TT\TT W_\TT\TT=\lambda(\TT,\TT)\notag\\
    &W_PP W_PP=\lambda(P,P)\notag\\
    &W_PP W_\mathcal{T}\TT = W_\TT\TT W_PP\notag\\
    &^{W_PPW_\TT\TT}\!\lambda_p=-\lambda_p
    \label{}
\end{align}
where $\lambda(\TT,\TT),\lambda(P,P)$ belongs to the global $Z_2$ IGG. And different choice of $\lambda$'s gives different SPT phases.

By definition, the action of symmetry on $W$'s and $\lambda$'s are
\begin{align}
    &^{\TT}\!W_R(i)=W_R^*(i)\notag\\
    &^{P}\!W_R(\alpha/\gamma)=W_R^\mathrm{t}(\gamma/\alpha)=(W_R^{-1}(\alpha/\gamma))^\mathrm{t}\notag\\
    &^{P}\!W_R(\beta/\delta)=W_R(\beta/\delta)
    \label{}
\end{align}
as well as
\begin{align}
    ^{\TT}\!\lambda_j=\lambda_j^*,\quad ^{P}\!\lambda_{l/r}=\lambda_{r/l}^{-1}, \quad^{P}\!\lambda_{u/d}=(\lambda_{u/d}^{-1})^\mathrm{t}
    \label{}
\end{align}

To realize these SPT phases, we start from $D=6$ PEPS. Without any constraint, a single tensor lives in a $dD^4=2592$ dimensional (complex) Hilbert space. IGG elements are chosen as
\begin{align}
    &\lambda=\sigma_0\oplus(-\sigma_0\otimes\sigma_0)\notag\\
    &\lambda_l=\sigma_z\oplus(\sigma_z\otimes\sigma_z),\quad \lambda_r=\sigma_z\oplus(-\sigma_z\otimes\sigma_z)\notag\\
    &\lambda_u=\sigma_z\oplus(\sigma_z\otimes\sigma_0),\quad \lambda_d=\sigma_z\oplus(-\sigma_z\otimes\sigma_0)
    \label{}
\end{align}

In the following, we discuss each class in $Z_2\times Z_2$ separately.
\begin{enumerate}
    \item $\lambda(\TT,\TT)$ and $\lambda(P,P)$ are both trivial. We get a trivial symmetric phase in this case.
    \item $\lambda(\TT,\TT)=\mathrm{I}$, $\lambda(P\TT,P\TT)$ is nontrivial. Time reversal and reflection symmetries on internal legs are represented as
        \begin{align}
            &W_\TT(i)=\sigma_x\oplus(\sigma_x\otimes\sigma_0)\notag\\
            &W_P(\alpha)=W_P(\beta)=\sigma_0\oplus(\sigma_0\otimes\ii\sigma_y)\notag\\
            &W_P(\gamma)=W_P(\delta)=\sigma_0\oplus(\sigma_0\otimes(-\ii\sigma_y))
            \label{}
        \end{align}
        The constrained sub-space is the direct sum of a 40 dimensional real Hilbert space and a 40 dimensional pure imaginary Hilbert space.
    \item $\lambda(P,P)=\mathrm{I}$, $\lambda(\TT,\TT)$ is nontrivial. Time reversal and reflection symmetries are represented as
        \begin{align}
            &W_\TT(\alpha)=W_\TT(\beta)=\sigma_x\oplus(\ii\sigma_y\otimes\sigma_0)\notag\\
            &W_\TT(\gamma)=W_\TT(\delta)=\sigma_x\oplus(\ii\sigma_y\otimes\sigma_0)\notag\\
            &W_P(i)=\sigma_0\oplus(\sigma_0\otimes\sigma_x)
            \label{}
        \end{align}
        The constrained sub-space is the direct sum of a 44 dimensional real Hilbert space and a 44 dimensional pure imaginary Hilbert space.
    \item $\lambda(\TT,\TT)$ and $\lambda(P\TT,P\TT)$ are both nontrivial. Time reversal and reflection symmetries are represented as
        \begin{align}
            &W_\TT(\alpha)=W_\TT(\beta)=\sigma_x\oplus(\ii\sigma_y\otimes\sigma_0)\notag\\
            &W_\TT(\gamma)=W_\TT(\delta)=\sigma_x\oplus(-\ii\sigma_y\otimes\sigma_0)\notag\\
            &W_P(\alpha)=W_P(\beta)=\sigma_0\oplus(\ii\sigma_0\otimes\sigma_y)\notag\\ 
            &W_P(\gamma)=W_P(\delta)=\sigma_0\oplus(-\ii\sigma_0\otimes\sigma_y)
            \label{}
        \end{align}
        The constrained sub-space is the direct sum of a 40 dimensional real Hilbert space and a 40 dimensional pure imaginary Hilbert space.
\end{enumerate}

\subsubsection{Weak SPT phases protected by lattice group}
In this part, we consider the interplay of translation with point group. It is known that in the presence of translation, there are more SPT phases, which are named as weak indices\cite{chen2013symmetry}. In Ref.\onlinecite{cheng2015translational}, the authors find that weak indices can be elegantly incorporated into the cohomology formulation by treating translation in the same way as the on-site symmetry. Weak indices can be explicitly calculated using K\"{u}nneth formula. In (2+1)D, assuming the symmetry group $SG=\mathbb{Z}^2\times G$, where $\mathbb{Z}^2$ denotes translational symmetry on the plane, the formula reads
\begin{align}
    H^3[\mathbb{Z}^2\times G,U(1)]=&H^3[G,U(1)]\times(H^2[G,U(1)])^2\times\notag\\
    &H^1[G,U(1)]
    \label{}
\end{align}
where $H^3[G,U(1)]$ classify the strong indices, $(H^2[G,U(1)])^2$ are weak indices capture (1+1)D SPT phases and $H^0[G,U(1)]$ simply captures different charges in a unit cell.

In our tensor construction of SPT phases, we show that it is indeed natural to treat lattice symmetry in the same way as on-site symmetry. Not surprising, the interplay between translation and point group leads to new ``weak SPT'' phases. 

\begin{figure}
    \includegraphics[width=0.45\textwidth]{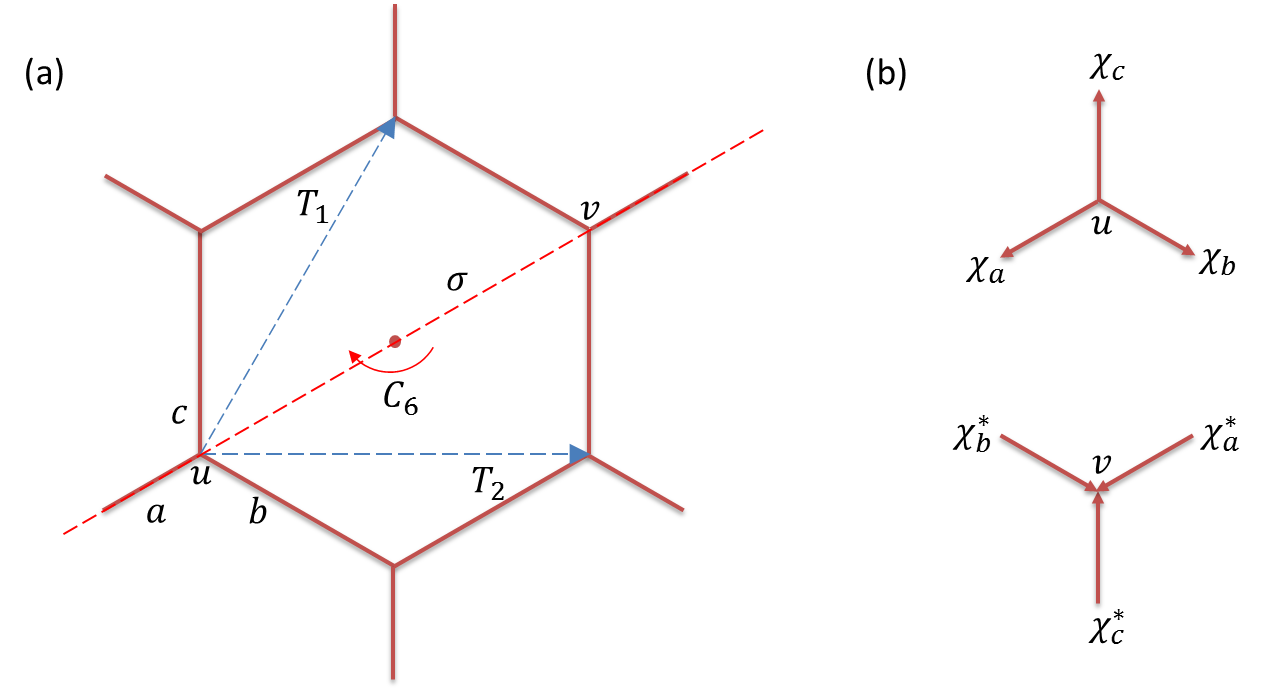}
    \caption{(a) The honeycomb lattice and generators the lattice symmetry group. $u,v$ labels sites while $a,b,c$ labels bonds in one unit cell. (b) The IGG element formed by phases. We require $\chi_a\cdot\chi_b\cdot\chi_c=1$.}
    \label{fig:honeycomb_lattice}
\end{figure}

Let us consider a spin system in a honeycomb lattice, as shown in Fig.\,\ref{fig:honeycomb_lattice}. In Ref.\onlinecite{kim2016featureless}, the authors obtain four classes of featureless insulators, which can be captured by two $Z_2$ indices $\chi_{C_6}$ and $\chi_\sigma$. The $Z_2\times Z_2$ classification can actually be understood as weak indices, which comes from the interplay between $C_6$, $\sigma$ and translation $T_1$, $T_2$. We leave the detailed calculation in Appendix\,\ref{app:weak_indices}.

\section{SPT phases in 3+1D}\label{sec:3D_tensor_SPT}
It is natural to generalize tensor construction of SPT phases to 3+1D. Before going into this higher dimensions, we would like to mention that in Appendix \ref{app:1D_SPT} we go to the lower dimensions and prove our results on 1+1D SPT. 

As the same in 2+1D, the symmetric tensor condition reads
\begin{align}
    W_gg\circ \mathbb{T}=\mathbb{T}
    \label{}
\end{align}
where $\mathbb{T}$ labels the 3+1D tensor network before contraction, and $W_g$ is the gauge transformation associated to symmetry $g$.

Then, $W_gg$ satisfies the group multiplication rules up to an IGG element:
\begin{align}
    W_{g_1}g_1W_{g_2}g_2=\lambda(g_1,g_2)W_{g_1g_2}g_1g_2
    \label{}
\end{align}
Due to associativity, $\lambda(g_1,g_2)$ satisfies the two cocycle condition:
\begin{align}
    \lambda(g_1,g_2)\lambda(g_1g_2,g_3)=\,{}^{W_{g_1}g_1}\!\lambda_(g_2,g_3)\lambda(g_1,g_2g_3)
    \label{}
\end{align}

In general, the nontrivial IGG leads to nontrivial topological order in 3+1D. In order to kill the topological order, we introduce cubic IGG $\{\lambda_c\}$, where $\lambda_c$ only acts nontrivially on the internal legs of cubic $c$. We further assume, any IGG element $\lambda$ can be decomposed to product of cubic IGG elements:
\begin{align}
    \lambda=\prod_c\lambda_c
    \label{}
\end{align}

Let us discuss the uniqueness of the above decomposition. We introduce the plaquette IGG $\{\xi_p\}$, which acts nontrivially only on legs belonging to plaquette $p$. Then, we can define a special kind of cubic IGG $\{\eta_c\}$, where any $\eta_c$ can be decomposed as multiplication of plaquette IGG elements,
\begin{align}
    \eta_c=\prod_{p\in c}\xi_p^c
    \label{}
\end{align}
If we further require $\xi_p^{c_1}=(\xi_p^{c_2})^{-1}$ for $p=c_1\cap c_2$, then, we get the decomposition of $\mathrm{I}$ as
\begin{align}
    \mathrm{I}=\prod_c \eta_c
    \label{}
\end{align}
In other words, the decomposition of a given IGG element $\lambda$ is not unique. We can always attach such kind of $\eta_c$ to get new decomposition. Then, roughly speaking, the cubic IGG element $\lambda_c(g_1,g_2)$ should satisfy a ``twist'' two cocycle condition, where the ``twist factors'' take value in $\{\eta_c\}$.

We can further prove $\eta_c(g_1,g_2,g_3)$ satisfies condition similar to three cocycles. We notice that the decomposition of $\eta_c$ to plaquette IGG elements $\xi_p$'s is also not unique, we can always attach some phase factor to $\xi_p$ such that the multiplication of $\xi_p$ is invariant. Then, $\xi_p$ should satisfy a ``twist'' three cocycle equation, where the ``twist factor'' is labeled as $\omega_p$. As shown in Appendix\,\ref{app:four_cohomology}, through some tedious calculations, we prove that $\omega_p$ satisfies the four cocycle condition, where time reversal and/or reflection symmetries are treated as antiunitary.
\begin{align}
    &\omega_p(g_1,g_2,g_3,g_4)\omega_p(g_1,g_2,g_3g_4,g_5)\omega_p(g_1g_2,g_3,g_4,g_5)=\notag\\
    &\,{}^{g_1}\!\omega_p(g_2,g_3,g_4,g_5)\omega_p(g_1,g_2g_3,g_4,g_5)\omega_p(g_1,g_2,g_3,g_4g_5)
    \label{}
\end{align}
and $\omega_p$ are defined up to coboundary.
\begin{align}
    &\omega_p(g_1,g_2,g_3,g_4)\sim\notag\\
    &\omega_p(g_1,g_2,g_3,g_4)\frac{\chi_p(g_1,g_2,g_3)\cdot \chi_p(g_1,g_2g_3,g_4)\cdot \,{}^{g_1}\!\chi_p(g_2,g_3,g_4)}{\chi_p(g_1g_2,g_3,g_4)\cdot \chi_p(g_1,g_2,g_3g_4)}
    \label{}
\end{align}

\section{Discussion}\label{sec:discussion}
In summary, by using tensor networks, we develop a general framework to (partially) classify bosonic SPT phases in any dimension, as well as construct generic tensor wavefunctions for each class. We find that for a general symmetry group $SG$, which include both on site symmetries as well as lattice symmetries, the cohomological bosonic SPT phases can be classified by $H^{d+1}(SG,U(1))$, where $d+1$ is the spacetime dimension. Here, time reversal and reflection symmetries should be treated as antiunitary. An important by-product is a generic relation between SET phases and SPT phases: SPT phases can be obtained from SET phases by condensing anyons carrying integer quantum numbers.

This work leaves several interesting future directions. On the conceptual side, it is known there are bosonic SPT phases beyond group cohomology classification. Famous examples include time reversal\cite{vishwanath2013physics,wang2013boson,burnell2014exactly} (or reflection\cite{song2016topological}) SPT phases in 3+1D, which has a $Z_2\times Z_2$ classification. However, group cohomology only capture a $Z_2$ class: $H^4(Z_2^\mathcal{T},U(1))=H^4(Z_2^P,U(1))=Z_2$. The other $Z_2$ is beyond our framework. It would be interesting to understand whether our framework can be further generalized to capture this missing index.

It is also interesting to generalize our formulation to construct generic wavefunctions for topological ordered phases as well as SET phases. We first point out that it is straightforward to ``(dynamically) gauge'' the on-site unitary discrete symmetries on tensor networks\cite{haegeman2015gauging}. Tensor networks invariant under symmetry $g$ satisfy the tensor equation $\mathbb{T}=W_gg\circ \mathbb{T}$. By gauging symmetry $g$, the new tensor equation becomes $\mathbb{T}=W_g\circ \mathbb{T}$, where $W_g$ is interpreted as gauge flux. Namely, for topological phases, we require additional global IGG elements, which cannot be decomposed into plaquette IGG elements.
By gauging onsite unitary symmetries of SPT phases\cite{levin2012braiding}, we are able to write down generic wavefunctions for Dijkgraaf-Witten type\cite{Dijkgraaf:1990p393} of topological ordered phases. Similarly, some SET phases can be obtained by gauging part of the symmetries\cite{Mesaros:2013p155115,hung2013quantized,cheng2016exactly,heinrich2016symmetry}.

As shown in Ref.\onlinecite{Buerschaper:2014p447,Williamson:2014p}, the SPT phases protected by onsite symmetries can also be classified by MPO injective PEPS. It would be interesting to see the connection between these two approaches.

As conjectured in Ref.\onlinecite{lin2014genralizations}, all topological ordered phases in 2+1D with gapped boundaries can be realized by exactly solvable models -- the string-net models, which have natural PEPS representations\cite{gu2009tensor,konig2009exact,sahinoglu2014characterizing,shukla2016boson,luo2016structure}. Our formulation is incapable to construct string-net models beyond the cohomological classes, which is due to the assumption that the only ambiguity of the decomposition to plaquette IGG element is the global phase ambiguity: $\lambda=\prod_p\lambda_p=\prod_p\chi_p\lambda_p$. In fact, it is possible to capture more phases by relaxing the assumption that we can hold matrix ambiguity, after which the cocycle condition becomes pentagon equations involving matrix. In addition, it would be interesting to generalize our formulation to fermionic cases using fermionic tensor network\cite{Corboz:2010p165104,Corboz:2009p165129,williamson2016fermionic}. We leave all these questions to future work.

On the practical side, it would be interesting to perform variational numerical simulations based on the symmetric tensor-network wavefunctions proposed here, and to test their performance. In particular, efficient gradient-based variational algorithms on tensor-network wavefunctions have been proposed\cite{vanderstraeten2016gradient}, which are exactly suitable to carry out these simulations.

We would like to thank Yuan-Ming Lu, Michael Hermele, Xie Chen, Chong Wang and Max Metlitski for helpful discussions, particularly on the ambiguities in the fluxon quantum numbers in symmetry defect arguments. This work is supported by National Science Foundation under Grant No. DMR-1151440. 

\bibliography{tensor_spt}

\clearpage

\begin{appendix}

\section{SPT phases in 1+1D}\label{app:1D_SPT}
In this part, we rederive the classification of 1D SPT\cite{Schuch:2011p165139,Chen:2011p35107,Chen:2011p235128,Pollmann:2012p75125} using the formulation we set in the main text. In particular, it is clear that time reversal and reflection symmetries act nontrivially on the two cohomology phase.

\begin{figure}
    \includegraphics[width=0.45\textwidth]{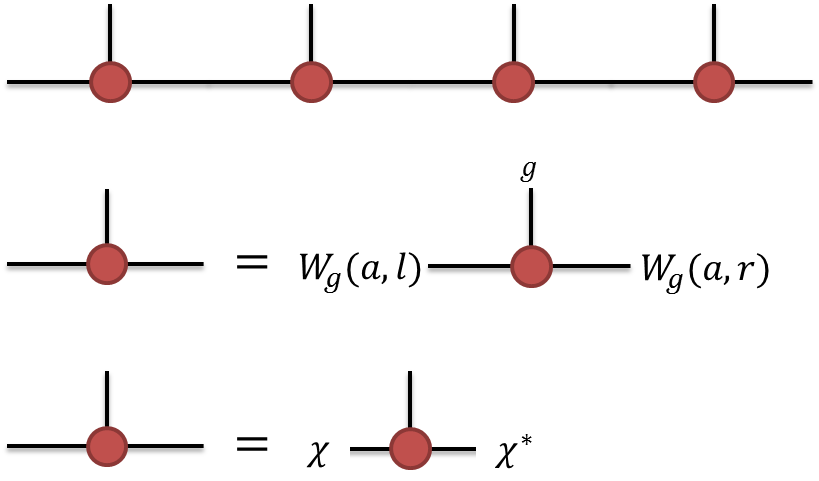}
    \caption{Symmetries and IGG in matrix product states.}
    \label{fig:1D_SPT}
\end{figure}

Consider an infinite MPS state with symmetry $SG$, then we can express the symmetric condition for a local tensor as
\begin{align}
    \mathbb{T}=W_gg\circ \mathbb{T}
    \label{}
\end{align}
where $\mathbb{T}$ represents a tensor network before contraction, $g\in SG$ and $W_g$ is the gauge transformation associated with $g$. 

Now, let us identify the IGG element. A single tensor is invariant if we multiply a phase $\chi$ to its left leg and $\chi^*$ to its right leg. Therefore, we at least have a $U(1)$ IGG for a generic MPS. In the following, we will focus on the $U(1)$ IGG.

Given the symmetry condition as well as the $U(1)$ IGG, we are able to list the tensor equation as following:
\begin{align}
    W_{g_1}g_1W_{g_2}g_2=\omega(g_1,g_2)W_{g_1g_2}g_1g_2
    \label{}
\end{align}
where $\omega(g_1,g_2)$ is an IGG element, which acts $\omega(g_1,g_2)$\,($\omega^*(g_1,g_2)$) on the left\,(right) leg. Due to associativity condition, we obtain the two cocycle condition for $\omega$ as
\begin{align}
    \omega(g_1,g_2)\omega(g_1g_2,g_3)=\,{}^{g_1}\!\omega(g_2,g_3)\omega(g_1,g_2g_3)
    \label{}
\end{align}
where $\,{}^{g_1}\!\omega\triangleq g_1\cdot\omega\cdot g_1^{-1}$. For onsite unitary $g_1$, the action is trivial. If $g_1$ is some anti-unitary operator, such as time reversal symmetry, $\,{}^{g_1}\!\omega=\omega^*$. For reflection symmetry $\sigma$, it maps the right\,(left) leg to the left\,(right) leg, so $\,{}^{\sigma}\!\omega=\omega^*$.

Notice, the symmetry operation is defined up to an IGG element. Namely, we have
\begin{align}
    \mathbb{T}=W_gg\circ \mathbb{T}=\epsilon(g)W_gg\circ \mathbb{T}
    \label{}
\end{align}
So, the equivalence condition for $\omega(g_1,g_2)$ is
\begin{align}
    \omega\sim\omega\cdot\frac{\epsilon(g_1g_2)}{\epsilon(g_1)\,{}^{g_1}\!\epsilon(g_2)}
    \label{}
\end{align}
In other words, $\omega$ is defined up to a coboundary. In summary, the 1D symmetric phase is classified by $H^2[SG,U(1)]$, where time reversal and reflection symmetries impose complex conjugation on the $U(1)$ phase factor.

\section{The three cohomology classification from tensor equations in 2+1D}\label{app:three_cohomology}
First, we discuss commutation relations between the IGG elements of plaquette $p_1$ and $p_2$ for later convenience:
\begin{align}
    \upsilon_{p_1p_2}\equiv(\lambda^1_{p_1})^{-1}(\lambda^2_{p_2})^{-1}\lambda^1_{p_1}\lambda^2_{p_2}
    \label{}
\end{align}
$\upsilon_{p_1p_2}$ still belongs to IGG according to the definition. Apparently, for the case where $p_1\cap p_2=\emptyset$ or they share only a common site, $\lambda^1_{p_1}$ and $\lambda^2_{p_2}$ commute. When $p_1$ and $p_2$ share a common edge $v$, $\upsilon_{p_1p_2}$ can only have nontrivial action on $v$. However, there is no such kind of nontrivial IGG, so $\lambda^1_{p_1}$ and $\lambda^2_{p_2}$ still commute. When $p_1=p_2\equiv p$, $\upsilon_{p}\equiv\upsilon_{p_1p_2}$ can act nontrivially on legs of $p$. So $\upsilon_{p}$ belongs to IGG of the plaquette $p$. To conclude, we have
\begin{align}
    (\lambda^1_{p_1})^{-1}(\lambda^2_{p_2})^{-1}\lambda^1_{p_1}\lambda^2_{p_2}=\lambda'_{p_1}\delta_{p_1p_2}
    \label{eq:plaquette_IGG_commutator}
\end{align}

As shown in the main text, $\lambda$'s satisfy the two cocycle relation:
\begin{align}
    \lambda(g_1,g_2)\lambda(g_1g_2,g_3)=\,{}^{W_{g_1}g_1}\!\lambda(g_2,g_3)\lambda(g_1,g_2g_3)
    \label{eq:app_IGG_two_cocycle}
\end{align}
According to Eq.(\ref{eq:IGG_phase_ambiguity}) and Eq.(\ref{eq:plaquette_IGG_commutator}), we can decompose IGG elements as
\begin{align}
    &\lambda(g_1,g_2)\lambda(g_1g_2,g_3)=\prod_p\lambda_p(g_1,g_2)\lambda_p(g_1g_2,g_3)\notag\\
    &\,{}^{W_{g_1}g_1}\!\lambda(g_2,g_3)\lambda(g_1,g_2g_3)=\prod_p\,{}^{W_{g_1}g_1}\!\lambda_p(g_2,g_3)\lambda_p(g_1,g_2g_3)
    \label{}
\end{align}
Further, due to the phase ambiguity in Eq.(\ref{eq:IGG_phase_ambiguity}), we conclude
\begin{align}
    &\lambda_p(g_1,g_2)\lambda_p(g_1g_2,g_3)=\notag\\
    &\omega_p(g_1,g_2,g_3)\,{}^{W_{g_1}g_1}\!\lambda_p(g_2,g_3)\lambda_p(g_1,g_2g_3)
    \label{eq:app_plq_IGG_twist_two_cocycle}
\end{align}

Now, we prove $\omega_p(g,g',g'')$ satisfies the 3-cocycle condition. We implement two ways to calculate the expression $\lambda_p(g_1,g_2)\lambda_p(g_1g_2,g_3)\lambda_p(g_1g_2g_3,g_4)$:
\begin{align}
    &\lambda_p(g_1,g_2)\lambda_p(g_1g_2,g_3)\lambda_p(g_1g_2g_3,g_4)\notag\\
    =&\omega_p(g_1,g_2,g_3)\,{}^{W_{g_1}g_1}\!\lambda_p(g_2,g_3)\lambda_p(g_1,g_2g_3)\lambda_p(g_1g_2g_3,g_4)\notag\\
    =&\omega_p(g_1,g_2,g_3)\,{}^{W_{g_1}g_1}\!\lambda_p(g_2,g_3)\omega_p(g_1,g_2g_3,g_4)\cdot\notag\\
    &\,{}^{W_{g_1}g_1}\!\lambda_p(g_2g_3,g_4)\lambda_p(g_1,g_2g_3g_4)\notag\\
    =&\omega_p(g_1,g_2,g_3)\omega_p(g_1,g_2g_3,g_4)\,{}^{g_1}\omega_p(g_2,g_3,g_4)\cdot\notag\\
    &\,{}^{W_{g_1}g_1W_{g_2}g_2}\!\lambda_p(g_3,g_4)\,{}^{W_{g_1}g_1}\lambda_p(g_2,g_3g_4)\lambda_p(g_1,g_2g_3g_4)
    \label{}
\end{align}
where we use Eq.(\ref{eq:app_plq_IGG_twist_two_cocycle}) to obtain the result. Notice that in the last line, we use the fact that $W_g$ always commutes with $\omega_p$, so ${}^{W_gg}\!\omega_p={}^g\!\omega_p$. Using another way to calculate, we get
\begin{align}
    &\lambda_p(g_1,g_2)\lambda_p(g_1g_2,g_3)\lambda_p(g_1g_2g_3,g_4)\notag\\
    =&\lambda_p(g_1,g_2)\omega_p(g_1g_2,g_3,g_4)\,{}^{W_{g_1g_2}g_1g_2}\!\lambda_p(g_3,g_4)\lambda_p(g_1g_2,g_3g_4)\notag\\
    =&\omega_p(g_1g_2,g_3,g_4)\,{}^{\lambda_p(g_1,g_2)W_{g_1g_2}g_1g_2}\!\lambda_p(g_3,g_4)\lambda_p(g_1,g_2)\cdot\notag\\
    &\lambda_p(g_1g_2,g_3g_4)\notag\\
    =&\omega_p(g_1g_2,g_3,g_4)\omega_p(g_1,g_2,g_3g_4)\,{}^{\lambda_p(g_1,g_2)W_{g_1g_2}g_1g_2}\!\lambda_p(g_3,g_4)\cdot\notag\\
    &\,{}^{W_{g_1}g_1}\!\lambda_p(g_2,g_3g_4)\lambda(g_1,g_2g_3g_4)\notag\\
    =&\omega_p(g_1g_2,g_3,g_4)\omega_p(g_1,g_2,g_3g_4)\,{}^{W_{g_1}g_1W_{g_2}g_2}\!\lambda_p(g_3,g_4)\cdot\notag\\
    &\,{}^{W_{g_1}g_1}\!\lambda_p(g_2,g_3g_4)\lambda(g_1,g_2g_3g_4)
    \label{}
\end{align}

Comparing the above results, we conclude $\omega_p$ satisfies three cocycle equation:
\begin{align}
    &\omega_p(g_1,g_2,g_3)\omega_p(g_1,g_2g_3,g_4)\,{}^{g_1}\omega_p(g_2,g_3,g_4)\notag\\
    &=\omega_p(g_1g_2,g_3,g_4)\omega_p(g_1,g_2,g_3g_4)
    \label{eq:app_omega_three_cocycle}
\end{align}
The action of $g$ on $\omega_p$ follows a very simple rule: for a leg $i$, we have $({}^g\!\omega_p)(i)=\omega^{s(g)}_{g^{-1}(p)}(g^{-1}(i))$, where $s(g)$ is trivial (complex conjugate) for unitary (anti-unitary) symmetry. 

According to Eq.(\ref{eq:IGG_phase_ambiguity}). We note that $\lambda_p(g,g')$ is defined up to a complex number. We can define $\lambda'_p(g,g')=\chi_p(g,g')\lambda_p(g,g')$. Then, we have
\begin{align}
    &\lambda'_p(g_1,g_2)\lambda'_p(g_1g_2,g_3)=\notag\\
    &\omega'_p(g_1,g_2,g_3)\,{}^{W_{g_1}g_1}\!\lambda'_p(g_2,g_3)\lambda'_p(g_1,g_2g_3)
    \label{}
\end{align}
Thus, we can always tune $\omega$ to be some $U(1)$ phase factor. In the following, we will restrict ourselves for the case where $\omega$'s and $\chi$'s are phase factors. Now, let us calculate $\omega'_p(g_1,g_2,g_3)$:
\begin{align}
    &\lambda'_p(g_1,g_2)\lambda'_p(g_1g_2,g_3)\notag\\
    =&\chi_p(g_1,g_2)\lambda_p(g_1,g_2)\chi_p(g_1g_2,g_3)\lambda_p(g_1g_2,g_3)\notag\\
    =&\chi_p(g_1,g_2)\chi_p(g_1g_2,g_3)\omega_p(g_1,g_2,g_3)\,{}^{W_{g_1}g_1}\!\lambda_p(g_2,g_3)\cdot\notag\\
    &\lambda_p(g_1,g_2g_3)\notag\\
    =&\frac{\chi_p(g_1,g_2)\chi_p(g_1g_2,g_3)}{{}^{g_1}\!\chi_p(g_2,g_3)\chi_p(g_1,g_2g_3)}\omega_p(g_1,g_2,g_3)\,{}^{W_{g_1}g_1}\!\lambda'_p(g_2,g_3)\cdot\notag\\
    &\lambda'_p(g_1,g_2g_3)
    \label{}
\end{align}
where we use the fact that ${}^{W_g}\!\chi_p=\chi_p$ in the last line. Comparing the above two equations, we conclude
\begin{align}
    \omega_p'(g_1,g_2,g_3)=\omega_p(g_1,g_2,g_3)\frac{\chi_p(g_1,g_2)\chi_p(g_1g_2,g_3)}{{}^{g_1}\!\chi_p(g_2,g_3)\chi_p(g_1,g_2g_3)}
    \label{}
\end{align}
It is straightforward to check that $\omega_p'$ also satisfies three cocycle condition in Eq.(\ref{eq:omega_three_cocycle}). In other words, the $\omega_p$ is well defined up 3-coboundary constructed by 2-cochain $\chi$. So, $\omega_p$ are classified by 3-cohomology $H^3(SG,U(1))$, where the symmetry group $SG$ may have nontrivial action on coefficient $U(1)$.

Notice that the physical wavefunction is invariant under gauge transformation $V$ as well as the IGG transformation $\widetilde{W}_g=\epsilon(g)W_g$, where $\epsilon(g)\in\mathrm{IGG}$. If $\omega_p$ classify the PEPS wavefunctions, $\omega_p$ should be invariant (up to coboundary) under these two kinds of transformations. 

For any gauge transformation $V$, $W_g\rightarrow VW_ggV^{-1}g^{-1}$. Then it is straightforward to prove that $\omega_p$ is invariant. 

Now, let us consider IGG transformation. For $\widetilde{W}_g=\epsilon(g)W_g$, we have
\begin{align}
    \widetilde{W}_{g_1}g_1\widetilde{W}_{g_2}g_2=\widetilde{\lambda}(g_1,g_2)\widetilde{W}_{g_1g_2}g_1g_2
    \label{eq:Wg_IGG_transformation}
\end{align}
where $\widetilde{\lambda}(g_1,g_2)=\epsilon(g_1)\,{}^{W_{g_1}g_1}\!\epsilon(g_2)\lambda(g_1,g_2)\epsilon^{-1}(g_1g_2)$.

Restrict to one plaquette, we calculate
\begin{align}
    &\widetilde{\lambda}_p(g_1,g_2)\widetilde{\lambda}_p(g_1g_2,g_3)\epsilon_p(g_1g_2g_3)\notag\\
    =&\epsilon_p(g_1)\,{}^{W_{g_1}g_1}\!\epsilon_p(g_2)\lambda_p(g_1,g_2)\,{}^{W_{g_1g_2}g_1g_2}\!\epsilon_p(g_3)\lambda_p(g_1g_2,g_3)\notag\\
    =&\epsilon_p(g_1)\,{}^{W_{g_1}g_1}\!\epsilon_p(g_2)\,{}^{W_{g_1}g_1W_{g_2}g_2}\!\epsilon_p(g_3)\lambda_p(g_1,g_2)\lambda_p(g_1g_2,g_3)\notag\\
    \label{}
\end{align}
where we use Eq.(\ref{eq:Wg_IGG_transformation}) several times. In second line, we have used the fact that ${}^{\lambda_p}\!\epsilon_p=\,{}^{\lambda}\!\epsilon_p$ as well as Eq.(\ref{eq:Wg_proj_rep}). On the other hand,
\begin{align}
    &{}^{\widetilde{W}_{g_1}g_1}\!\widetilde{\lambda}_p(g_2,g_3)\widetilde{\lambda}_p(g_1,g_2g_3)\epsilon_p(g_1g_2g_3)\notag\\
    =&{}^{\epsilon_p(g_1)W_{g_1}g_1}\!(\epsilon_p(g_2)\,{}^{W_{g_2}g_2}\!\epsilon_p(g_3)\lambda_p(g_2,g_3)\epsilon_p^{-1}(g_2g_3))\epsilon_p(g_1)\cdot\notag\\
    &\,{}^{W_{g_1}g_1}\!\epsilon_p(g_2,g_3)\lambda_p(g_1,g_2g_3)\notag\\
    =&\epsilon_p(g_1)\,{}^{W_{g_1}g_1}\!\epsilon_p(g_2)\,{}^{W_{g_1}g_1W_{g_2}g_2}\!\epsilon_p(g_3)\,{}^{W_{g_1}g_1}\!\lambda_p(g_2,g_3)\lambda_p(g_1,g_2g_3)\notag\\
    \label{}
\end{align}
According to Eq.(\ref{eq:plq_IGG_twist_two_cocycle}), we conclude that
\begin{align}
    &\widetilde{\lambda}_p(g_1,g_2)\widetilde{\lambda}_p(g_1g_2,g_3)=\notag\\
    &\omega_p(g_1,g_2,g_3)\,{}^{\widetilde{W}_{g_1}g_1}\!\widetilde{\lambda}_p(g_2,g_3)\widetilde{\lambda}_p(g_1,g_2g_3)
    \label{}
\end{align}
So, one obtains the same 3-cocycle for $\epsilon_p$ transformation.

We now make a general remark: our tensor construction for SPT phases in 2+1D is related to \emph{crossed module extension} known in the mathematical literature. 

Let us first review the SPT phases in 1+1D with symmetry group $SG$, which are classified by different projective representations of $SG$, or equivalently, by different central extensions of $SG$:
\begin{align}
    1\rightarrow U(1)\rightarrow E\rightarrow SG\rightarrow 1
    \label{}
\end{align}
In the tensor network construction, the center $U(1)$ is mapped to the $U(1)$ phase IGG, and symmetry actions on all legs of the tensor network $W_gg$ together with the $U(1)$ IGG form the extended group $E$. So, the construction of 1+1D SPT phases by MPS can be viewed as a realization of the central extension.

A crossed module extension is an exact sequence:
\begin{align}
    1\rightarrow U(1)\rightarrow N\xrightarrow{\varphi} E\rightarrow SG\rightarrow 1
    \label{eq:crossed_module_extension}
\end{align}
with a left action of $E$ on $N$, represented by $n\mapsto \,{}^{e}\!n$, such that $\,{}^{\varphi(n)}\!n'=nn'n^{-1}$ as well as $\varphi\left( \,{}^{e}\!n \right)=e\varphi(n)e^{-1}$, for all $n,n'\in N$ and $e\in E$. It is well known\cite{holt1979interpretation,huebschmann1980crossed,thomas2009third,brown2012cohomology,else2014classifying} that the crossed module extensions of $SG$ by $U(1)$ are classified by $H^3(SG,U(1))$, which is the same object classifies the 2+1D SPT phases protected by $SG$. As in the 1+1D case, our construction can be viewed as a realization of a crossed module extension by tensor networks. Namely, given a crossed module extension characterized by a three cohomology $[\omega]$, we can write down tensor equations realize this crossed module extension and construct generic tensor wavefunctions for the SPT phase characterized by $[\omega]$. This fact also indicates that our tensor constructions are able to capture all cohomological bosonic SPT phases in 2+1D.

Now, let us describe the procedure to obtain tensor equations from a crossed module extension. Given a crossed module extension in Eq.(\ref{eq:crossed_module_extension}), one can decompose it to two short exact sequences as following:
\begin{align}
    &1\rightarrow U(1)\rightarrow N\xrightarrow{\phi} M\rightarrow 1\notag\\
    &1\rightarrow M\xrightarrow{i} E\rightarrow SG\rightarrow 1
    \label{eq:decompose_crossed_module}
\end{align}
where $M$ is identified as $\varphi(N)$, and $i:M\hookrightarrow E$ is an inclusion map. Apparently $\varphi=i\circ\phi$.

We can write down tensor equations to realize these two short exact sequence. As shown in Eq.(\ref{eq:Wg_proj_rep}), symmetry actions on all legs of tensor networks $\{W_gg|\forall g\in SG\}$ form a projective representation with coefficient in group $\left\{ \lambda \right\}$, which we identify as $M$. In the anyon condensation context, $M$ is the gauge group characterizing the topological order before condensation. $M$ together with $\left\{ W_gg|\forall g \right\}$ form the extended group $E$, which captures the SET physics before anyon condensation. According to the assumption, $\forall\lambda\in M$ can be decomposed to plaquette IGG elements: $\lambda=\prod_p\lambda_p$. An element $n\in N$ is identified as a set of plaquette IGG elements: $n=\left\{ \lambda_p|\forall p \right\}$, which satisfies $\prod_p\lambda_p=\lambda$. Then, $N=\left\{ \{\lambda_p|\forall p\}|\prod_p\lambda_p=\lambda\in M \right\}$. And mapping $\phi$ is defined as 
\begin{align}
    &\phi:N\mapsto M, \notag\\
    &\phi(n)=\prod_p\lambda_p 
    \label{}
\end{align}
It is easy to see that the kernel of $\phi$ forms a $U(1)$ group: $\left\{ \left\{ \chi_p|\forall p \right\}|\prod_p\chi_p=I \right\}\cong U(1)$.

Now, let us consider the action of $E$ on $N$. Set $n=\left\{ \lambda_p|\forall p \right\}$, $n'=\left\{ \lambda_p'|\forall p \right\}$ and $e=\lambda^{(e)}W_gg\in E$, we define the action as
\begin{align}
    &\,{}^{\varphi(n)}\!n'\triangleq\left\{\,{}^{\lambda}\!\lambda_p'|\forall p\right\}=\left\{\lambda_p\cdot\lambda_p'\cdot\lambda_p^{-1}|\forall p\right\}=n\cdot n'\cdot n^{-1}\notag\\
    &\varphi(\,{}^{e}\!n)=\prod_p\,{}^{\lambda^{(e)}W_{g}g}\!\lambda_p=\,{}^{\lambda^{(e)}W_{g}g}\!\lambda=e\cdot\varphi(n)\cdot e^{-1}
    \label{}
\end{align}
which indeed satisfies the crossed module condition. In summary, from a crossed module extension, we are able to construct tensor equations for SPT phases and vice versa.


\section{Examples for SPT phases with $Z_2\times Z_2^\TT$}
In this part, we write down examples for SPT phases protected by $Z_2\times Z_2^\TT$ symmetry on square lattice. We label group elements as $\left\{ \mathrm{I},g,\TT,g\TT \right\}$, where $g=\sigma_x$ and the time reversal acts as complex conjugation. For simplicity, we assume translationally invariant tensors.

First, we have $H^3(Z_2\times Z_2^\TT,U(1))=Z_2\times Z_2$. The tensor equations for these SPT phases are
\begin{align}
    &W_g^2=\lambda(g,g)\notag\\
    &W_\TT\TT W_\TT \TT=\lambda(\TT,\TT)\notag\\
    &W_gW_\TT\TT=W_\TT\TT W_g\notag\\
    &\,{}^{W_{g}g}\!\lambda_p=-\lambda_p
    \label{}
\end{align}
Here $\lambda(g,g)$ and $\lambda(\TT,\TT)$ are IGG elements, which take value in $\left\{ \mathrm{I}, \lambda \right\}$. So, we totally get $Z_2\times Z_2$ classification.

We list the results in the following.
\begin{enumerate}
    \item $\lambda(g,g)=\lambda(\TT,\TT)=\mathrm{I}$. This case corresponds to trivial phase.
    \item $\lambda(g,g)=\lambda$ and $\lambda(\TT,\TT)=\mathrm{I}$. This class can be realized by $D=4$ PEPS, where the local Hilbert space for a single tensor is $dD^4=512$ dimensional.

        Elements of IGG are represented as
        \begin{align}
            &\lambda=\sigma_0\oplus(-\sigma_0)\notag\\
            &\lambda_l=\lambda_u=\sigma_z\oplus\sigma_z\notag\\
            &\lambda_r=\lambda_d=\sigma_z\oplus(-\sigma_z)
            \label{}
        \end{align}
        Symmetries are represented as
        \begin{align}
            &W_g(\alpha)=W_g(\beta)=W_g^{-1}(\gamma)=W_g^{-1}(\delta)=\sigma_x\oplus\ii\sigma_y\notag\\
            &W_\TT(i)=\sigma_0\oplus\sigma_0
            \label{}
        \end{align}
        The constrained sub-space is a 16 dimensional real Hilbert space.
    \item $\lambda(g,g)=\lambda(\TT,\TT)=\lambda$. This class can be realized by $D=4$ PEPS, where the local Hilbert space for a single tensor is $dD^4=512$ dimensional. 

        IGG elements are represented as
        \begin{align}
            &\lambda=\sigma_0\oplus(-\sigma_0)\notag\\
            &\lambda_l=\lambda_u=\sigma_z\oplus\sigma_z\notag\\
            &\lambda_r=\lambda_d=\sigma_z\oplus(-\sigma_z)
            \label{}
        \end{align}
        Symmetries are represented as
        \begin{align}
            &W_g(\alpha)=W_g(\beta)=W_g^{-1}(\gamma)=W_g^{-1}(\delta)=\sigma_x\oplus\ii\sigma_y\notag\\
            &W_\TT(\alpha)=W_\TT(\beta)=W_\TT^{-1}(\gamma)=W_\TT^{-1}(\delta)=\sigma_x\oplus\ii\sigma_y\notag\\
            \label{}
        \end{align}
        The constrained sub-space is a 16 dimensional real Hilbert space. Notice, this phase is related to the previous case by relabelling $g\TT$ as $\TT$.
    \item $\lambda(g,g)=\mathrm{I}$ and $\lambda(\TT,\TT)=\lambda$. This class can be realized by $D=6$ PEPS, where the local Hilbert space for a single tensor is $dD^4=2592$ dimensional.
        
        IGG elements are
        \begin{align}
            &\lambda=\sigma_0\oplus(-\sigma_0\otimes\sigma_0)\notag\\
            &\lambda_l=\lambda_u=\sigma_z\oplus(\sigma_z\otimes\sigma_z)\notag\\
            &\lambda_r=\lambda_d=\sigma_z\oplus(-\sigma_z\otimes\sigma_z)
            \label{}
        \end{align}
        Symmetries are represented as
        \begin{align}
            &W_g(i)=\sigma_x\oplus(\sigma_x\otimes\sigma_0)\notag\\
            &W_\TT(\alpha)=W_\TT(\beta)=W_\TT^{-1}(\gamma)=W_\TT^{-1}(\delta)=\sigma_0\oplus(\sigma_0\otimes\ii\sigma_y)\notag\\
            \label{}
        \end{align}
        The constrained sub-space is the direct sum of an 18 dimensional real Hilbert space and 16 dimensional pure imaginary Hilbert space.
\end{enumerate}

\section{Examples on ``weak indices''}\label{app:weak_indices}
In this part, we discuss weak SPT indices due to interplay of point group and translation on honeycomb lattice. The lattice group is defined by group generators $T_1, T_2, C_6, \sigma$, as in Fig. \ref{fig:honeycomb_lattice}(a). And the relation between this generators are listed as following:
\begin{align}
  T_2^{-1}T_1^{-1}T_2T_1&=\mathrm{e}\notag\\
  C_6^{-1}T_2^{-1}C_6 T_1&=\mathrm{e}\notag\\
  C_6^{-1}T_2^{-1}T_1C_6T_2&=\mathrm{e}\notag\\
  \sigma^{-1}T_1^{-1}\sigma T_2&=\mathrm{e}\notag\\
  \sigma^{-1}T_2^{-1}\sigma T_1&=\mathrm{e}\notag\\
  \sigma C_6\sigma C_6&=\mathrm{e}\notag\\
  C_6^{6}=\sigma^2=\mathcal{T}^2&=\mathrm{e}
  \label{eq:honeycomb_sym_group}
\end{align}

Consider group relation $T_2^{-1}T_1^{-1}T_2T_1=\mathrm{e}$, the corresponding tensor equation takes the form
\begin{align}
  &W_{T_2}^{-1}(T_2(x,y,i))W_{T_1}^{-1}(T_1T_2(x,y,i))W_{T_2}(T_1T_2(x,y,i))\cdot\notag\\
  &W_{T_1}(T_2^{-1}T_1T_2(x,y,i))=\chi_{12}(x,y,i)
  \label{eq:chi12_virtual_leg}
\end{align}
where $i\in\{a,b,c\}$ labels bonds. And $\chi_{12}(x,y,i)\triangleq\chi_{12}(x,y,u,i)$ denotes the IGG element action on leg $i$ of site $(x,y,u)$. By tuning the phase ambiguity $W_{T_i}\rightarrow \epsilon_{T_i}W_{T_i}$, we are able to set $\chi_{12}=1$. Further, we can set $W_{T_1}=W_{T_2}=\mathrm{I}$ by gauge transformation $V$. Then, the remaining gauge transformation $V$ should be translational invariant: $V(x,y,i)=V(i)$.

Now, let us add rotation symmetry $C_6$. For relation $C_6^{-1}T_2^{-1}C_6 T_1=\mathrm{e}$, we have
\begin{align}
    W_{C_6}^{-1}(x,y,i)W_{C_6}(x,y+1,i)=\chi_{C_6T_1}(C_6^{-1}(x,y,i))
    \label{}
\end{align}
where we use $W_{T_i}=\mathrm{I}$. It is easy to see, by choosing proper $\epsilon_{C_6}$, and redefine $W_{C_6}\rightarrow\epsilon_{C_6}W_{C_6}$, we can set $\chi_{C_6T_1}$ to be identity. Thus, $W_{C_6}$ are equal in the $y$ direction: 
\begin{align}
    W_{C_6}(x,y+1,i)=W_{C_6}(x,y,i)=W_{C_6}(x,0,i)
    \label{}
\end{align}
The remaining $\epsilon_{C_6}$ should also be $y$-invariant:  $\epsilon_{C_6}(x,y+1,i)=\epsilon_{C_6}(x,y,i)$. Further, as loops of phases, $\epsilon_{C_6}$ should satisfy the Gauss law:
\begin{align}
    &\epsilon_{C_6}(x,y,a)\cdot\epsilon_{C_6}(x,y,b)\cdot\epsilon_{C_6}(x,y,c)=1\notag\\
    &\epsilon_{C_6}^*(x,y,a)\cdot\epsilon_{C_6}^*(x-1,y,b)\cdot\epsilon_{C_6}^*(x,y-1,c)=1
    \label{eq:epsilon_C6_Gauss_law}
\end{align}
Since $\epsilon_{C_6}(x,y-1,c)=\epsilon_{C_6}(x,y,c)$, we conclude the remaining $\epsilon_{C_6}(x,y,b)$ are also $x$-invariant. Then, we left with the $\epsilon_{C_6}$ redundancy satisfying
\begin{align}
    &\epsilon_{C_6}(x,y,a/c)=\epsilon_{C_6}(x,0,a/c)\notag\\
    &\epsilon_{C_6}(x,y,b)=\epsilon_{C_6}(b)
    \label{eq:epsilon_C6_b_invariance}
\end{align}

We consider another relation $C_6^{-1}T_2^{-1}T_1C_6T_2=\mathrm{e}$, and the tensor equation reads
\begin{align}
    W_{C_6}^{-1}(x,y,i)W_{C_6}(x-1,y+1,i)=\chi_{C_6T_2}(C_6^{-1}(x,y,i))
    \label{}
\end{align}
Since $W_{C_6}$ is $y$ independent, $\chi_{C_6T_2}(C_6^{-1}(x,y,i))=\chi_{C_6T_2}(C_6^{-1}(x,0,i))$. Under $\epsilon_{C_6}$ transformation, 
\begin{align}
    &\chi_{C_6T_2}(C_6^{-1}(x,0,i))\rightarrow\notag\\
    &\chi_{C_6T_2}(C_6^{-1}(x,0,i))\cdot\epsilon_{C_6}(x,0,i)\cdot\epsilon_{C_6}^*(x-1,0,i)
    \label{}
\end{align}
We can tune $\epsilon_{C_6}(x,0,c)$ to make $\chi_{C_6T_2}(C_6^{-1}(x,y,c))=1$. The remaining $\epsilon_{C_6}(x,y,i)=\epsilon_{C_6}(i)$, which do not affect $\chi_{C_6T_1}$. 

Since $\chi_{T_2C_6}$ are also loops of phases satisfying Gauss law, $\chi_{T_2C_6}(x,y,a)=\chi_{T_2C_6}^*(x,y,b)\equiv\chi_{C_6}$. Then, we have
\begin{align}
    &W_{C_6}(x,y,a)=\chi_{C_6}^xW_{C_6}(a)\notag\\
    &W_{C_6}(x,y,b)=(\chi_{C_6}^*)^xW_{C_6}(b)\notag\\
    &W_{C_6}(x,y,c)=W_{C_6}(c)
    \label{eq:honeycomb_WC6_transl}
\end{align}

For relation $C_6^6=\mathrm{e}$, we have
\begin{align}
    &\chi_{C_6}^{4x+2y-3}\cdot W_{C_6}(b)\cdot[W_{C_6}^{-1}(c)]^\mathrm{t}\cdot W_{C_6}(a)\cdot[W_{C_6}^{-1}(b)]^\mathrm{t}\cdot \notag\\
    &W_{C_6}(c)\cdot [W_{C_6}^{-1}(a)]^\mathrm{t}\notag\\
    =&\chi_{C_6^6}(x,y,a)=\chi_{C_6^6}^*(-x+1,-y+1,a)\notag\\
    =&\chi_{C_6^6}(-x-y+1,x,u,b)=\chi_{C_6^6}^*(x+y-1,-x+1,u,b)\notag\\
    =&\chi_{C_6^6}(y,-x-y+1,u,c)=\chi_{C_6^6}^*(-y+1,x+y-1,u,c)
    \label{}
\end{align}
where we use Eq.(\ref{eq:honeycomb_WC6_transl}) to get the above equation. Then, we conclude 
\begin{align}
    &\chi_{C_6^6}(x,y,u,a)=\chi_{C_6}^{4x+2y-3}\cdot\xi_{C_6}\notag\\
    &\chi_{C_6^6}(x,y,u,b)=\chi_{C_6}^{-2x+2y-1}\cdot\xi_{C_6}\notag\\
    &\chi_{C_6^6}(x,y,u,c)=\chi_{C_6}^{-2x+4y+1}\cdot\xi_{C_6}
    \label{}
\end{align}
where $\xi_{C_6}\triangleq W_{C_6}(b)\cdot[W_{C_6}^{-1}(c)]^\mathrm{t}\cdot W_{C_6}(a)\cdot[W_{C_6}^{-1}(b)]^\mathrm{t}\cdot W_{C_6}(c)\cdot [W_{C_6}^{-1}(a)]^\mathrm{t}=\pm1$. Using Gauss law on site $(x,y,u/v)$, we obtain
\begin{align}
    \chi_{C_6}^3=\xi
    \label{}
\end{align}

Under gauge transformation $V(i)$,
\begin{align}
    W_{C_6}(i)\rightarrow V(i)\cdot W_{C_6}(i)\cdot V^\mathrm{t}(C_6^{-1}(i))
    \label{}
\end{align}
So, we can make 
\begin{align}
    &W_{C_6}(x,y,a)=\chi_{C_6}^x\notag\\
    &W_{C_6}(x,y,a)=(\chi_{C_6}^*)^x\notag\\
    &W_{C_6}(x,y,c)=W_{C_6}(c)=\chi_{C_6}^3W_{C_6}^\mathrm{t}(c)
    \label{eq:honeycomb_WC6}
\end{align}
The remaining $V$ satisfies $V(a)=V(c)=[V^{-1}(b)]^\mathrm{t}$.

Now, let us consider reflection symmetry $\sigma$. For relation $\sigma^{-1}T_2^{-1}\sigma T_1=\mathrm{e}$, we have
\begin{align}
    W_\sigma^{-1}(x,y,i)W_\sigma(x,y+1,i)=\chi_{\sigma T_1}(\sigma^{-1}(x,y,i))
    \label{}
\end{align}
Using $\epsilon_\sigma$ ambiguity, $\chi_{\sigma T_1}$ can be tuned to identity. Thus, $W_\sigma$ is $y$ independent. And the remaining $\epsilon_\sigma$ satisfies
\begin{align}
    &\epsilon_\sigma(x,y,a/c)=\epsilon_\sigma(x,0,a/c)\notag\\
    &\epsilon_\sigma(x,y,b)=\epsilon_\sigma(b)
    \label{}
\end{align}

For relation $\sigma^{-1}T_1^{-1}\sigma T_2=\mathrm{e}$, we get
\begin{align}
    W_\sigma^{-1}(x,0,i)W_\sigma(x+1,0,i)=\chi_{\sigma T_2}(\sigma^{-1}(x,y,i))
    \label{}
\end{align}
Then, by tuning $\epsilon_\sigma$, we are able to set
\begin{align}
    &\chi_{\sigma T_2}(\sigma^{-1}(x,y,a))=\chi_{\sigma T_2}^*(\sigma^{-1}(x,y,b))\equiv\chi_\sigma\notag\\
    &\chi_{\sigma T_2}(\sigma^{-1}(x,y,c))=1
    \label{}
\end{align}
with remaining $\epsilon_\sigma(x,y,i)=\epsilon_\sigma(i)$. And for $W_\sigma$, we get
\begin{align}
    &W_\sigma(x,y,a)=\chi_\sigma^x W_{\sigma}(a)\notag\\
    &W_\sigma(x,y,b)=(\chi_\sigma^*)^x W_{\sigma}(b)\notag\\
    &W_\sigma(x,y,c)=W_\sigma(c)
    \label{eq:honeycomb_Wsigma_trans}
\end{align}

Consider relation $\sigma^2=\mathrm{e}$, we have
\begin{align}
    W_\sigma(x,y,i)W_\sigma(\sigma(x,y,i))=\chi_{\sigma^2}(x,y,i)=\chi_{\sigma^2}(\sigma(x,y,i))
    \label{eq:honeycomb_Wsigma_sq}
\end{align}
By setting $(x,y)=(0,0)$, we have $\chi_{\sigma^2}(b)=\chi_{\sigma^2}(c)$. Further we can set $\chi_{\sigma^2}(a)=\chi_{\sigma^2}(b)=\chi_{\sigma^2}(c)=1$ by using $\epsilon_\sigma$ ambiguity. Then, we have the remaining $\epsilon_\sigma(x,y,a)=1$ and $\epsilon_\sigma(x,y,b)\cdot\epsilon_\sigma(x,y,c)=1$. 

Inserting Eq.(\ref{eq:honeycomb_Wsigma_trans}) to Eq.(\ref{eq:honeycomb_Wsigma_sq}), we get
\begin{align}
    &\chi_{\sigma^2}(x,y,a)=\chi_\sigma^{x+y}\notag\\
    &\chi_{\sigma^2}(x,y,b)=(\chi_\sigma^*)^x\notag\\
    &\chi_{\sigma^2}(x,y,c)=(\chi_\sigma^*)^y
    \label{}
\end{align}
Use Gauss law of $\chi_{\sigma^2}$:
\begin{align}
    \chi_{\sigma^2}^*(x,y,a)\cdot\chi_{\sigma^2}^*(x-1,y,b)\cdot\chi_{\sigma^2}^*(x,y-1,c)=1
    \label{}
\end{align}
we conclude $\chi_\sigma=\pm1$.

For relation $\sigma C_6\sigma C_6=\mathrm{e}$, we have
\begin{align}
    &W_\sigma(x,y,s,i)W_{C_6}(\sigma(x,y,s,i))W_\sigma(\sigma C_6(x,y,s,i))\cdot \notag\\
    &W_{C_6}(C_6(x,y,i))=\chi_{\sigma C_6}(x,y,i)
    \label{}
\end{align}
Combine with Eq.(\ref{eq:honeycomb_WC6}) and Eq.(\ref{eq:honeycomb_Wsigma_trans}), we get
\begin{align}
    &\chi_{\sigma}^{y-1}\chi_{C_6}^y\cdot W_\sigma(a\cdot)[W_\sigma^{-1}(b)]^\mathrm{t}[W_{C_6}^{-1}(c)]^\mathrm{t}=\chi_{\sigma C_6}(x,y,a)\notag\\
    =&\chi^*_{\sigma C_6}(x+y+1,-y,a)
\end{align}
and
\begin{align}
    &\chi_\sigma^{y+1}\chi_{C_6}^yW_\sigma(b)W_{C_6}(c)[W_{\sigma}^{-1}(a)]^\mathrm{t}=\chi_{\sigma C_6}(x,y,b)\notag\\
    =&\chi_{\sigma C_6}^{*}(x+y+1,-y,a)
\end{align}
as well as
\begin{align}
    &\chi_{C_6}^{-2y}[W_\sigma(c)][W_\sigma^{-1}(c)]^\mathrm{t}=\chi_{\sigma C_6}(x,y,c)=\chi_{\sigma C_6}^*(x+y,-y,c)
    \label{}
\end{align}
Then, using Gauss law for $\chi_{\sigma C_6}(x,y,i)$, we obtain
\begin{align}
    &\chi_{C_6}^2=1\notag\\
    &W_\sigma(c)=W_\sigma^\mathrm{t}(c)
    \label{}
\end{align}
So, we conclude $\chi_{\sigma C_6}(x,y,c)=1$. By using remaining $\epsilon_\sigma$ ambiguity, we can set $\chi_{\sigma C_6}(a)=\chi_{\sigma C_6}(b)$. Namely
\begin{align}
    W_\sigma(a)[W_\sigma^{-1}(b)]^\mathrm{t}[W_{C_6}^{-1}(c)]^\mathrm{t}=1
    \label{}
\end{align}
By using $V$ ambiguity, we can set $W_{\sigma}(c)=W_{\sigma}(b)=\mathrm{I}$. We are left with overall gauge transformation $V$, with $V\cdot V^\mathrm{t}=\mathrm{I}$.

Now, let us summarize the result. For a spin system in honeycomb lattice, we construct four types of phases labeled by two $Z_2$ indices $\chi_{C_6}$ and $\chi_\sigma$. These phases can be viewed as ``weak SPT'' for point group, which are caused by the interplay between translational symmetries together with point group symmetries. By choosing gauge, the symmetry transformation rules on internal legs are:
\begin{align}
    &W_{T_1}(x,y,i)=W_{T_2}(x,y,i)=1\notag\\
    &W_{C_6}(x,y,a)=W_{C_6}(x,y,b)=\chi_{C_6}^x\notag\\
    &W_{C_6}(x,y,c)=W_{C_6}(c)=\chi_{C_6}W_{C_6}^\mathrm{t}(c)\notag\\
    &W_{\sigma}(x,y,a)=\chi_\sigma^x W_{\sigma}(a)=\chi_\sigma^x\chi_{C_6}W_{C_6}^\mathrm{t}\notag\\
    &W_\sigma(x,y,b)=\chi_\sigma^x\notag\\
    &W_\sigma(x,y,c)=\mathrm{I}
    \label{eq:honeycomb_weak_indices_Wg}
\end{align}

\begin{figure}
    \includegraphics[width=0.45\textwidth]{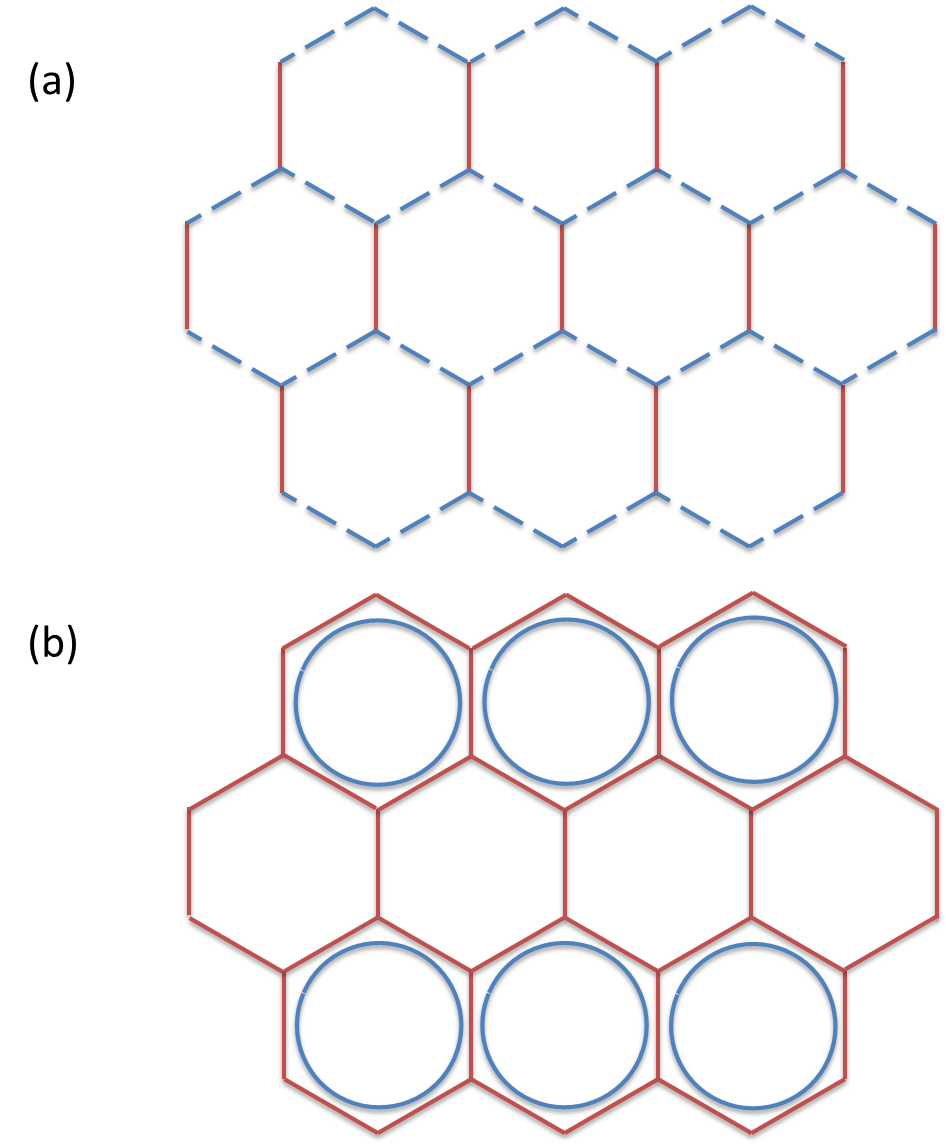}
    \caption{(a) One configuration of ``global'' IGG element $\lambda(T_2C_6,T_1)$, where the dashed blue line means $\chi_{C_6}$ action. It is invariant under translation symmetries. (b) The decomposition into multiplication of plaquette IGG element $\lambda_p(T_2C_6,T_1)$. The plaquette IGG element is no longer translationally invariant if $\chi_{C_6}=-1$.}
    \label{fig:honeycomb_decompose_IGG}
\end{figure}

To see the relation to the cohomology explicitly, let us consider $\omega(T_2,C_6,C_6)$. According to Eq.(\ref{eq:plq_IGG_twist_two_cocycle}), we get
\begin{align}
    &\lambda_p(T_2,C_6)\lambda_p(T_2C_6,T_1)\notag\\
    =&\omega(T_2,C_6,T_1)\,{}^{W_{T_2}T_2}\!\lambda_p(C_6,T_1)\lambda_p(T_2,C_6T_1)
    \label{}
\end{align}
To get plaquette IGG elements $\lambda_p$'s, we need to figure out the global IGG elements $\lambda$'s before decomposition first. We set the convention that the group elements is written as $g=T_1^{s_1}T_2^{s_2}C_6^{s_3}\sigma^{s_4}$, where $s_1,s_2\in\mathbb{Z}$, $s_3\in\mathbb{Z}_6$ and $s_4\in\mathbb{Z}_2$. The induced action on tensor networks (both physical and internal legs) is defined as
\begin{align}
    W_g=(W_{T_1}T_1)^{s_1}(W_{T_2}T_2)^{s_2}(W_{C_6}C_6)^{s_3}(W_\sigma\sigma)^{s_4}
    \label{}
\end{align}
Then, using Eq.\ref{eq:honeycomb_weak_indices_Wg}, we can calculate $\lambda$'s related to $\omega(T_2,C_6,C_6)$ as
\begin{align}
    &\lambda(T_2,C_6)=\lambda(T_2,C_6T_1)=\mathrm{I}\notag\\
    &\lambda(T_2C_6,T_1)(x,y,a/b)=\lambda(C_6,T_1)(x,y,a/b)=\chi_{C_6}\notag\\
    &\lambda(T_2C_6,T_1)(x,y,c)=\lambda(C_6,T_1)(x,y,c)=\mathrm{I}
    \label{}
\end{align}
We draw the configuration of $\lambda(T_2C_6,T_1)$ in Fig\,\ref{fig:honeycomb_decompose_IGG}. It is easy to verify the ``two cocycle'' condition for $\lambda$'s,
\begin{align}
    \lambda(T_2,C_6)\lambda(T_2C_6,T_1)=\,{}^{W_{T_2}T_2}\!\lambda(C_6,T_1)\lambda(T_2,C_6T_1)
    \label{}
\end{align}
where we use the fact that the action of $T_2$ on $\lambda(C_6,T_1)$ is trivial, as shown in Fig.\,\ref{fig:honeycomb_decompose_IGG}(a). However, when we decompose global IGG elements $\lambda$'s to plaquette IGG elements $\lambda_p$'s, $\lambda_p(C_6,T_1)$ is no longer invariant under $T_2$ transformation if $\chi_{C_6}$ is nontrivial, as shown in Fig.\,\ref{fig:honeycomb_decompose_IGG}(b). Instead, we get $\,{}^{W_{T_2}T_2}\!\lambda_p(C_6,T_1)=-\chi_{C_6}\lambda_p(C_6,T_1)$. Namely, $\omega(T_2,C_6,T_1)=\chi_{C_6}$. 

We can perform similar calculation for reflection, where we get $\omega(T_2,\sigma,T_1)=\chi_{\sigma}$. Thus, $\omega$ belongs to some nontrivial cohomology class for nontrivial $\chi_{C_6}$ and/or $\chi_\sigma$.

\section{The four cohomology classification from tensor equations in 3+1D}\label{app:four_cohomology}
In this part, we formulate the framework for tensor construction of SPT phases in 3+1D. 

If an infinity 3D tensor network is symmetric under group $SG$, we have 
\begin{align}
    W_gg\circ \mathbb{T}=\mathbb{T},\forall g\in SG
    \label{}
\end{align} 
where $W_g$ is the gauge transformation associated to group element $g\in SG$.

Now, let us turn to the IGG of 3D tensor networks. We consider the case where \emph{all elements of the IGG can be decomposed to some cubic IGG elements $\lambda_c$'s}: 
\begin{align}
    \lambda=\prod_c\lambda_c   
    \label{}
\end{align}
Here $\lambda_c$ only acts nontrivially on legs belonging to cubic $c$. In general, $\lambda_c$'s belonging to different cubic $c$ do not commute, so we should keep track of the multiplication order.

For later convenience, we also introduce the plaquette IGG $\{\xi_p\}$, where $\xi_p$ acts nontrivially only on legs belonging to plaquette $p$. 

Let us discuss commutation relations between different elements of IGG. For cubic IGG elements $\lambda_c$'s, we consider the expression $\varsigma_{cc'}=(\lambda_{c})^{-1}(\lambda_{c'}')^{-1}\lambda_{c}\lambda_{c'}'$. According to the definition, $\varsigma_{cc'}$ belongs to IGG. 
\begin{enumerate}
    \item When $c\cap c'=\emptyset$ or they only share a common site, $\varsigma_{cc'}=\mathrm{I}$.
    \item When $c\cap c'=v$, where $v$ is the common edge,  $\varsigma_v\equiv\varsigma_{cc'}$ only has nontrivial action on legs of $v$. Then, $\varsigma_v=\mathrm{I}$ in this case.
    \item When $c\cap c'=p$, where $p$ is the common plaquette, according to the definition, we conclude $\varsigma_{cc'}$ is a plaquette IGG element.
    \item When $c=c'$, then $\varsigma_{cc'}\in\{\lambda_c\}$.
\end{enumerate}  
To summarize, we have
\begin{align}
    \lambda_c^{-1}(\lambda_{c'}')^{-1}\lambda_c\lambda_{c'}'=
    \begin{cases}
        \widetilde{\xi}_p & c\cap c'=p\\
        \widetilde{\lambda}_c & c\equiv c'\\
        I & \text{otherwise}
    \end{cases}
    \label{eq:3D_cubic_IGG_commutation}
\end{align}

For the plaquette IGG, similar to the 2D case, we have 
\begin{align}
    (\xi_{p})^{-1}(\xi_{p'}')^{-1}\xi_{p}\xi_{p'}'=\widetilde{\xi}_{p}\delta_{pp'}
    \label{eq:3D_plq_IGG_commutation}
\end{align}
It is also straightforward to see the commutator between a plaquette IGG element and a cubic IGG element
\begin{align}
    \lambda_c^{-1}\xi_p^{-1}\lambda_c\xi_p=
    \begin{cases}
        \mathrm{I} & p\not\in c\\
        \widetilde{\xi}_p & p\in c\\
    \end{cases}
    \label{eq:3D_cubic_plq_IGG_commutation}
\end{align}

Now, we define a special kind of cubic IGG, which are formed by multiplication of plaquette IGG elements $\eta_c\equiv\prod_{p\in c}\xi^c_p$. For a set of $\left\{ \eta_c \right\}$, if we further require $\xi^c_p(i)=(\xi^{c'}_p(i))^{-1}$ for cubic $c$ and $c'$ sharing the same plaquette $p$\,($i$ labels a leg belonging to $p$), we have 
\begin{align}
    \prod_c\eta_c=\mathrm{I}
    \label{}
\end{align}
We assume \emph{it is the only way to decompose identity to cubic IGG}.

Notice that $\eta_c$ is defined as multiplication of plaquette IGG elements $\xi^c_p$, so there are phase ambiguities when we decompose $\eta_c$ to $\xi^c_p$,
\begin{align}
    \eta_c=\prod_{p\in c}\xi^c_p=\prod_{p\in c}\chi^c_p\xi^c_p
    \label{eq:3D_plq_IGG_ambiguity}
\end{align}
As we will show later, the phase ambiguities are essential ingredients to get the four cohomology classification of 3D SPT phases.

Let us define $\Lambda_c=\prod_{c'\le c}\lambda_{c'}$. Then, given $\lambda$, due to the $\eta_c$ ambiguity, we get
\begin{align}
    \lambda&=\prod_c\lambda_c=\prod_c\Lambda_{c-1}^{-1}\cdot\Lambda_{c}\notag\\
    &=\prod_c\Lambda_{c-1}^{-1}\cdot\eta_c\cdot\Lambda_{c} 
    \label{eq:3D_cubic_IGG_ambiguity}
\end{align}

According to Eq.(\ref{eq:3D_cubic_IGG_commutation}) and Eq.(\ref{eq:3D_cubic_plq_IGG_commutation}), we conclude 
\begin{align}
    \lambda_c^{-1}(\Lambda_{c'}')^{-1}\lambda_c\Lambda_{c'}'=
    \begin{cases}
        \widetilde{\lambda}_c & c\le c'\\
        \widetilde{\eta}_c & c>c'\\
    \end{cases}
    \label{eq:3D_cubic_CUBIC_IGG_commutation}
\end{align}
where $\widetilde{\lambda}_c$ belongs to cubic IGG, and $\widetilde{\eta}_c$ is the special IGG element formed by multiplication of plaquette IGG elements. 

Now, let us consider the symmetry equation on tensor networks. According to the definition of IGG, we have
\begin{align}
    W_{g_1}g_1W_{g_2}g_2=\lambda(g_1,g_2)W_{g_1,g_2}g_1g_2, \forall g_1,g_2\in SG
    \label{}
\end{align}
where $\lambda$'s belong to global IGG, which can always be decomposed into cubic IGG elements due to our assumption. 

From associativity
\begin{align}
    (W_{g_1}W_{g_2})W_{g_3}=W_{g_1}(W_{g_2}W_{g_3})
    \label{}
\end{align}
we get
\begin{align}
    \lambda(g_1,g_2)\lambda(g_1g_2,g_3)=\,{}^{W_{g_1}g_1}\!\lambda(g_2,g_3)\lambda(g_1,g_2g_3)
    \label{}
\end{align}
\begin{widetext}
We then write the above equation in terms of $\Lambda_c$'s,
\begin{align}
    &\prod_c [\Lambda_{c-1}(g_1,g_2)\Lambda_{c-1}(g_1g_2,g_3)]^{-1}\cdot[\Lambda_{c}(g_1,g_2)\Lambda_{c}(g_1g_2,g_3)]\notag\\
    =&\prod_c [\,{}^{W_{g_1}g_1}\!\Lambda_{c-1}(g_2,g_3)\Lambda_{c-1}(g_1,g_2g_3)]^{-1}\cdot [\,{}^{W_{g_1}g_1}\!\Lambda_{c}(g_2,g_3)\Lambda_{c}(g_1,g_2g_3)]
    \label{}
\end{align}
According to Eq.(\ref{eq:3D_cubic_IGG_ambiguity}), we conclude that
\begin{align}
    &[\Lambda_{c-1}(g_1,g_2)\Lambda_{c-1}(g_1g_2,g_3)]^{-1}\cdot[\Lambda_{c}(g_1,g_2)\Lambda_{c}(g_1g_2,g_3)]\notag\\
    =&[\,{}^{W_{g_1}g_1}\!\Lambda_{c-1}(g_2,g_3)\Lambda_{c-1}(g_1,g_2g_3)]^{-1}\cdot[\eta_c(g_1,g_2,g_3)]\cdot[\,{}^{W_{g_1}g_1}\!\Lambda_{c}(g_2,g_3)\Lambda_{c}(g_1,g_2g_3)]
    \label{eq:3D_single_cubic_twist_two_cocycle}
\end{align}
Further, we get
\begin{align}
    \Lambda_c(g_1,g_2)\Lambda_c(g_1g_2,g_3)&=\prod_{c'\le c}[\Lambda_{c'-1}(g_1,g_2)\Lambda_{c'-1}(g_1g_2,g_3)]^{-1}\cdot[\Lambda_{c'}(g_1,g_2)\Lambda_{c'}(g_1g_2,g_3)]\notag\\
    &=\prod_{c'\le c}[\,{}^{W_{g_1}g_1}\!\Lambda_{c'-1}(g_2,g_3)\Lambda_{c'-1}(g_1,g_2g_3)]^{-1}\cdot\eta_{c'}(g_1,g_2,g_3)\cdot[\,{}^{W_{g_1}g_1}\!\Lambda_{c'}(g_2,g_3)\Lambda_{c'}(g_1,g_2g_3)]\notag\\
    &=H_c(g_1,g_2,g_3)\cdot\,{}^{W_{g_1}g_1}\!\Lambda_c(g_2,g_3)\Lambda_c(g_1,g_2g_3)    
    \label{}
\end{align}
where $H_{c}\triangleq\prod_{c'\le c}\eta_{c'}$. So, we have
\begin{align}
    \Lambda_c(g_1,g_2)\Lambda_c(g_1g_2,g_3)=H_c(g_1,g_2,g_3)\cdot\,{}^{W_{g_1}g_1}\!\Lambda_c(g_2,g_3)\Lambda_c(g_1,g_2g_3)    
    \label{eq:3D_cubic_IGG_twist_two_cocycle}
\end{align}

Now, let us determine the condition for $H_c$. Similar to 2D case, we calculate
\begin{align}
    &\Lambda_c(g_1,g_2)\Lambda_c(g_1g_2,g_3)\Lambda_c(g_1g_2g_3,g_4)\notag\\
    =&H_c(g_1,g_2,g_3)\cdot\,{}^{W_{g_1}g_1}\!\Lambda_c(g_2,g_3)\Lambda_c(g_1,g_2g_3)\Lambda_c(g_1g_2g_3,g_4)\notag\\
    =&H_c(g_1,g_2,g_3)\cdot\,{}^{W_{g_1}g_1}\!\Lambda_c(g_2,g_3)\cdot H_c(g_1,g_2g_3,g_4)\cdot \,{}^{W_{g_1}g_1}\!\Lambda_c(g_2g_3,g_4)\Lambda_c(g_1,g_2g_3g_4)\notag\\
    =&H_c(g_1,g_2,g_3)\cdot\left[\,{}^{W_{g_1}g_1}\!\Lambda_c(g_2,g_3)\circ H_c(g_1,g_2g_3,g_4)\right]\cdot \,{}^{W_{g_1}g_1}\!H_c(g_2,g_3,g_4)\cdot \,{}^{W_{g_1}g_1W_{g_2}g_2}\!\Lambda_c(g_3,g_4)\,{}^{W_{g_1}g_1}\!\Lambda_c(g_2,g_3g_4)\Lambda_c(g_1,g_2g_3g_4)
    \label{eq:3D_pentagon_first_way}
\end{align}
where we define $a\circ b\triangleq a\cdot b\cdot a^{-1}$. We calculate the above equation in another way as
\begin{align}
    &\Lambda_c(g_1,g_2)\Lambda_c(g_1g_2,g_3)\Lambda_c(g_1g_2g_3,g_4)\notag\\
    =&\Lambda_c(g_1,g_2)H_c(g_1g_2,g_3,g_4)\,{}^{W_{g_1g_2}g_1g_2}\!\Lambda_c(g_3,g_4)\Lambda_c(g_1g_2,g_3g_4)\notag\\
    =&\left[\Lambda_c(g_1,g_2)\circ H_c(g_1g_2,g_3,g_4)\right]\cdot\big[ \Lambda_c(g_1,g_2)W_{g_1g_2}g_1g_2\circ \Lambda_c(g_3,g_4) \big]\cdot H_c(g_1,g_2,g_3g_4)\cdot \,{}^{W_{g_1}g_1}\!\Lambda_c(g_2,g_3g_4)\cdot \Lambda_c(g_1,g_2g_3g_4)\notag\\
    =&\left[\Lambda_c(g_1,g_2)\circ H_c(g_1g_2,g_3,g_4)\right]\cdot\widetilde{H}_c(g_1,g_2,g_3,g_4)\cdot \left[\,{}^{W_{g_1}g_1W_{g_2}g_2}\!\Lambda_c(g_3,g_4)\circ H_c(g_1,g_2,g_3g_4)\right]\cdot \,{}^{W_{g_1}g_1W_{g_2}g_2}\!\Lambda_c(g_3,g_4)\,{}^{W_{g_1}g_1}\!\Lambda_c(g_2,g_3g_4)\cdot \notag\\
    &\Lambda_c(g_1,g_2g_3g_4)
    \label{eq:3D_pentagon_second_way}
\end{align}
In the last line, we use the following relation:
\begin{align}
    &\Lambda_c(g_1,g_2)W_{g_1g_2}g_1g_2\circ\Lambda_c(g_3,g_4)\notag\\
    =&\Lambda_c(g_1,g_2)V_{c+1}^{-1}(g_1,g_2)\Lambda_c^{-1}(g_1,g_2)\circ\,{}^{W_{g_1}g_1W_{g_2}g_2}\!\Lambda_c(g_3,g_4)\notag\\
    =&\widetilde{H}_c(g_1,g_2,g_3,g_4)\cdot\,{}^{W_{g_1}g_1W_{g_2}g_2}\!\Lambda_c(g_3,g_4)
    \label{}
\end{align}
where we define $V_c\triangleq\prod_{c'\ge c}\lambda_{c'}$, so $\lambda=\Lambda_c\cdot V_{c+1}$. 

Comparing Eq.(\ref{eq:3D_pentagon_first_way}) and Eq.(\ref{eq:3D_pentagon_second_way}), we conclude
\begin{align}
    &H_c(g_1,g_2,g_3)\cdot\left[\,{}^{W_{g_1}g_1}\!\Lambda_c(g_2,g_3)\circ H_c(g_1,g_2g_3,g_4)\right]\cdot \,{}^{W_{g_1}g_1}\!H_c(g_2,g_3,g_4)\notag\\
    =&\left[\Lambda_c(g_1,g_2)\circ H_c(g_1g_2,g_3,g_4)\right]\cdot \widetilde{H}_c(g_1,g_2,g_3,g_4)\cdot\left[ \,{}^{W_{g_1}g_1W_{g_2}g_2}\!\Lambda_c(g_3,g_4)\circ H_c(g_1,g_2,g_3g_4)\right]
    \label{eq:3D_three_cocycle}
\end{align}

Notice, $H_c$ can be expressed as $H_c=\prod_{p}\xi_p$, where $p$ takes value in $\left(\bigcup_{c'\le c}c'\right)\cap\left(\bigcup_{c''>c}c''\right)$. When decomposing $H_c$, there is an associated phase ambiguity $\chi_p(i)$, where $\chi_p(i)=\chi_{p'}^*(i)$ if $p$ and $p'$ share the same leg $i$. We apply this observation to Eq.(\ref{eq:3D_three_cocycle}) and choose $p=c_0\cap c_1$, where $c_0\le c$ and $c_1>c$, then we get
\begin{align}
    &\omega_p^{-1}(g_1,g_2,g_3,g_4)\cdot\xi_p(g_1,g_2,g_3)\cdot\left[\,{}^{W_{g_1}g_1}\!\Lambda_c(g_2,g_3)\circ \xi_p(g_1,g_2g_3,g_4)\right]\cdot\,{}^{W_{g_1}g_1}\!\xi_p(g_2,g_3,g_4)\notag\\
    =&\left[\Lambda_c(g_1,g_2)\circ \xi_p(g_1g_2,g_3,g_4)\right]\cdot\widetilde{\xi}_p(g_1,g_2,g_3,g_4)\cdot\left[ \,{}^{W_{g_1}g_1W_{g_2}g_2}\!\Lambda_c(g_3,g_4)\circ \xi_p(g_1,g_2,g_3g_4)\right]
    \label{}
\end{align}
Further, we get $\Lambda_c\circ\xi_p=\lambda_{c_0}\circ\xi_p$. And there is a canonical choice for $\widetilde{\xi}_p(g_1,g_2,g_3,g_4)$, which reads
\begin{align}
    \widetilde{\xi}_p(g_1,g_2,g_3,g_4)=\,{}^{\lambda_{c_0}(g_1,g_2)}\!\lambda_{c_1}^{-1}(g_1,g_2)\cdot\,{}^{W_{g_1}g_1W_{g_2}g_2}\!\lambda_{c_0}(g_3,g_4)\cdot\,{}^{\lambda_{c_0}(g_1,g_2)}\!\lambda_{c_1}(g_1,g_2)\cdot\,{}^{W_{g_1}g_1W_{g_2}g_2}\!\lambda_{c_0}^{-1}(g_3,g_4)
    \label{}
\end{align}
So the equation becomes
\begin{align}
    &\omega_p^{-1}(g_1,g_2,g_3,g_4)\cdot\xi_p(g_1,g_2,g_3)\cdot\left[\,{}^{W_{g_1}g_1}\!\lambda_{c_0}(g_2,g_3)\circ \xi_p(g_1,g_2g_3,g_4)\right]\cdot\,{}^{W_{g_1}g_1}\!\xi_p(g_2,g_3,g_4)\notag\\
    =&\left[\lambda_{c_0}(g_1,g_2)\circ \xi_p(g_1g_2,g_3,g_4)\right]\cdot\widetilde{\xi}_p(g_1,g_2,g_3,g_4)\cdot\left[ \,{}^{W_{g_1}g_1W_{g_2}g_2}\!\lambda_{c_0}(g_3,g_4)\circ\xi_p(g_1,g_2,g_3g_4)\right]\notag\\
    =&\left[\lambda_{c_0}(g_1,g_2)\circ \xi_p(g_1g_2,g_3,g_4)\right]\cdot\left[ \,{}^{\lambda_{c_0}(g_1,g_2)W_{g_1g_2}g_1g_2}\!\lambda_{c_0}(g_3,g_4)\circ\xi_p(g_1,g_2,g_3g_4)\right]\cdot\widetilde{\xi}_p(g_1,g_2,g_3,g_4)\notag\\
    \label{eq:3D_twist_three_cocycle}
\end{align}

Further, we have
\begin{align}
    &\widetilde{\xi}_p(g_1,g_2,g_3,g_4)\cdot\xi_p=\left[ \,{}^{\lambda_{c_0}(g_1,g_2)W_{g_1g_2}g_1g_2}\!\lambda_{c_0}(g_3,g_4) \,{}^{W_{g_1}g_1W_{g_2}g_2}\!\lambda_{c_0}^{-1}(g_3,g_4) \circ\xi_p \right]\cdot \widetilde{\xi}_p(g_1,g_2,g_3,g_4) \notag\\
    &\xi_p\cdot\widetilde{\xi}_p(g_1,g_2,g_3,g_4)=\widetilde{\xi}_p(g_1,g_2,g_3,g_4)\cdot\left[ \,{}^{W_{g_1}g_1W_{g_2}g_2}\!\lambda_{c_0}(g_3,g_4)\,{}^{\lambda_{c_0}(g_1,g_2)W_{g_1g_2}g_1g_2}\!\lambda_{c_0}(g_3,g_4)^{-1}\circ\xi_p \right]
    \label{}
\end{align}
where we use
\begin{align}
    \,{}^{W_{g_1}g_1W_{g_2}g_2}\!\xi_p=\,{}^{\lambda(g_1,g_2)W_{g_1g_2}g_1g_2}\!\xi_p=\,{}^{\lambda_{c_0}(g_1,g_2)\lambda_{c_1}(g_1,g_2)W_{g_1g_2}g_1g_2}\!\xi_p
    \label{}
\end{align}


Let us calculate the following expression.
\begin{align}
    &\left[ \xi(g_1,g_2,g_3) \right]\cdot\left[ \,{}^{W_{g_1}g_1}\!\lambda_0(g_2,g_3)\circ\xi(g_1,g_2g_3,g_4) \right]\cdot\left[ \,{}^{W_{g_1}g_1}\!\xi(g_2,g_3,g_4) \right]\cdot \big[\,{}^{W_{g_1}g_1W_{g_2}g_2}\!\lambda_0(g_3,g_4)\,{}^{W_{g_1}g_1}\!\lambda_0(g_2,g_3g_4)\circ \notag\\
    &\xi(g_1,g_2g_3g_4,g_5) \big]\cdot \left[ \,{}^{W_{g_1}g_1W_{g_2}g_2}\!\lambda_0(g_3,g_4)\circ\,{}^{W_{g_1}g_1}\!\xi(g_2,g_3g_4,g_5) \right]\cdot \left[ \,{}^{W_{g_1}g_1W_{g_2}g_2}\!\xi(g_3,g_4,g_5) \right] \notag\\
    \notag\\
    =&\omega(g_1,g_2,g_3,g_4)\omega(g_1,g_2,g_3g_4,g_5)\cdot \left[ \lambda_0(g_1,g_2)\circ\xi(g_1g_2,g_3,g_4) \right]\cdot [\widetilde{\xi}(g_1,g_2,g_3,g_4)]\cdot \notag\\
    &\left[ \,{}^{W_{g_1}g_1W_{g_2}g_2}\!\lambda_0(g_3,g_4)\lambda_0(g_1,g_2)\circ\xi(g_1g_2,g_3g_4,g_5) \right]\cdot \left[ \,{}^{W_{g_1}g_1W_{g_2}g_2}\!\lambda_0(g_3,g_4)\,{}^{\lambda_0(g_1,g_2)W_{g_1g_2}g_1g_2}\!\lambda_0(g_3g_4,g_5)\circ\xi(g_1,g_2,g_3g_4g_5) \right]\cdot \notag\\
    &\left[ \,{}^{W_{g_1}g_1W_{g_2}g_2}\!\lambda_0(g_3,g_4)\circ\widetilde{\xi}(g_1,g_2,g_3g_4,g_5) \right]\cdot \left[ \,{}^{W_{g_1}g_1W_{g_2}g_2}\!\xi(g_3,g_4,g_5) \right] \notag\\
    \notag\\
    =&\omega(g_1,g_2,g_3,g_4)\omega(g_1,g_2,g_3g_4,g_5)\cdot \left[ \lambda_0(g_1,g_2)\circ\xi(g_1g_2,g_3,g_4) \right]\cdot \left[ \lambda_0(g_1,g_2)\,{}^{W_{g_1g_2}g_1g_2}\!\lambda_0(g_3,g_4)\circ\xi(g_1g_2,g_3g_4,g_5) \right]\cdot \notag\\
    &\left[ \,{}^{\lambda_0(g_1,g_2)W_{g_1g_2}g_1g_2}\!\left(\lambda_0(g_3,g_4)\lambda_0(g_3g_4,g_5)\right)\circ\xi(g_1,g_2,g_3g_4g_5) \right]\cdot [\widetilde{\xi}(g_1,g_2,g_3,g_4)]\cdot \left[ \,{}^{W_{g_1}g_1W_{g_2}g_2}\!\lambda_0(g_3,g_4)\circ\widetilde{\xi}(g_1,g_2,g_3g_4,g_5) \right]\notag\\
    &\left[ \,{}^{W_{g_1}g_1W_{g_2}g_2}\!\xi(g_3,g_4,g_5) \right] \notag\\
    \notag\\
    =&\omega(g_1,g_2,g_3,g_4)\omega(g_1,g_2,g_3g_4,g_5)\cdot \left[ \lambda_0(g_1,g_2)\circ\xi(g_1g_2,g_3,g_4) \right]\cdot \left[ \lambda_0(g_1,g_2)\,{}^{W_{g_1g_2}g_1g_2}\!\lambda_0(g_3,g_4)\circ\xi(g_1g_2,g_3g_4,g_5) \right]\cdot \notag\\
    &\left[ \lambda_0(g_1,g_2)W_{g_1g_2}g_1g_2\circ\xi(g_3,g_4,g_5) \right]\cdot \left[ \,{}^{\lambda_0(g_1,g_2)W_{g_1g_2}g_1g_2}\!\left(\,{}^{W_{g_3}g_3}\!\lambda_0(g_4,g_5)\lambda_0(g_3,g_4g_5)\right)\circ\xi(g_1,g_2,g_3g_4g_5) \right]\cdot \notag\\
    &\left[ \lambda_0(g_1,g_2)W_{g_1g_2}g_1g_2\circ\xi^{-1}(g_3,g_4,g_5) \right]\cdot [\widetilde{\xi}(g_1,g_2,g_3,g_4)]\cdot \left[ \,{}^{W_{g_1}g_1W_{g_2}g_2}\!\lambda_0(g_3,g_4)\circ\widetilde{\xi}(g_1,g_2,g_3g_4,g_5) \right]\cdot \notag\\
    &\left[ \,{}^{W_{g_1}g_1W_{g_2}g_2}\!\xi(g_3,g_4,g_5) \right] \notag\\
    \notag\\
    =&\omega(g_1,g_2,g_3,g_4)\omega(g_1,g_2,g_3g_4,g_5)\omega(g_1g_2,g_3,g_4,g_5)\cdot \left[ \lambda_0(g_1,g_2)\lambda_0(g_1g_2,g_3)\circ\xi(g_1g_2g_3,g_4,g_5) \right]\cdot \notag\\
    &\left[ \lambda_0(g_1,g_2)\,{}^{\lambda_{0}(g_1g_2,g_3)W_{g_1g_2g_3}g_1g_2g_3}\!\lambda(g_4,g_5)\circ\xi(g_1g_2,g_3,g_4g_5) \right]\cdot \big[ \,{}^{\lambda_0(g_1,g_2)\lambda_0(g_1g_2,g_3)W_{g_1g_2g_3}g_1g_2g_3}\!\lambda_0(g_4,g_5)\cdot \notag\\
    &\,{}^{\lambda_0(g_1,g_2)W_{g_1g_2}g_1g_2}\!\lambda_0(g_3,g_4g_5)\circ\xi(g_1,g_2,g_3g_4g_5) \big]\cdot \left[ \lambda_0(g_1,g_2)\circ\widetilde{\xi}(g_1g_2,g_3,g_4,g_5) \right]\cdot \left[ \lambda_0(g_1,g_2)W_{g_1g_2}g_1g_2\circ\xi^{-1}(g_3,g_4,g_5) \right]\cdot \notag\\
    &[\widetilde{\xi}(g_1,g_2,g_3,g_4)]\cdot \left[ \,{}^{W_{g_1}g_1W_{g_2}g_2}\!\lambda_0(g_3,g_4)\circ\widetilde{\xi}(g_1,g_2,g_3g_4,g_5) \right]\cdot \left[ \,{}^{W_{g_1}g_1W_{g_2}g_2}\!\xi(g_3,g_4,g_5) \right] \notag\\
    \label{}
\end{align}

We use another way to calculate the above expression in the following.
\begin{align}
    &\left[ \xi(g_1,g_2,g_3) \right]\cdot\left[ \,{}^{W_{g_1}g_1}\!\lambda_0(g_2,g_3)\circ\xi(g_1,g_2g_3,g_4) \right]\cdot\left[ \,{}^{W_{g_1}g_1}\!\xi(g_2,g_3,g_4) \right]\cdot \big[\,{}^{W_{g_1}g_1W_{g_2}g_2}\!\lambda_0(g_3,g_4)\,{}^{W_{g_1}g_1}\!\lambda_0(g_2,g_3g_4)\circ \notag\\
    &\xi(g_1,g_2g_3g_4,g_5) \big]\cdot \left[ \,{}^{W_{g_1}g_1W_{g_2}g_2}\!\lambda_0(g_3,g_4)\circ\,{}^{W_{g_1}g_1}\!\xi(g_2,g_3g_4,g_5) \right]\cdot \left[ \,{}^{W_{g_1}g_1W_{g_2}g_2}\!\xi(g_3,g_4,g_5) \right] \notag\\
    \notag\\
    =&\left[ \xi(g_1,g_2,g_3) \right]\cdot \left[ \,{}^{W_{g_1}g_1}\!\lambda_0(g_2,g_3)\circ\xi(g_1,g_2g_3,g_4) \right]\cdot \left[ \,{}^{W_{g_1}g_1}\!\lambda_0(g_2,g_3)\,{}^{W_{g_1}g_1}\!\lambda_0(g_2g_3,g_4)\circ\xi(g_1,g_2g_3g_4,g_5) \right]\cdot \notag\\
    &\left[ \,{}^{W_{g_1}g_1}\!\xi(g_2,g_3,g_4) \right]\cdot \left[ W_{g_1}g_1\,{}^{W_{g_2}g_2}\!\lambda_0(g_3,g_4)\circ \xi(g_2,g_3g_4,g_5) \right]\cdot \left[ \,{}^{W_{g_1}g_1W_{g_2}g_2}\!\xi(g_3,g_4,g_5) \right] \notag\\
    \notag\\
    =&\,{}^{g_1}\!\omega(g_2,g_3,g_4,g_5)\cdot \left[ \xi(g_1,g_2,g_3) \right]\cdot \left[ \,{}^{W_{g_1}g_1}\!\lambda_0(g_2,g_3)\circ\xi(g_1,g_2g_3,g_4) \right]\cdot \left[ \,{}^{W_{g_1}g_1}\!\lambda_0(g_2,g_3)\,{}^{W_{g_1}g_1}\!\lambda_0(g_2g_3,g_4)\circ\xi(g_1,g_2g_3g_4,g_5) \right]\cdot \notag\\
    &\left[ \,{}^{W_{g_1}g_1}\!\lambda_0(g_2,g_3)\circ\,{}^{W_{g_1}g_1}\!\xi(g_2g_3,g_4,g_5) \right]\cdot \left[ W_{g_1}g_1\,{}^{\lambda_{0}(g_2,g_3)W_{g_2g_3}g_2g_3}\!\lambda_0(g_4,g_5)\circ\xi(g_2,g_3,g_4g_5) \right]\cdot \left[ \,{}^{W_{g_1}g_1}\!\widetilde{\xi}_p(g_2,g_3,g_4,g_5) \right] \notag\\
    \notag\\
    =&\,{}^{g_1}\!\omega(g_2,g_3,g_4,g_5)\omega(g_1,g_2g_3,g_4,g_5)\cdot \left[ \xi(g_1,g_2,g_3) \right]\cdot \left[ \,{}^{W_{g_1}g_1}\!\lambda_0(g_2,g_3)\lambda_0(g_1,g_2g_3)\circ\xi(g_1g_2g_3,g_4,g_5) \right]\cdot \notag\\
    &\left[ \,{}^{W_{g_1}g_1}\!\lambda_0(g_2,g_3)\,{}^{\lambda_{0}(g_1,g_2g_3)W_{g_1g_2g_3}g_1g_2g_3}\!\lambda_0(g_4,g_5)\circ\xi(g_1,g_2g_3,g_4g_5) \right]\cdot \left[ \,{}^{W_{g_1}g_1}\!\lambda_0(g_2,g_3)\circ\widetilde{\xi}(g_1,g_2g_3,g_4,g_5) \right]\cdot \notag\\
    &\left[ \,{}^{W_{g_1}g_1}\!\lambda_{0}(g_2,g_3)\,{}^{W_{g_1}g_1W_{g_2g_3}g_2g_3}\!\lambda_0(g_4,g_5)\,{}^{W_{g_1}g_1}\!\lambda_0^{-1}(g_2,g_3)\circ\,{}^{W_{g_1}g_1}\!\xi(g_2,g_3,g_4g_5) \right]\cdot \left[ \,{}^{W_{g_1}g_1}\!\widetilde{\xi}_p(g_2,g_3,g_4,g_5) \right] \notag\\
    \notag\\
    =&\,{}^{g_1}\!\omega(g_2,g_3,g_4,g_5)\omega(g_1,g_2g_3,g_4,g_5)\cdot \left[ \lambda_0(g_1,g_2)\lambda_0(g_1g_2,g_3)\circ\xi(g_1g_2g_3,g_4,g_5) \right]\cdot \Big\{ \,{}^{\lambda_0(g_1,g_2)\lambda_{0}(g_1g_2,g_3)W_{g_1g_2g_3}g_1g_2g_3}\!\lambda_0(g_4,g_5)\circ \notag\\
    &\left[ \xi(g_1,g_2,g_3) \right]\cdot \left[ \,{}^{W_{g_1}g_1}\!\lambda_0(g_2,g_3)\circ\xi(g_1,g_2g_3,g_4g_5) \right]\cdot \left[ \,{}^{W_{g_1}g_1}\!\xi(g_2,g_3,g_4g_5) \right]\cdot \left[ \xi^{-1}(g_1,g_2,g_3) \right] \Big\}\cdot \left[ \xi(g_1,g_2,g_3) \right]\cdot \notag\\
    &\left[ \,{}^{W_{g_1}g_1}\!\lambda_0(g_2,g_3)\circ\widetilde{\xi}(g_1,g_2g_3,g_4,g_5) \right]\cdot \left[ \,{}^{W_{g_1}g_1}\!\widetilde{\xi}_p(g_2,g_3,g_4,g_5) \right]\notag\\
    \notag\\
    =&\,{}^{g_1}\!\omega(g_2,g_3,g_4,g_5)\omega(g_1,g_2g_3,g_4,g_5)\omega(g_1,g_2,g_3,g_4g_5)\cdot \left[ \lambda_0(g_1,g_2)\lambda_0(g_1g_2,g_3)\circ\xi(g_1g_2g_3,g_4,g_5) \right]\cdot \notag\\
    &\left[ \lambda_0(g_1,g_2)\,{}^{\lambda_{0}(g_1g_2,g_3)W_{g_1g_2g_3}g_1g_2g_3}\!\lambda_0(g_4,g_5)\circ\xi(g_1g_2,g_3,g_4g_5) \right]\cdot \notag\\
    &\left[ \,{}^{\lambda_0(g_1,g_2)\lambda_{0}(g_1g_2,g_3)W_{g_1g_2g_3}g_1g_2g_3}\!\lambda_0(g_4,g_5)\,{}^{\lambda_0(g_1,g_2)W_{g_1g_2}g_1g_2}\!\lambda_0(g_3,g_4g_5)\circ\xi(g_1,g_2,g_3g_4g_5) \right]\cdot \left[ \xi(g_1,g_2,g_3) \right]\cdot \notag\\
    &\left[ \,{}^{W_{g_1}g_1}\!\lambda_0(g_2,g_3)\circ\widetilde{\xi}(g_1,g_2g_3,g_4,g_5) \right]\cdot \left[ \,{}^{W_{g_1}g_1}\!\widetilde{\xi}_p(g_2,g_3,g_4,g_5) \right]\cdot \left[ \,{}^{W_{g_1}g_1W_{g_2}g_2W_{g_3}g_3}\!\lambda_0(g_4,g_5)\circ\xi^{-1}(g_1,g_2,g_3) \right]\cdot \notag\\
    &\left[ \,{}^{W_{g_1}g_1W_{g_2}g_2W_{g_3}g_3}\!\lambda_0(g_4,g_5)\circ\widetilde{\xi}(g_1,g_2,g_3,g_4g_5) \right] \notag\\
    \label{}
\end{align}

Next, let us prove the following equation:
\begin{align}
    &\left[ \lambda_0(g_1,g_2)\circ\widetilde{\xi}(g_1g_2,g_3,g_4,g_5) \right]\cdot \left[ \lambda_0(g_1,g_2)W_{g_1g_2}g_1g_2\circ\xi^{-1}(g_3,g_4,g_5) \right]\cdot [\widetilde{\xi}(g_1,g_2,g_3,g_4)]\cdot \notag\\
    &\left[ \,{}^{W_{g_1}g_1W_{g_2}g_2}\!\lambda_0(g_3,g_4)\circ\widetilde{\xi}(g_1,g_2,g_3g_4,g_5) \right]\cdot \left[ \,{}^{W_{g_1}g_1W_{g_2}g_2}\!\xi(g_3,g_4,g_5) \right] \notag\\
    =&\left[ \xi(g_1,g_2,g_3) \right]\cdot \left[ \,{}^{W_{g_1}g_1}\!\lambda_0(g_2,g_3)\circ\widetilde{\xi}(g_1,g_2g_3,g_4,g_5) \right]\cdot \left[ \,{}^{W_{g_1}g_1}\!\widetilde{\xi}_p(g_2,g_3,g_4,g_5) \right]\cdot \left[ \,{}^{W_{g_1}g_1W_{g_2}g_2W_{g_3}g_3}\!\lambda_0(g_4,g_5)\circ\xi^{-1}(g_1,g_2,g_3) \right]\cdot \notag\\
    &\left[ \,{}^{W_{g_1}g_1W_{g_2}g_2W_{g_3}g_3}\!\lambda_0(g_4,g_5)\circ\widetilde{\xi}(g_1,g_2,g_3,g_4g_5) \right]
    \label{}
\end{align}
Before that, let us mention some useful relations. First, we have
\begin{align}
    \,{}^{\lambda_{c_0}\lambda_{c_1}}\!\lambda_{c_2}=\,{}^{\lambda_{c_1}\lambda_{c_0}}\!\lambda_{c_2},\quad \text{if}\,c_0\neq c_1\neq c_2
    \label{eq:3D_three_lambda_commutation}
\end{align}
Then, we conclude 
\begin{align}
    &\widetilde{\xi}(g_1,g_2,g_3,g_4)=\left[ \,{}^{\lambda_0(g_1,g_2)}\!\lambda_1^{-1}(g_1,g_2)\circ \,{}^{W_{g_1}g_1W_{g_2}g_2}\!\lambda_0(g_3,g_4) \right]\cdot \left[ \,{}^{W_{g_1}g_1W_{g_2}g_2}\!\lambda_{c_0}^{-1}(g_3,g_4) \right]\notag\\
    =&\left[ \,{}^{\lambda_0(g_1,g_2)}\!\lambda_1^{-1}(g_1,g_2)\lambda(g_1,g_2)\circ \,{}^{W_{g_1g_2}g_1g_2}\!\lambda_0(g_3,g_4) \right]\cdot \left[ \,{}^{W_{g_1}g_1W_{g_2}g_2}\!\lambda_{c_0}^{-1}(g_3,g_4) \right]\notag\\
    =&\left[ \,{}^{\lambda_{\slashed{c}_1}(g_1,g_2)W_{g_1g_2}g_1g_2}\!\lambda_0(g_3,g_4) \right]\cdot \left[ \,{}^{W_{g_1}g_1W_{g_2}g_2}\!\lambda_0^{-1}(g_3,g_4) \right]\notag
    \label{}
\end{align}
where $\lambda_{\slashed{c}_1}\triangleq\Lambda_{c_1-1}V_{c_1+1}$.

In order to proceed, we consider the following relation according to Eq.(\ref{eq:3D_single_cubic_twist_two_cocycle}):
\begin{align}
    &\left[ \Lambda_{c_0-1}^{-1}(g_1g_2,g_3)\circ\lambda_0(g_1,g_2) \right]\cdot \left[ \lambda_0(g_1g_2,g_3) \right]\notag\\
    =&\left[ \left( \,{}^{W_{g_1}g_1}\!\Lambda_{c_0-1}(g_2,g_3)\Lambda_{c_0-1}(g_1,g_2g_3) \right)^{-1}\circ\eta_0(g_1,g_2,g_3) \right]\cdot \left[ \Lambda_{c_0-1}^{-1}(g_1,g_2g_3)\circ\,{}^{W_{g_1}g_1}\!\lambda_0(g_2,g_3) \right]\circ \left[ \lambda_0(g_1,g_2g_3) \right]
    \label{}
\end{align}
According to the commutation relation Eq.(\ref{eq:3D_cubic_IGG_commutation}), Eq.(\ref{eq:3D_plq_IGG_commutation}) and Eq.(\ref{eq:3D_cubic_plq_IGG_commutation}), we conclude
\begin{align}
    \lambda_0(g_1,g_2)\lambda_0(g_1g_2,g_3)=\eta'_0(g_1,g_2,g_3)\,{}^{W_{g_1}g_1}\!\lambda_0(g_2,g_3)\lambda_0(g_1,g_2g_3)
    \label{}
\end{align}
where 
\begin{align}
    \eta_0'(g_1,g_2,g_3)=\left( \prod_{p\in\{c'<c_0\}\cap c_0}\xi'_p(g_1,g_2,g_3) \right)\cdot \left( \prod_{p\in\{c'>c_0\}\cap c_0}\xi_p(g_1,g_2,g_3) \right)
    \label{}
\end{align}
where the prime label is due to nontrivial commutation relation.

Further, we have
\begin{align}
    &\lambda_{\slashed{c}_1}(g_1,g_2)\lambda_{\slashed{c}_1}(g_1g_2,g_3)\circ\lambda_0\notag\\
    =&\left[ \left( \Lambda_{c_0-1}(g_1,g_2)\circ\lambda_{c_0}(g_1,g_2) \right) \left( \lambda_{\slashed{c}_0\slashed{c}_1}(g_1,g_2)\Lambda_{c_0-1}(g_1g_2,g_3)\circ\lambda_0(g_1g_2,g_3) \right) \left( \prod_{c\neq c_0,c_1}\lambda_c(g_1,g_2)\lambda_c(g_1g_2,g_3) \right) \circ\lambda_0 \right]\notag\\
    =&\left[ \eta''_{c_0}(g_1,g_2,g_3)\,{}^{W_{g_1}g_1}\!\lambda_{c_0}(g_2,g_3)\lambda_{c_0}(g_1,g_2g_3) \left( \eta'_c(g_1,g_2,g_3)\prod_{c\neq c_0,c_1}\,{}^{W_{g_1}g_1}\!\lambda_c(g_2,g_3)\lambda_c(g_1,g_2g_3) \right) \circ\lambda_0 \right]\notag\\
    =&\xi_{c_0\cap c_1}(g_1,g_2,g_3)\,{}^{W_{g_1}g_1}\!\lambda_{\slashed{c}_1}(g_2,g_3)\lambda_{\slashed{c}_1}(g_1,g_2g_3)\circ\lambda_0 
    \label{}
\end{align}
In the second line, we use Eq.(\ref{eq:3D_three_lambda_commutation}). In the third line, we have 
\begin{align}
    \eta_{c_0}''(g_1,g_2,g_3)=\xi_{c_0\cap c_1}(g_1,g_2,g_3)\cdot\prod_{p\in c_0\cap\{c\neq c_1\}}\xi_{c_0\cap c}''(g_1,g_2,g_3)
    \label{}
\end{align}
The last line is from the observation that $\lambda_{\slashed{c}_1}(g_1,g_2)\lambda_{\slashed{c}_1}(g_1g_2,g_3)=\eta_{c_1}\,{}^{W_{g_1}g_1}\!\lambda_{\slashed{c}_1}(g_2,g_3)\lambda_{\slashed{c}_1}(g_1,g_2g_3)$ for some $\eta_{c_1}$. And only $\xi_{c_0\cap c_1}$ has nontrivial action on some $\lambda_0$. This plaquette IGG element should equal to $\xi(g_1,g_2,g_3)$ up to some phase factor, due to the above derivation.

Now let us calculate the following expression
\begin{align}
    &\left[ \lambda_{\slashed{c}_1}(g_1,g_2)\lambda_{\slashed{c}_1}(g_1g_2,g_3)W_{g_1g_2g_3}g_1g_2g_3\circ\lambda_0(g_4,g_5) \right]\cdot \left[ \lambda_{\slashed{c}_1}(g_1,g_2)W_{g_1g_2}g_1g_2\circ\lambda_{0}(g_3,g_4g_5) \right]\notag\\
    =&\left[ \lambda_{\slashed{c}_1}(g_1,g_2)\circ\widetilde{\xi}(g_1g_2,g_3,g_4,g_5) \right]\cdot \left[ \lambda_{\slashed{c}_1}(g_1,g_2)W_{g_1g_2}g_1g_2W_{g_3}g_3\circ\lambda_0(g_4,g_5) \right]\cdot \left[ \lambda_{\slashed{c}_1}(g_1,g_2)W_{g_1g_2}g_1g_2\circ\lambda_0(g_3,g_4g_5) \right]\notag\\
    =&\left[ \lambda_{\slashed{c}_1}(g_1,g_2)\circ\widetilde{\xi}(g_1g_2,g_3,g_4,g_5) \right]\cdot \left[ \lambda_{\slashed{c}_1}(g_1,g_2)W_{g_1g_2}g_1g_2\circ(\eta'_0)^{-1}(g_3,g_4,g_5) \right]\cdot \left[ \lambda_{\slashed{c}_1}(g_1,g_2)W_{g_1g_2}g_1g_2\circ\lambda_0(g_3,g_4) \right]\cdot \notag\\
    &\left[ \lambda_{\slashed{c}_1}(g_1,g_2)W_{g_1g_2}g_1g_2\circ\lambda_0(g_3g_4,g_5) \right]\notag\\
    =&\left[ \lambda_{\slashed{c}_1}(g_1,g_2)\circ\widetilde{\xi}(g_1g_2,g_3,g_4,g_5) \right]\cdot \left[ \lambda_{\slashed{c}_1}(g_1,g_2)W_{g_1g_2}g_1g_2\circ(\eta'_0)^{-1}(g_3,g_4,g_5) \right]\cdot \left[ \widetilde{\xi}(g_1,g_2,g_3,g_4) \right]\cdot \left[ W_{g_1}g_1W_{g_2}g_2\circ\lambda_0(g_3,g_4) \right]\cdot \notag\\
    &\left[ \widetilde{\xi}(g_1,g_2,g_3g_4,g_5) \right]\cdot \left[ \,{}^{W_{g_1}g_1W_{g_2}g_2}\!\lambda_0(g_3g_4,g_5) \right]\notag\\
    =&\left[ \lambda_{\slashed{c}_1}(g_1,g_2)\circ\widetilde{\xi}(g_1g_2,g_3,g_4,g_5) \right]\cdot \left[ \lambda_{\slashed{c}_1}(g_1,g_2)W_{g_1g_2}g_1g_2\circ(\eta'_0)^{-1}(g_3,g_4,g_5) \right]\cdot \left[ \widetilde{\xi}(g_1,g_2,g_3,g_4) \right]\cdot \big[ \,{}^{W_{g_1}g_1W_{g_2}g_2}\!\lambda_0(g_3,g_4)\circ\notag\\
    &\widetilde{\xi}(g_1,g_2,g_3g_4,g_5) \big]\cdot \left[ W_{g_1}g_1W_{g_2}g_2\circ\eta_0'(g_3,g_4,g_5) \right]\cdot \left[ \,{}^{W_{g_1}g_1W_{g_2}g_2W_{g_3}g_3}\!\lambda_0(g_4,g_5) \right]\cdot \left[ \,{}^{W_{g_1}g_1W_{g_2}g_2}\!\lambda_0(g_3,g_4g_5) \right]\notag\\
    =&\left[ \lambda_{c_0}(g_1,g_2)\circ\widetilde{\xi}(g_1g_2,g_3,g_4,g_5) \right]\cdot \left[ \lambda_{c_0}(g_1,g_2)W_{g_1g_2}g_1g_2\circ\xi^{-1}(g_3,g_4,g_5) \right]\cdot \left[ \widetilde{\xi}(g_1,g_2,g_3,g_4) \right]\cdot \big[ \,{}^{W_{g_1}g_1W_{g_2}g_2}\!\lambda_0(g_3,g_4)\circ \notag\\
    &\widetilde{\xi}(g_1,g_2,g_3g_4,g_5) \big]\cdot \left[ W_{g_1}g_1W_{g_2}g_2\circ\xi(g_3,g_4,g_5) \right]\cdot \left[ \,{}^{W_{g_1}g_1W_{g_2}g_2W_{g_3}g_3}\!\lambda_0(g_4,g_5) \right]\cdot \left[ \,{}^{W_{g_1}g_1W_{g_2}g_2}\!\lambda_0(g_3,g_4g_5) \right]
    \label{}
\end{align}

We can calculate the above equation in another way as
\begin{align}
    &\left[ \lambda_{\slashed{c}_1}(g_1,g_2)\lambda_{\slashed{c}_1}(g_1g_2,g_3)W_{g_1g_2g_3}g_1g_2g_3\circ\lambda_0(g_4,g_5) \right]\cdot \left[ \lambda_{\slashed{c}_1}(g_1,g_2)W_{g_1g_2}g_1g_2\circ\lambda_{0}(g_3,g_4g_5) \right]\notag\\
    =&\left[ \xi(g_1,g_2,g_3)\,{}^{W_{g_1}g_1}\!\lambda_{\slashed{c}_1}(g_2,g_3)\lambda_{\slashed{c}_1}(g_1,g_2g_3)W_{g_1g_2g_3}g_1g_2g_3\circ\lambda_0(g_4,g_5) \right]\cdot \left[ \widetilde{\xi}(g_1,g_2,g_3,g_4g_5) \right]\cdot \left[ \,{}^{W_{g_1}g_1W_{g_2}g_2}\!\lambda_0(g_3,g_4g_5) \right]\notag\\
    =&\left[ \xi(g_1,g_2,g_3) \right]\cdot \left[ \,{}^{W_{g_1}g_1}\!\lambda_{\slashed{c}_1}(g_2,g_3)\circ\widetilde{\xi}(g_1,g_2g_3,g_4,g_5) \right]\cdot \left[ W_{g_1}g_1\lambda_{\slashed{c}_1}(g_2,g_3)W_{g_2g_3}g_2g_3\circ\lambda_0(g_4,g_5) \right]\cdot \left[ \xi^{-1}(g_1,g_2,g_3) \right]\cdot \notag\\
    &\left[ \widetilde{\xi}(g_1,g_2,g_3,g_4g_5) \right]\cdot \left[ \,{}^{W_{g_1}g_1W_{g_2}g_2}\!\lambda_0(g_3,g_4g_5) \right]\notag\\
    =&\left[ \xi(g_1,g_2,g_3) \right]\cdot \left[ \,{}^{W_{g_1}g_1}\!\lambda_{\slashed{c}_1}(g_2,g_3)\circ\widetilde{\xi}(g_1,g_2g_3,g_4,g_5) \right]\cdot \left[ \,{}^{W_{g_1}g_1}\widetilde{\xi}(g_2,g_3,g_4,g_5) \right]\cdot \left[ \,{}^{W_{g_1}g_1W_{g_2}g_2W_{g_3}g_3}\lambda_0(g_4,g_5) \right]\cdot \left[ \xi^{-1}(g_1,g_2,g_3) \right]\cdot \notag\\
    &\left[ \widetilde{\xi}(g_1,g_2,g_3,g_4g_5) \right]\cdot \left[ \,{}^{W_{g_1}g_1W_{g_2}g_2}\!\lambda_0(g_3,g_4g_5) \right]\notag\\
    =&\left[ \xi(g_1,g_2,g_3) \right]\cdot \left[ \,{}^{W_{g_1}g_1}\!\lambda_{c_0}(g_2,g_3)\circ\widetilde{\xi}(g_1,g_2g_3,g_4,g_5) \right]\cdot \left[ \,{}^{W_{g_1}g_1}\widetilde{\xi}(g_2,g_3,g_4,g_5) \right]\cdot \left[ \,{}^{W_{g_1}g_1W_{g_2}g_2W_{g_3}g_3}\lambda_0(g_4,g_5)\circ \xi^{-1}(g_1,g_2,g_3) \right]\cdot \notag\\
    &\left[ \,{}^{W_{g_1}g_1W_{g_2}g_2W_{g_3}g_3}\lambda_0(g_4,g_5)\circ \widetilde{\xi}(g_1,g_2,g_3,g_4g_5) \right]\cdot \left[ \,{}^{W_{g_1}g_1W_{g_2}g_2W_{g_3}g_3}\lambda_0(g_4,g_5) \right]\cdot \left[ \,{}^{W_{g_1}g_1W_{g_2}g_2}\!\lambda_0(g_3,g_4g_5) \right]
    \label{}
\end{align}

According to the above discussion, we prove that $\omega$ satisfies four cocycle condition:
\begin{align}
    \omega_p(g_1,g_2,g_3,g_4)\omega_p(g_1,g_2,g_3g_4,g_5)\omega_p(g_1g_2,g_3,g_4,g_5)=\,{}^{g_1}\!\omega_p(g_2,g_3,g_4,g_5)\omega_p(g_1,g_2g_3,g_4,g_5)\omega_p(g_1,g_2,g_3,g_4g_5)
    \label{}
\end{align}
Here the action of symmetry operator $g$ is similar as the case in 2+1D. Namely, for time reversal symmetry as well as reflection\,(inversion) symmetry, $g$ acts antiunitary on the $U(1)$ phase.

Further, since $\xi_p$'s are defined up to $U(1)$ phases $\chi_p$, one can show $\omega$'s are equivalent up to coboundary according to Eq.(\ref{eq:3D_twist_three_cocycle}):
\begin{align}
    \omega_p(g_1,g_2,g_3,g_4)\sim \omega_p(g_1,g_2,g_3,g_4)\frac{\chi_p(g_1,g_2,g_3)\cdot \chi_p(g_1,g_2g_3,g_4)\cdot \,{}^{g_1}\!\chi_p(g_2,g_3,g_4)}{\chi_p(g_1g_2,g_3,g_4)\cdot \chi_p(g_1,g_2,g_3g_4)}
    \label{}
\end{align}

\end{widetext}

\end{appendix}

\end{document}